\documentclass[11pt]{article}

\usepackage[margin=1in]{geometry}
\usepackage{amsmath,amssymb,amsfonts,bm,mathtools}
\usepackage{graphicx}
\usepackage[table]{xcolor}
\usepackage{booktabs}
\usepackage{array}
\usepackage{siunitx}
\usepackage{enumitem}
\usepackage{microtype}
\usepackage{placeins}
\usepackage{tikz}
\usetikzlibrary{arrows.meta,decorations.pathreplacing,calc,positioning}
\usepackage[numbers,sort&compress]{natbib}
\usepackage[colorlinks=true,linkcolor=blue!55!black,citecolor=blue!55!black,urlcolor=blue!55!black]{hyperref}

\newcommand{\given}{\,|\,}
\newcommand{\Pcal}{\mathcal P}
\newcommand{\VL}{V_L}
\newcommand{\CL}{C_L}
\newcommand{\DL}{D_L}
\newcommand{\mat}{\mathrm{mat}}

\newcommand{\E}{\mathbb{E}}
\newcommand{\Var}{\operatorname{Var}}
\newcommand{\Cov}{\operatorname{Cov}}
\newcommand{\Prob}{\mathbb{P}}
\newcommand{\ee}{\mathrm e}

\newcommand{\scal}{\mathrm{sc}}
\newcommand{\full}{\mathrm{full}}
\newcommand{\missingfigure}[1]{%
  \fbox{\parbox[c][0.42\textheight][c]{0.90\textwidth}{\centering
  \textbf{Figure file not supplied}\\[0.75em]
  \texttt{\detokenize{#1}}\\[0.75em]
  This placeholder keeps the manuscript compilable. Replace it with the original figure file before submission.}}}

\title{What sets the critical genome length for sympatric speciation? A closed form and asymptotic theory}
\author{\parbox{0.92\textwidth}{\centering
Dan Braha\textsuperscript{1,3} \quad Marcus A. M. de Aguiar\textsuperscript{1,2,*} \quad Vitor M. Marquioni\textsuperscript{4}\\[0.75em]
\small \textsuperscript{1}New England Complex Systems Institute, Cambridge, Massachusetts, United States of America\\
\small \textsuperscript{2}Instituto de F\'isica Gleb Wataghin, Universidade Estadual de Campinas,\\
\small Unicamp 13083-970, Campinas, SP, Brazil\\
\small \textsuperscript{3}University of Massachusetts Dartmouth, Dartmouth, Massachusetts, United States of America\\
\small \textsuperscript{4}Laboratoire de Chimie Bact\'erienne (UMR7283), Turing Centre for Living Systems,\\
\small IMM, IM2B, CNRS, Aix Marseille Universit\'e, Marseille, France\\
\small \textsuperscript{*}Corresponding author}}
\date{}

\begin{document}
\maketitle

\begin{abstract}
In the Derrida--Higgs model of sympatric speciation, a sexually reproducing population with finite binary genomes can mate only when the genetic overlap between two individuals exceeds a threshold \(q_{\min}\). Depending on the genome length \(L\), the population either remains genetically connected or fragments into reproductively isolated species. A central problem is therefore to predict the critical genome length \(L_c\) at which fragmentation begins. At finite \(L\), fluctuations broaden the overlap distribution and allow genetically distant regions of the population to remain connected. A previously proposed transient variance criterion captures this effect, but its evaluation requires numerical iteration of coupled moment equations. Here we first correct the unrestricted moment equations by removing a previously implicit assumption and then derive an explicit closed form expression for \(L_c\). The resulting formula shows that the critical genome length is determined, at the time the mean overlap reaches \(q_{\min}\), by the competition between deterministic separation from the unrestricted equilibrium and the transient genealogical variance of the overlap distribution. This expression permits a systematic asymptotic analysis. When the deterministic contribution dominates, \(L_c\) becomes essentially independent of the population size \(M\) and scales as \(\mu^{-2}\). When the transient genealogical variance dominates, \(L_c\) grows as \(M^{3/2}\) or as \(\sqrt{M}/\mu\), depending on how \(M\) and the mutation rate \(\mu\) jointly vary.  Simulations support all predicted behaviors. Our results identify transient genealogical variance as the principal mechanism linking finite genome fluctuations to the onset of reproductive fragmentation and provide a practical analytical prediction for the critical genome length.
\end{abstract}

\noindent\textbf{Keywords:} Derrida--Higgs model; sympatric speciation; finite size scaling; transient variance; genome length; overlap distribution.

\section{Introduction}
\label{sec:introduction}

A long-standing question in evolutionary biology is how sympatric
speciation can occur, with reproductive isolation arising while
populations remain in the same geographic region.  Because unrestricted mating continually restores gene flow, models of sympatric divergence require a mechanism that reduces reproduction among sufficiently different individuals.  Proposed mechanisms include assortative mating, temporal separation in reproduction, and ecological competition.  The theoretical plausibility, empirical frequency, and most appropriate definition of sympatric speciation have therefore remained subjects of sustained discussion \citep{Gavrilets2003,BolnickFitzpatrick2007,FitzpatrickEtAl2008,RichardsEtAl2019}.  Section~\ref{sec:biological-context} places the present work within this broader literature.

The Derrida--Higgs model provides a particularly transparent setting in which to study this problem.  Individuals carry binary genomes and mate only with partners whose genetic similarity exceeds a threshold \(q_{\min}\) \citep{HiggsDerrida1991,HiggsDerrida1992}.  In the homogeneous population model, where mating is unrestricted, the mean overlap between two genomes relaxes to a natural equilibrium value \(q_0\).  In the species formation model, mating is restricted by the minimum similarity \(q_{\min}\).  In the classical infinite genome limit, the population can fragment
into reproductively isolated groups when
\begin{equation}
q_{\min}>q_0 .
\label{eq:classical-condition}
\end{equation}
This condition has a direct interpretation: mutation and recombination drive the homogeneous population toward overlaps below the threshold, while the mating rule prevents sufficiently dissimilar individuals from producing offspring.

For finite genomes, however, Eq.~\eqref{eq:classical-condition} does not work as a criterion for  fragmentation  \citep{Aguiar2017,MarquioniAguiar2025}.  Even when the unrestricted mean overlap approaches the mating threshold, the distribution of pairwise overlaps can remain broad enough that many compatible pairs persist.  Those pairs can keep the compatibility graph connected and delay the formation of reproductively isolated groups.  A narrow distribution near the threshold has the opposite effect.  The finite genome problem is therefore controlled by the motion of the mean overlap together with the transient width of the overlap distribution.

Marquioni and de Aguiar \citep{MarquioniAguiar2025} developed a heuristic theory of this transition by deriving moment equations for the unrestricted Derrida--Higgs dynamics and using their values at the threshold crossing time to estimate the critical genome length \(L_c\).  That construction is effective, but it requires numerical iteration of a coupled system of difference equations and does not expose the dependence of \(L_c\) on the population size \(M\), mutation rate \(\mu\), and mating threshold \(q_{\min}\).  We refer to its central idea as the transient variance criterion: the unrestricted moments are followed until the mean overlap reaches the mating threshold, and the finite genome width at that time is used to estimate the onset of fragmentation.

The present paper develops this idea into an explicit analytical theory.  We first rederive the coupled moment equations without a simplifying assumption used previously.  From the corrected system and an exact decomposition of the finite genome overlap variance, we obtain a closed form matrix expression for \(L_c\).  The expression separates the deterministic distance between the mating threshold and the unrestricted equilibrium from the transient genealogical width of the overlap distribution.  A scalar reduction then permits a systematic asymptotic analysis.  The theory predicts that \(L_c\) becomes essentially independent of \(M\) at fixed positive \(\mu\), grows as \(M^{3/2}\) when \(M\mu\) remains of order one, grows as \(\sqrt{M}/\mu\) in an intermediate weak mutation range, and diverges approximately as \((q_{\min}-q_0)^{-1}\) near the infinite genome boundary.  Numerical calculations using the full moment hierarchy and direct stochastic simulations of the restricted species formation model test both the quantitative formula and these scaling predictions.

The remainder of the paper is organized as follows.  Section~\ref{sec:biological-context} reviews the biological and theoretical literature that leads to the finite genome question.  Section~\ref{sec:dh-models} defines the Derrida--Higgs models and the notation used throughout.  Section~\ref{sec:tv-criterion} formulates the transient variance criterion, derives the deterministic crossing time, and states the exact variance decomposition.  Section~\ref{sec:matrix-closed-form} solves the reduced matrix system and derives the matrix formula for the critical genome length.  Section~\ref{sec:scalar-reduction} develops the scalar approximation and its relation to the matrix result.  Section~\ref{sec:asymptotic-regimes} analyzes the limiting behavior of the formula.  Section~\ref{sec:numerical-tests} evaluates the theory against the full moment hierarchy and direct stochastic simulations.  Section~\ref{sec:discussion} discusses the implications, limitations, and possible extensions.

\section{Biological and theoretical background}
\label{sec:biological-context}

\subsection{Sympatric speciation and mechanisms that reduce gene flow}
\label{subsec:sympatric-background}

Sympatric speciation is the evolution of reproductive isolation between groups arising from a single ancestral population without geographic separation.  Its status has long been debated because recombination and mating within a shared habitat tend to oppose genetic divergence \citep{Gavrilets2003,KirkpatrickNuismer2004,BolnickFitzpatrick2007,FitzpatrickEtAl2008,RichardsEtAl2019}.  Mathematical models have nevertheless shown that reproductive isolation can arise when gene flow is reduced by assortative mating, disruptive ecological selection, mating preferences, or interactions among several evolving traits \citep{Udovic1980,Felsenstein1981,KondrashovMina1986,KondrashovShpak1998,KondrashovKondrashov1999,DieckmannDoebeli1999,Doebeli2005,Gavrilets2006}.  Empirical examples are more difficult to establish because the geographic history of a lineage and the timing of reproductive isolation are rarely observed directly \citep{CoyneOrr2004,Nosil2012,RichardsEtAl2019}.  The central theoretical issue is therefore not whether divergence is imaginable, but which mechanisms can reduce gene flow strongly enough, for long enough, to create distinct reproductive groups.

The Derrida--Higgs model addresses this issue through a minimal compatibility rule.  Two individuals can mate only when their genomes are sufficiently similar.  The rule is a direct form of assortative mating: it does not invoke ecological differences, spatial separation, or selection, but it can still reduce gene flow once the population becomes genetically heterogeneous.  This simplicity makes the model useful for separating the effects of mutation, recombination, genetic drift, population size, and genome length.

\subsection{From genotype frequency theories to finite population sequence models}
\label{subsec:frequency-to-finite-population}

Early theories of sequence evolution often described change through deterministic frequencies of all possible genotypes.  The quasispecies framework developed by Eigen and collaborators is a prominent example.  It studies mutation and selection in large populations and identifies error thresholds that separate localized and broadly distributed genotype frequencies \citep{EigenEtAl1989}.  Comparisons of sexual and asexual stationary states were also used to discuss the evolution of sexual reproduction.  Such arguments assume populations large enough for genotype frequencies to behave deterministically.  In population genetics, however, the number of possible sequences grows exponentially with genome length and can greatly exceed the number of individuals.  Sampling fluctuations, extinction of lineages, and random genealogy then become essential.

Derrida and Peliti introduced a finite population model with asexual reproduction in which binary genomes evolve through mutation and random ancestry \citep{DerridaPeliti1991}.  The model permitted exact results for genealogical statistics and genetic variability.  Higgs and Derrida extended this finite population sequence approach to sexual reproduction \citep{HiggsDerrida1991,HiggsDerrida1992}.  In the unrestricted sexual model, recombination and drift lead to a genetically homogeneous population whose mean pairwise similarity approaches an equilibrium value \(q_0\).  When mating is restricted to pairs with similarity at least \(q_{\min}\), the model can instead separate into groups that no longer interbreed.  In the infinite genome limit the change from one connected population to several reproductive groups is sharp, and the condition \(q_{\min}>q_0\) identifies the parameter region in which such reproductive fragmentation is possible.

The finite population formulation is important for two reasons.  First, it retains genealogical fluctuations that disappear from deterministic genotype frequency theories.  Second, it describes individuals by explicit genomes, so the number of loci can itself constrain the transition.  These features distinguish the Derrida--Higgs setting from classical frequency based models and make the critical genome length a natural quantity to study.

\subsection{Finite genomes, spatial structure, and alternative restrictions on mating}
\label{subsec:dh-extensions}

The infinite genome criterion does not automatically extend to finite genomes.  Direct simulations by \citet{Aguiar2017} showed that a finite number of loci can prevent fragmentation even when \(q_{\min}>q_0\), thereby creating a genuine genome length barrier to sympatric speciation.  When genomes are short, overlap fluctuations maintain links between otherwise distinct regions of the population, and the average similarity can remain close to the unrestricted value.  Additional restrictions on mating can then become important.

Spatial versions of the model restrict mating through local proximity as well as genetic similarity.  They have been used to study global diversity patterns and the structure of phylogenetic trees \citep{AguiarEtAl2009,CostaEtAl2019}.  Many early simulations used parameter values chosen mainly for computational tractability.  More recent work has shown that realistic dispersal can connect the model to speciation patterns in natural butterfly populations and can permit results from smaller simulated populations to be related to much larger biological populations \citep{Nelson2024}.  These studies demonstrate that geography and dispersal can alter the parameter ranges in which reproductive groups form.  The present paper instead focuses on the well mixed model so that the role of genome length can be isolated analytically.

Other variants reduce gene flow without imposing the same fixed similarity threshold.  \citet{CaetanoEtAl2020} considered individuals with a limited random mating pool, each of which chooses the most similar available partner.  Genetic clusters can form even though mating between clusters is never completely forbidden.  Geographic extensions of the Derrida--Higgs model have also considered two populations linked by migration.  The original infinite genome treatment showed that geographic separation makes speciation easier than in the sympatric model \citep{ManzoPeliti1994}.  Finite genome studies with continuous and intermittent migration again found that genome length strongly affects diversity and the timing of speciation \citep{PrincepeEtAl2022,PrincepeEtAl2024}.  In the asexual setting, Higgs and Woodcock studied Muller's ratchet, mutation accumulation, and genealogical trees under selection, including the distribution of times to the most recent common ancestor \citep{HiggsWoodcock1995}.  Together, these extensions show that the Derrida--Higgs framework is broad enough to separate several mechanisms that limit gene flow, but they also reinforce the need for a clear baseline theory of the well mixed finite genome model.

\subsection{Connectivity, adaptive landscapes, and related threshold models}
\label{subsec:connectivity-context}

The compatibility rule also connects the model to broader ideas about connectivity in genotype space.  Gavrilets formulated a theory of evolution on holey adaptive landscapes, where viable genotypes form connected regions of a high dimensional sequence space and reproductive success depends on genetic compatibility \citep{Gavrilets1999}.  The theory also shows how subdivision into many small populations can promote rapid divergence even when migration continues.  More recently, the structural properties of networks built from random bit strings have been studied directly, with links joining strings whose Hamming distance lies below a threshold \citep{SchneiderZanette2025}.  Degree distributions, clustering, assortativity, path lengths, and the existence of a giant connected component can all be analyzed in such networks.  The compatibility graph in the Derrida--Higgs model is dynamic rather than sampled once from a fixed distribution, but the same connectivity question is central: species formation corresponds to the loss of paths between groups of individuals.

Similarity thresholds also appear outside population genetics.  In bounded confidence models of opinion dynamics, agents interact only when their opinions are sufficiently close \citep{DeffuantEtAl2000,HegselmannKrause2005}.  Axelrod's model of cultural dissemination similarly makes interaction more likely when agents already share traits, producing local convergence together with global diversity \citep{Axelrod1997,BarbosaFontanari2009}.  A more direct mathematical connection between population genetics and opinion dynamics was established by \citet{BrahaAguiar2026}, who mapped a multiallelic Moran process with mutation to a multicandidate voter model and showed that the two models share stationary distributions and corresponding diversity thresholds under the appropriate parameter correspondence.  This exact mapping is different from the similarity threshold considered here, but it illustrates how a model from opinion dynamics can provide a useful representation of a population genetics process.  These connections do not imply that biological speciation and social influence are the same process.  They do show that a threshold on similarity can transform local mixing into large scale fragmentation, and that the location of this change depends on both average similarity and fluctuations around it.

\subsection{The analytical gap addressed in this paper}
\label{subsec:analytical-gap}

The literature establishes three points that motivate the present analysis.  First, the infinite genome Derrida--Higgs model has a clear condition for possible fragmentation.  Second, finite genomes can prevent that fragmentation, making genome length a central control parameter.  Third, spatial structure, restricted mating pools, and migration can change the outcome, but the well mixed finite genome model remains the natural reference case against which those extensions should be understood.

What has been missing is an explicit analytical prediction for the critical genome length.  The simulations of de Aguiar \citep{Aguiar2017} established the barrier numerically.  Marquioni and de Aguiar \citep{MarquioniAguiar2025} then introduced a transient variance criterion that relates the barrier to the overlap distribution of the unrestricted population.  Their criterion requires numerical iteration of coupled moment equations and therefore does not directly reveal how the critical length depends on the population size \(M\), mutation rate \(\mu\), and mating threshold \(q_{\min}\).  The goal of the present work is to close this gap.  We derive a closed form expression, identify the competing deterministic and genealogical contributions, and test the resulting predictions against both the full moment hierarchy and the restricted stochastic model.

\section{Derrida--Higgs models and the finite genome transition}
\label{sec:dh-models}

With the broader biological and theoretical context established, we now define the model and notation used in the analysis.  We begin with the common representation of genomes, overlaps, and mutation.  We then describe the one parent model, the unrestricted homogeneous population model, and the species formation model.  The purpose is to isolate the ingredients that enter the finite genome transition: relaxation of the mean overlap, the mating threshold \(q_{\min}\), and the finite genome width of the overlap distribution.

\subsection{Genomes, overlaps, and mutation}
\label{subsec:genomes-overlaps-mutation}

We consider a population of fixed size \(M\). At generation \(t\), each individual
\(\alpha\in\{1,\ldots,M\}\) carries a haploid biallelic genome of length \(L\),
\begin{equation}
S_t^\alpha=\bigl(s_{t,1}^\alpha,\ldots,s_{t,L}^\alpha\bigr),
\qquad 
s_{t,i}^\alpha\in\{-1,+1\}.
\label{eq:genome-definition}
\end{equation}
The spin notation is mathematically convenient, but it may also be interpreted biologically as a binary representation of allelic states at \(L\) loci. The genetic similarity between two individuals \(\alpha\) and \(\beta\) is measured by their normalized overlap
\begin{equation}
q_t^{\alpha\beta}
=
\frac{1}{L}\sum_{i=1}^L s_{t,i}^\alpha s_{t,i}^\beta .
\label{eq:pairwise-overlap}
\end{equation}
Thus \(q_t^{\alpha\beta}=1\) if the two genomes are identical, while smaller values indicate increasing genetic dissimilarity. In particular, the diagonal elements satisfy \(q_t^{\alpha\alpha}=1\) for every individual \(\alpha\). Equivalently, if \(d_t^{\alpha\beta}\) denotes the Hamming distance between the two genomes, then
\begin{equation}
d_t^{\alpha\beta}
=
\frac{L}{2}\bigl(1-q_t^{\alpha\beta}\bigr),
\qquad
q_t^{\alpha\beta}
=
1-\frac{2d_t^{\alpha\beta}}{L}.
\label{eq:hamming-overlap-relation}
\end{equation}
The overlap \(q_t^{\alpha\beta}\) is the basic state variable for the theory: the homogeneous population model is described by the evolution of the distribution of overlaps between distinct individuals, while the species formation model imposes a mating constraint by comparing \(q_t^{\alpha\beta}\) with a threshold \(q_{\min}\).

It is useful to collect the off diagonal pairwise overlaps, that is, the overlaps between distinct individuals, into the empirical overlap distribution
\begin{equation}
P_t(q)
=
\frac{2}{M(M-1)}
\sum_{1\leq \alpha<\beta\leq M}
\delta\!\left(q-q_t^{\alpha\beta}\right),
\label{eq:empirical-overlap-distribution}
\end{equation}
where \(\delta\) denotes the Dirac delta, understood here as a point mass in the empirical distribution. The diagonal self overlaps \(q_t^{\alpha\alpha}=1\) are not included. The corresponding mean overlap is
\begin{equation}
\bar q_t
=
\frac{2}{M(M-1)}
\sum_{1\leq \alpha<\beta\leq M}
q_t^{\alpha\beta},
\label{eq:mean-overlap}
\end{equation}
and the corresponding variance is
\begin{equation}
\sigma_{L,t}^2
=
\frac{2}{M(M-1)}
\sum_{1\leq \alpha<\beta\leq M}
\left(q_t^{\alpha\beta}-\bar q_t\right)^2 .
\label{eq:overlap-variance}
\end{equation}
Thus \(\bar q_t\) and \(\sigma_{L,t}^2\) describe the empirical distribution of overlaps between distinct individuals, not the full \(M\times M\) overlap matrix including its diagonal entries.

Strictly speaking, \(\bar q_t\) is also computed from genomes of length \(L\). We suppress this dependence in the notation because, in the unrestricted homogeneous population model, the expected evolution of the mean overlap has no explicit dependence on \(L\). By contrast, the width of the overlap distribution does depend explicitly on genome length, and we therefore write \(\sigma_{L,t}^2\). This finite genome width is the quantity used in the transient variance criterion of Marquioni and de Aguiar \citep{MarquioniAguiar2025} and in the closed form approximation derived below.

Following Marquioni and de Aguiar \citep{MarquioniAguiar2025}, the moment equations used below also involve higher order overlap quantities. In particular, for four individuals \(\alpha,\beta,\gamma,\delta\), define the four genome overlap
\begin{equation}
q_t^{\alpha\beta\gamma\delta}
=
\frac{1}{L}
\sum_{i=1}^L
s_{t,i}^{\alpha}
s_{t,i}^{\beta}
s_{t,i}^{\gamma}
s_{t,i}^{\delta}.
\label{eq:four-genome-overlap}
\end{equation}
This quantity appears naturally when deriving the second moment of the pairwise overlap distribution, because products of two pairwise similarities involve products of four spin variables at the same locus.

The model variants considered below differ in how parents are chosen, but they share the same basic mutation rule. After inheritance, each locus mutates independently. We write \(\mu\) for the mutation rate and
\begin{equation}
u=\frac{1-\ee^{-2\mu}}{2}
\label{eq:mutation-flip-probability}
\end{equation}
for the probability that a copied allele changes sign during one generation. Thus, if an offspring inherits allele \(s\in\{-1,+1\}\) before mutation, then after mutation the allele equals \(s\) with probability \(1-u\) and \(-s\) with probability \(u\). Equivalently,
\begin{equation}
\E[s_{\mathrm{after}}\mid s_{\mathrm{before}}]
=
(1-2u)s_{\mathrm{before}}
=
\ee^{-2\mu}s_{\mathrm{before}}.
\label{eq:mutation-expectation}
\end{equation}
For weak mutation, \(u\simeq \mu\). Since two independently inherited alleles each acquire a factor \(\ee^{-2\mu}\) in expectation, mutation contributes a factor \(\ee^{-4\mu}\) to the expected pairwise overlap from one generation to the next. This factor is the source of the exponential terms that appear in the homogeneous population recurrence relations below.

\subsection{The one parent model}
\label{subsec:one-parent-model}

The first Derrida--Higgs variant is the one parent model, an asexual neutral process in which each individual in generation \(t+1\) descends from a single individual in generation \(t\) \citep{DerridaPeliti1991,HiggsDerrida1991}. The model is not the species formation model studied below, because it contains neither sexual reproduction nor a mating threshold. It is nevertheless useful as a reference case: it shows how random ancestry and mutation alone generate structure in the overlap distribution, and it provides a natural contrast with the mixing effects of sexual reproduction in the homogeneous population model introduced below in Sec.~\ref{subsec:homogeneous-population-model}.

For each offspring \(\alpha\in\{1,\ldots,M\}\), let
\[
G_t(\alpha)\in\{1,\ldots,M\}
\]
denote its parent in the previous generation. In the one parent model, the parent \(G_t(\alpha)\) is chosen uniformly at random, independently for each offspring. Conditional on this parent, each locus is copied with mutation according to the mutation rule introduced in Sec.~\ref{subsec:genomes-overlaps-mutation}:
\begin{equation}
\Prob\!\left(
s_{t+1,i}^{\alpha}=s_{t,i}^{G_t(\alpha)}
\right)
=1-u
=
\frac{1+\ee^{-2\mu}}{2},
\qquad
\Prob\!\left(
s_{t+1,i}^{\alpha}=-s_{t,i}^{G_t(\alpha)}
\right)
=u
=
\frac{1-\ee^{-2\mu}}{2}.
\label{eq:opm-copying-rule}
\end{equation}
Thus the population size remains fixed, generations do not overlap, and all diversity is introduced through mutation.

The overlap dynamics have a simple form. Conditional on the parent choices and on the genomes in generation \(t\), the expected overlap between two distinct offspring \(\alpha\neq\beta\) is
\begin{equation}
\E\!\left[
q_{t+1}^{\alpha\beta}
\,\middle|\,
G_t(\alpha),G_t(\beta),\{S_t^\gamma\}_{\gamma=1}^M
\right]
=
\ee^{-4\mu}
q_t^{G_t(\alpha)G_t(\beta)} ,
\label{eq:opm-conditional-overlap}
\end{equation}
where \(q_t^{aa}=1\) for every \(a\in\{1,\ldots,M\}\). The factor \(\ee^{-4\mu}\) is the two lineage version of Eq.~\eqref{eq:mutation-expectation}: each copied allele contributes a factor \(\ee^{-2\mu}\) in expectation, so the product of two independently copied alleles contributes \(\ee^{-4\mu}\).

For finite \(L\), Eq.~\eqref{eq:opm-conditional-overlap} gives the conditional mean of the realized overlap, which is an average over \(L\) loci. The realized overlap fluctuates about this conditional mean with typical size of order \(L^{-1/2}\). In the infinite genome limit these fluctuations vanish, and the overlap matrix evolves deterministically according to
\begin{equation}
q_{t+1}^{\alpha\beta}
=
\ee^{-4\mu}
q_t^{G_t(\alpha)G_t(\beta)},
\qquad
\alpha\neq\beta,
\qquad
q_{t+1}^{\alpha\alpha}=1 .
\label{eq:opm-overlap-update}
\end{equation}

Averaging Eq.~\eqref{eq:opm-conditional-overlap} over the random parent choices gives the expected evolution of the empirical mean overlap. Since parents are drawn independently, $G_t(\alpha)$ and $G_t(\beta)$ coincide with probability \(1/M\), in which case their overlap is \(1\), and are distinct with probability \(1-1/M\), in which case their expected overlap is \(\bar q_t\). Hence

\begin{equation}
\E\!\left[
\bar q_{t+1}
\,\middle|\,
\{q_t^{ab}\}_{a,b=1}^M
\right]
=
\ee^{-4\mu}
\left[
\frac{1}{M}
+
\left(1-\frac{1}{M}\right)\bar q_t
\right].
\label{eq:opm-mean-overlap}
\end{equation}
Consequently, the fixed point of the mean overlap recursion satisfies
\begin{equation}
q_0
=
\frac{1}{M\ee^{4\mu}-(M-1)}
\simeq
\frac{1}{1+4M\mu},
\label{eq:opm-stationary-mean}
\end{equation}
where the approximation uses \(\ee^{4\mu}\simeq 1+4\mu\). The same characteristic overlap \(q_0\) will appear below in the homogeneous population model and in the infinite genome speciation criterion.

The important feature of the one parent model, however, is not only its mean overlap but the structure of the full overlap distribution. If two individuals share a common ancestor \(T\) generations in the past, their expected overlap is approximately \(\ee^{-4\mu T}\). Thus overlaps encode genealogical branching times. Because different branches of the population expand, contract, and disappear under genetic drift, the empirical distribution \(P_t(q)\) remains structured and not self averaging even in the large genome limit. This behavior contrasts with the homogeneous population model, where random sexual mixing concentrates the overlap distribution around \(q_0\). It is precisely this contrast between genealogical structure and sexual mixing that motivates the sequence of models reviewed here: the homogeneous population model provides the unrestricted baseline, while the species formation model reintroduces structure through the compatibility constraint \(q_t^{\alpha\beta}\ge q_{\min}\).

\subsection{The homogeneous population model}
\label{subsec:homogeneous-population-model}

The homogeneous population model is the unrestricted sexual counterpart of the one parent model \citep{HiggsDerrida1991,HiggsDerrida1992}. Each individual in generation \(t+1\) has two distinct parents in generation \(t\), and every pair of distinct individuals is eligible to mate. Thus there is sexual reproduction, but no compatibility constraint. This model will serve as the unrestricted reference case for the thresholded sexual model introduced in Sec.~\ref{subsec:species-formation-model}, where mating is allowed only between sufficiently similar genomes, \(q_t^{\alpha\beta}\ge q_{\min}\).

For each offspring \(\alpha\in\{1,\ldots,M\}\), let
\[
G_{1,t}(\alpha),G_{2,t}(\alpha)\in\{1,\ldots,M\},
\qquad
G_{1,t}(\alpha)\neq G_{2,t}(\alpha),
\]
denote its two parents in generation \(t\). The ordered pair
\((G_{1,t}(\alpha),G_{2,t}(\alpha))\) is chosen uniformly from all ordered pairs of distinct individuals in the previous generation, independently for each offspring. At each locus, the offspring inherits the allele from either parent with probability \(1/2\), followed by the mutation rule introduced in Sec.~\ref{subsec:genomes-overlaps-mutation}. Equivalently, if \(B_{t,i}(\alpha)\in\{1,2\}\) denotes the parental source used at locus \(i\), then
\[
\Prob\!\left(B_{t,i}(\alpha)=1\right)
=
\Prob\!\left(B_{t,i}(\alpha)=2\right)
=
\frac{1}{2},
\]
independently across loci and offspring, and the copied allele then mutates with flip probability \(u=(1-\ee^{-2\mu})/2\).

The expected overlap between two distinct offspring has a simple four parent form. Conditional on the parent choices and on the genomes in generation \(t\),
\begin{align}
&\E\!\left[
q_{t+1}^{\alpha\beta}
\,\middle|\,
G_{1,t}(\alpha),G_{2,t}(\alpha),
G_{1,t}(\beta),G_{2,t}(\beta),
\{S_t^\gamma\}_{\gamma=1}^M
\right]
\notag\\
&\qquad =
\frac{\ee^{-4\mu}}{4}
\sum_{r=1}^2\sum_{s=1}^2
q_t^{G_{r,t}(\alpha)G_{s,t}(\beta)},
\qquad \alpha\neq\beta .
\label{eq:hpm-conditional-overlap}
\end{align}
The average over the four terms appears because, at a given locus, offspring \(\alpha\) may inherit from either of its two parents and offspring \(\beta\) may inherit from either of its two parents. The factor \(\ee^{-4\mu}\) is again the two lineage mutation factor from Eq.~\eqref{eq:mutation-expectation}.

For finite \(L\), Eq.~\eqref{eq:hpm-conditional-overlap} gives the conditional mean of the realized overlap. The realized overlap still fluctuates about this value because it is an average over finitely many loci. In the infinite genome limit, these locus level fluctuations vanish. The overlap matrix then evolves, conditional on the parent choices, according to
\begin{equation}
q_{t+1}^{\alpha\beta}
=
\frac{\ee^{-4\mu}}{4}
\sum_{r=1}^2\sum_{s=1}^2
q_t^{G_{r,t}(\alpha)G_{s,t}(\beta)},
\qquad
\alpha\neq\beta,
\qquad
q_{t+1}^{\alpha\alpha}=1 .
\label{eq:hpm-overlap-update}
\end{equation}
This is the homogeneous population analogue of the one parent overlap update in Eq.~\eqref{eq:opm-overlap-update}.

Averaging Eq.~\eqref{eq:hpm-conditional-overlap} over the random parent choices gives the expected evolution of the empirical mean overlap. For a given locus in two distinct offspring, the two source parents are marginally independent uniform draws from the \(M\) individuals in generation \(t\). Therefore the two source labels coincide with probability \(1/M\), in which case their overlap is \(1\), and are distinct with probability \(1-1/M\), in which case their expected overlap is \(\bar q_t\). Hence
\begin{equation}
\E\!\left[
\bar q_{t+1}
\,\middle|\,
\{q_t^{ab}\}_{a,b=1}^M
\right]
=
\ee^{-4\mu}
\left[
\frac{1}{M}
+
\left(1-\frac{1}{M}\right)\bar q_t
\right].
\label{eq:hpm-mean-overlap-conditional}
\end{equation}
Thus the expected mean overlap obeys the same scalar recursion as in the one parent model. The difference between the two models is not the mean recursion itself, but the structure of the full overlap distribution: random sexual mixing in the homogeneous population model continually averages over parental lineages and concentrates the distribution around its mean.

Let
\begin{equation}
m_t=\E[\bar q_t]
\label{eq:hpm-mt-def}
\end{equation}
denote the expected off diagonal mean overlap in the homogeneous population model. Taking expectations in Eq.~\eqref{eq:hpm-mean-overlap-conditional} gives
\begin{equation}
m_{t+1}
=
\ee^{-4\mu}
\left[
\frac{1}{M}
+
\left(1-\frac{1}{M}\right)m_t
\right].
\label{eq:hpm-mean-overlap-recursion}
\end{equation}
It is useful to write
\begin{equation}
a=\ee^{-4\mu},
\qquad
r=a\left(1-\frac{1}{M}\right).
\label{eq:hpm-a-r-def}
\end{equation}
Then Eq.~\eqref{eq:hpm-mean-overlap-recursion} becomes
\begin{equation}
m_{t+1}
=
\frac{a}{M}+r m_t .
\label{eq:hpm-mean-overlap-recursion-ar}
\end{equation}

The stationary mean overlap \(q_0\) is the fixed point of this recursion. Setting \(m_{t+1}=m_t=q_0\) gives
\begin{equation}
q_0
=
\frac{1}{M\ee^{4\mu}-(M-1)}
\simeq
\frac{1}{1+4M\mu},
\label{eq:hpm-stationary-overlap}
\end{equation}
where the approximation uses \(\ee^{4\mu}\simeq 1+4\mu\). This value is the characteristic overlap of an unrestricted well mixed sexual population. In the thresholded model introduced next, \(q_0\) provides the natural infinite genome reference point: if the mating threshold \(q_{\min}\) lies below this homogeneous population overlap scale, the unrestricted population already remains sufficiently compatible, whereas if \(q_{\min}\) lies above it, the compatibility constraint can change the long term dynamics.

The transient solution of Eq.~\eqref{eq:hpm-mean-overlap-recursion-ar} will be used below in the finite genome criterion. Since \(q_0\) satisfies
\[
q_0=\frac{a}{M}+r q_0,
\]
subtracting this fixed point identity from Eq.~\eqref{eq:hpm-mean-overlap-recursion-ar} gives
\[
m_{t+1}-q_0=r(m_t-q_0).
\]
Therefore,
\begin{equation}
m_t
=
q_0+(m_0-q_0)r^t .
\label{eq:hpm-transient-mean-general}
\end{equation}
For the usual initial condition in which all genomes are identical, \(m_0=1\), and hence
\begin{equation}
m_t
=
q_0+(1-q_0)r^t .
\label{eq:hpm-transient-mean}
\end{equation}
This formula describes the relaxation of the expected unrestricted mean overlap from \(1\) toward \(q_0\). In the transient variance construction below, the relevant time scale is the first time at which this unrestricted mean reaches the mating threshold \(q_{\min}\).

The homogeneous population model therefore plays two roles in the analysis. Biologically, it describes a well mixed sexual population before reproductive compatibility constraints are imposed. Mathematically, it supplies the baseline trajectory \(m_t\), the stationary overlap \(q_0\), and the reference overlap distribution whose finite genome width is compared with the deterministic drift through the mating threshold.

\subsection{The species formation model}
\label{subsec:species-formation-model}

The species formation model is obtained from the homogeneous population model by adding a reproductive compatibility constraint. This thresholded sexual model is the central object motivating the finite genome problem studied in this paper. Individuals are still represented by spin genomes, reproduction is still sexual, and each offspring still inherits each locus from one of two parents followed by mutation. The difference is that not every pair of individuals is allowed to mate. Two distinct individuals are compatible at generation \(t\) only if their overlap is at least the threshold \(q_{\min}\):
\[
q_t^{\alpha\beta}\ge q_{\min}.
\]
Thus \(q_{\min}\) is the minimum genetic similarity required for successful reproduction.

The overlap threshold defines an instantaneous compatibility graph. Its adjacency matrix is
\begin{equation}
A_t^{\alpha\beta}
=
\begin{cases}
1, & \alpha\neq\beta \ \text{and}\ q_t^{\alpha\beta}\geq q_{\min},\\
0, & \text{otherwise}.
\end{cases}
\label{eq:compatibility-adjacency}
\end{equation}
The graph with adjacency matrix \(A_t\) records which pairs are allowed to mate. In the homogeneous population model this constraint is absent, so all distinct pairs are compatible. In the species formation model, by contrast, the loss of enough compatible edges can disconnect the graph into reproductively isolated components. The finite genome transition studied in this paper is the transition between these two behaviors as the genome length \(L\) is varied.

For each offspring \(\alpha\), the first parent \(G_{1,t}(\alpha)\) is chosen uniformly from the set of individuals with at least one compatible partner. Equivalently, one may choose a first parent uniformly from the whole population and redraw if it has no compatible partner. Conditional on the first parent \(G_{1,t}(\alpha)\), the second parent \(G_{2,t}(\alpha)\) is chosen uniformly from the individuals compatible with it, that is, from the individuals \(b\neq G_{1,t}(\alpha)\) satisfying
\[
q_t^{G_{1,t}(\alpha)b}\ge q_{\min}.
\]
Once a compatible pair has been selected, inheritance and mutation proceed as in the homogeneous population model: at each locus the offspring copies one of the two parental alleles with probability \(1/2\), and the copied allele then mutates with flip probability \(u=(1-\ee^{-2\mu})/2\).

Conditional on the parent choices and on the genomes in generation \(t\), the expected overlap between two distinct offspring has the same four parent form as in the homogeneous population model:
\begin{align}
&\E\!\left[
q_{t+1}^{\alpha\beta}
\,\middle|\,
G_{1,t}(\alpha),G_{2,t}(\alpha),
G_{1,t}(\beta),G_{2,t}(\beta),
\{S_t^\gamma\}_{\gamma=1}^M
\right]
\notag\\
&\qquad =
\frac{\ee^{-4\mu}}{4}
\sum_{r=1}^2\sum_{s=1}^2
q_t^{G_{r,t}(\alpha)G_{s,t}(\beta)},
\qquad \alpha\neq\beta .
\label{eq:sfm-conditional-overlap}
\end{align}
For finite \(L\), this is the conditional mean of the realized overlap. In the infinite genome limit, the locus level fluctuations vanish and the realized overlap matrix evolves, conditional on the sampled compatible parents, according to
\begin{equation}
q_{t+1}^{\alpha\beta}
=
\frac{\ee^{-4\mu}}{4}
\sum_{r=1}^2\sum_{s=1}^2
q_t^{G_{r,t}(\alpha)G_{s,t}(\beta)},
\qquad
\alpha\neq\beta,
\qquad
q_{t+1}^{\alpha\alpha}=1 .
\label{eq:sfm-overlap-update}
\end{equation}
The algebraic form of the overlap update is therefore the same as in Eq.~\eqref{eq:hpm-overlap-update}. The essential difference is that the parents are no longer sampled independently of the overlap matrix. They are sampled from the compatibility graph \(A_t\) defined in Eq.~\eqref{eq:compatibility-adjacency}. This dependence makes the species formation model nonlinear.

The graph interpretation is useful. If the compatibility graph \(A_t\) is connected, mating can in principle mix genetic material across the whole population. If the graph separates into components, mating is restricted within those components. These components are the model's reproductive clusters, or species. Within a component, reproduction resembles a homogeneous population model on a smaller effective population; between components, there is no direct mating, and their mutual overlaps tend to drift downward under mutation. This mechanism explains why the overlap distribution in the species formation model can develop multiple moving peaks rather than a single narrow peak: the peaks correspond to within species and between species overlap classes.

\subsection{The infinite genome criterion and the finite genome problem}
\label{subsec:infinite-genome-finite-genome-problem}

The homogeneous population model provides the natural reference point for the
species formation model. In the unrestricted model, random sexual mixing drives the
mean overlap toward the stationary value \(q_0\) in
Eq.~\eqref{eq:hpm-stationary-overlap}. In the infinite genome limit, the overlap distribution of the homogeneous population model is sharply concentrated about this value. The classical Derrida--Higgs intuition is therefore based only on the mean: if
\(q_{\min}<q_0\), typical individuals remain compatible and the mating constraint has
little effect; if \(q_{\min}>q_0\), typical pairs in the unrestricted population fall below
the mating threshold, so the compatibility constraint can fragment the population into
reproductively isolated species.

This criterion can be written as
\begin{equation}
q_{\min}>q_0
\label{eq:infinite-genome-criterion}
\end{equation}
or, equivalently, using Eq.~\eqref{eq:hpm-stationary-overlap},
\begin{equation}
q_{\min}
>
\frac{1}{M\ee^{4\mu}-(M-1)}
\simeq
\frac{1}{1+4M\mu}.
\label{eq:infinite-genome-criterion-explicit}
\end{equation}
This is an infinite genome condition because it treats the homogeneous population overlap distribution as concentrated at its mean.

For finite genomes, however, the transition cannot be determined from the mean overlap alone \citep{Aguiar2017,MarquioniAguiar2025}. Even when the homogeneous population mean lies below \(q_{\min}\), the overlap distribution at finite \(L\) has a nonzero width. Some pairs may therefore remain above the threshold and continue to
provide compatible mating edges \citep{Aguiar2017,MarquioniAguiar2025}. Conversely, when the width is too small, the thresholded compatibility graph loses enough edges
that reproductive connectivity can no longer be maintained.

The finite genome problem is therefore not simply to ask whether the mean overlap has crossed \(q_{\min}\). It is to ask whether the fluctuations at finite \(L\) about the homogeneous population mean are large enough to preserve a connected compatibility
graph. The remainder of the paper develops this idea by using the homogeneous population model as a reference process. We first identify the deterministic time at which the unrestricted mean overlap reaches the threshold \(q_{\min}\). We then estimate the finite genome variance of the overlap distribution at that crossing time and use it to derive a critical genome length \(L_c\). This critical length marks the point at which finite genome fluctuations are no longer large enough to maintain sufficient compatibility above the threshold.

\section{The transient variance criterion}
\label{sec:tv-criterion}

The preceding section explains why the transition in genomes of finite length cannot be predicted from the mean overlap alone.  The purpose of the present section is to turn that observation into a quantitative criterion.  We follow the central idea of Marquioni and de Aguiar~\citep{MarquioniAguiar2025}: use the unrestricted homogeneous population model as a reference process, follow its mean overlap until it reaches the mating threshold \(q_{\min}\), and evaluate the width of the unrestricted overlap distribution at that transient time. This width measures the spread of pairwise similarities near the threshold and therefore indicates how many compatible mating pairs may still remain when the deterministic mean has just reached \(q_{\min}\).

Our formulation separates three ingredients.  First, the deterministic trajectory of the homogeneous population model gives the time at which the threshold is reached.  Second, the overlap variance for genomes of finite length admits an exact decomposition into a genealogical contribution that remains in the infinite genome limit and a correction caused by finite genome length.  Third, the critical genome length is obtained by comparing two quantities defined below: the deterministic distance over which the mean passes through the threshold in one generation, and the extra width contributed by finite genome length.  Later sections provide explicit approximations for the variance terms that enter this criterion.  In this section, the variance decomposition and the algebra leading to the critical genome length condition are stated independently of those later approximations.

\subsection{Mean overlap and deterministic crossing time}
\label{subsec:mean-overlap-crossing-time}

The transient variance construction uses the unrestricted homogeneous population model as its reference dynamics.  We recall from Section~\ref{subsec:homogeneous-population-model} that
\(m_t=\E[\bar q_t]\) denotes the expected off diagonal mean overlap, and that the homogeneous population recursion can be written as
\begin{equation}
 m_{t+1}=\frac{a}{M}+r m_t,
 \qquad
 a=\ee^{-4\mu},
 \qquad
 r=a\left(1-\frac1M\right).
\label{eq:tv-mean-recursion}
\end{equation}
Its fixed point is the overlap \(q_0\) of the homogeneous population model in Eq.~\eqref{eq:hpm-stationary-overlap}.  For an initially clonal population, meaning that all genomes are identical at \(t=0\), Eq.~\eqref{eq:hpm-transient-mean} gives
\begin{equation}
 m_t=q_0+(1-q_0)r^t .
\label{eq:tv-mean-solution}
\end{equation}
Thus the unrestricted mean overlap decreases monotonically from \(m_0=1\) toward \(q_0\), because \(0<r<1\).

The transient variance criterion for the critical genome length below is relevant in the regime
\begin{equation}
q_{\min}>q_0,
\label{eq:tv-regime-qmin}
\end{equation}
where the unrestricted mean overlap eventually falls below the mating threshold.  We define the deterministic crossing time \(\tau\) by
\begin{equation}
m_\tau=q_{\min}.
\label{eq:tau-definition}
\end{equation}
Using Eq.~\eqref{eq:tv-mean-solution}, this gives
\begin{equation}
q_{\min}=q_0+(1-q_0)r^\tau,
\end{equation}
and therefore
\begin{equation}
r^\tau=\frac{q_{\min}-q_0}{1-q_0}.
\label{eq:r-tau}
\end{equation}
Taking logarithms yields
\begin{equation}
\tau
=
\frac{\log\!\left[(q_{\min}-q_0)/(1-q_0)\right]}{\log r}.
\label{eq:tau-formula}
\end{equation}
Both logarithms are negative when \(q_0<q_{\min}<1\), so \(\tau>0\).  In a simulation with discrete generations, one may use the first integer generation at which the unrestricted mean has crossed the threshold.  For the analytical formulas below, it is more convenient to use the value of \(\tau\) in Eq.~\eqref{eq:tau-formula}.

Marquioni and de Aguiar's criterion also requires the deterministic distance over which the unrestricted mean passes through the threshold in one generation.  This distance can be obtained directly from the same recursion.  Since \(m_\tau=q_{\min}\), the mean one generation earlier is
\begin{equation}
m_{\tau-1}
=
q_0+(1-q_0)r^{\tau-1}
=
q_0+\frac{q_{\min}-q_0}{r},
\label{eq:mean-one-step-before-crossing}
\end{equation}
where Eq.~\eqref{eq:r-tau} was used in the last equality.  Hence the deterministic drop in the unrestricted mean over the crossing generation is
\begin{align}
\delta q
&=m_{\tau-1}-m_\tau  \\
&=\left(q_0+\frac{q_{\min}-q_0}{r}\right)-q_{\min} \\
&=\frac{1-r}{r}\,(q_{\min}-q_0).
\label{eq:deltaq-first-form}
\end{align}
This is the same quantity obtained by asking how far above \(q_{\min}\) the mean can be one generation before crossing and still fall below the threshold in the next generation.  Using
\[
r=a\frac{M-1}{M},
\qquad
(1-r)q_0=\frac{a}{M},
\]
we have
\begin{equation}
\frac{1-r}{r}=\frac{1}{q_0(M-1)}.
\label{eq:one-minus-r-over-r}
\end{equation}
Therefore the deterministic crossing scale is
\begin{equation}
\delta q
=
\frac{q_{\min}-q_0}{q_0(M-1)}
=
\frac{q_{\min}/q_0-1}{M-1}.
\label{eq:delta-q-definition}
\end{equation}
This quantity is small when the threshold lies only slightly above the overlap \(q_0\) of the homogeneous population model, and it is also suppressed by population size.  It will be compared below with the additional overlap width caused by finite genome length.

\subsection{Exact variance decomposition for finite genomes}
\label{subsec:exact-variance-decomposition}

The second ingredient in the criterion is the variance of the unrestricted overlap distribution at the crossing time.  Let \(Q_L(t)\) denote the overlap of a randomly chosen pair of distinct individuals in the unrestricted homogeneous population model:
\begin{equation}
Q_L(t)=q_t^{\alpha\beta},
\qquad \alpha\neq \beta,
\label{eq:QL-def}
\end{equation}
where the pair is sampled uniformly from all distinct pairs and the randomness also includes the reproduction and mutation process.  Define
\begin{equation}
\sigma_L^2(t)=\Var(Q_L(t)).
\label{eq:sigmaL-variance-def}
\end{equation}
This variance is the population level counterpart of the empirical width obtained from the off diagonal overlaps in a single realization.  In simulations, one population realization gives an empirical variance over all distinct pairs; \(\sigma_L^2(t)\) is estimated by averaging this empirical quantity, or equivalently by repeatedly sampling a distinct pair, over independent runs.

Marquioni and de Aguiar showed that the overlap variance for genomes of finite length can be written in the form \(\Lambda_1(t)+\Lambda_2(t)/L\).  Here we give a direct derivation of this decomposition, which also provides a direct interpretation of the two coefficients.  We write the overlap of the selected pair as an average over loci,
\begin{equation}
Q_L(t)=\frac1L\sum_{i=1}^L X_i(t),
\qquad
X_i(t)=s_{t,i}^{\alpha}s_{t,i}^{\beta}.
\label{eq:QL-as-locus-average}
\end{equation}
Each \(X_i(t)\) takes the values \(\pm1\).  By exchangeability of loci,
\begin{equation}
\E[X_i(t)]=m_t,
\qquad
\Var(X_i(t))=1-m_t^2
\label{eq:Xi-mean-var}
\end{equation}
for every locus \(i\).  Similarly, the covariance between two distinct loci is the same for every pair \(i\neq j\).  We denote this common covariance between distinct loci by
\begin{equation}
v_\infty(t)=\Cov(X_i(t),X_j(t)),
\qquad i\neq j.
\label{eq:vinfty-as-locus-cov}
\end{equation}
The reason for the notation \(v_\infty(t)\) will become clear from the finite genome variance decomposition below.  Using Eq.~\eqref{eq:QL-as-locus-average},
\begin{align}
\sigma_L^2(t)
&=
\Var\left(\frac1L\sum_{i=1}^L X_i(t)\right) \\
&=
\frac{1}{L^2}
\left[
\sum_{i=1}^L \Var(X_i(t))
+
\sum_{\substack{i,j=1\\ i\neq j}}^L \Cov(X_i(t),X_j(t))
\right].
\label{eq:variance-average-expanded}
\end{align}
Substituting Eqs.~\eqref{eq:Xi-mean-var} and \eqref{eq:vinfty-as-locus-cov} gives
\begin{align}
\sigma_L^2(t)
&=
\frac{1}{L^2}
\left[
L(1-m_t^2)+L(L-1)v_\infty(t)
\right] \\
&=
v_\infty(t)
+
\frac{1-m_t^2-v_\infty(t)}{L}.
\label{eq:exact-variance-decomposition-main}
\end{align}
Equivalently, writing
\begin{equation}
v_1(t)=1-m_t^2-v_\infty(t),
\label{eq:v1-definition}
\end{equation}
we obtain
\begin{equation}
\sigma_L^2(t)=v_\infty(t)+\frac{v_1(t)}{L}.
\label{eq:exact-variance-decomposition}
\end{equation}
Equation~\eqref{eq:exact-variance-decomposition} explains the notation \(v_\infty(t)\).  Although \(v_\infty(t)\) enters the derivation as the covariance between the two locus level overlap variables \(X_i(t)\) and \(X_j(t)\) for distinct loci, it is also the variance that remains in the infinite genome limit:
\[
\lim_{L\to\infty}\sigma_L^2(t)=v_\infty(t).
\]
Thus \(v_\infty(t)\) should be read in two equivalent ways.  At the level of individual loci, it is the common covariance between distinct loci.  At the level of whole genome overlaps, it is the infinite genome genealogical variance, namely the width that remains because different sampled pairs have different genealogical relationships even after finite locus sampling noise has averaged away.

The relation in Eq.~\eqref{eq:exact-variance-decomposition} is useful because it shows that once the infinite genome genealogical variance \(v_\infty(t)\) is known, the coefficient \(v_1(t)\) of the \(1/L\) correction in that equation is fixed by Eq.~\eqref{eq:v1-definition}.  In the notation of Marquioni and de Aguiar~\citep{MarquioniAguiar2025}, Eq.~\eqref{eq:exact-variance-decomposition} corresponds to writing the overlap variance as
\[
\sigma_L^2(t)=\Lambda_1(t)+\frac{\Lambda_2(t)}{L},
\]
with \(\Lambda_1(t)=v_\infty(t)\) and \(\Lambda_2(t)=v_1(t)\).  The derivation here emphasizes the interpretation of \(v_\infty(t)\) and \(v_1(t)\), rather than treating them as independent coefficients obtained through moment recursions.

We note that the identity in Eq.~\eqref{eq:exact-variance-decomposition} is not a large \(L\) approximation.  It is an exact consequence of writing the overlap \(Q_L(t)\) in Eq.~\eqref{eq:QL-as-locus-average} as an average over exchangeable loci.  The approximation enters later, when \(v_\infty(t)\) and \(v_1(t)\) are evaluated using two approximation schemes.

\subsection{The critical genome length condition}
\label{subsec:critical-length-condition}

The transient variance criterion below compares the deterministic crossing scale \(\delta q\) in Eq.~\eqref{eq:delta-q-definition} with the extra width of the unrestricted overlap distribution caused by finite genome length at the crossing time \(\tau\)~\citep{MarquioniAguiar2025}.  More specifically, the total standard deviation at genome length \(L\) is
\[
\sqrt{\sigma_L^2(\tau)}
=
\sqrt{v_\infty(\tau)+\frac{v_1(\tau)}{L}},
\]
whereas the infinite genome standard deviation is \(\sqrt{v_\infty(\tau)}\).  We define the excess width as
\begin{equation}
\Delta_L
=
\sqrt{v_\infty(\tau)+\frac{v_1(\tau)}{L}}
-
\sqrt{v_\infty(\tau)}.
\label{eq:DeltaL-definition}
\end{equation}
Thus the finite locus correction \(v_1(\tau)/L\) is the additional variance in Eq.~\eqref{eq:exact-variance-decomposition}, whereas the excess width \(\Delta_L\) is the corresponding increase in standard deviation.  It decreases as \(L\) increases and vanishes in the infinite genome limit.

The critical genome length is obtained by imposing the equality
\begin{equation}
\Delta_L=\delta q.
\label{eq:critical-equality}
\end{equation}
The logic of this criterion is as follows.  The quantity \(\delta q\) is the deterministic overlap distance traversed by the unrestricted mean as it crosses the threshold in one generation.  The quantity \(\Delta_L\) is the extra overlap width supplied by finite genome length at that same transient time.  When \(\Delta_L>\delta q\), broadening due to finite genome length is still large enough to leave a substantial compatible tail above the threshold.  When \(\Delta_L<\delta q\), that extra broadening is too small to compensate for the deterministic crossing.  In other words, speciation becomes possible when the deterministic crossing scale is larger than the finite genome buffer.  Thus Eq.~\eqref{eq:critical-equality} marks the boundary at which the finite genome contribution just matches the deterministic crossing scale.  Along this critical boundary, the infinite genome limit is recovered when \(q_{\min}\to q_0\); in that limiting case both \(\Delta_L\) and \(\delta q\) vanish.

Substituting Eq.~\eqref{eq:DeltaL-definition} into Eq.~\eqref{eq:critical-equality} gives
\begin{equation}
\sqrt{v_\infty(\tau)+\frac{v_1(\tau)}{L}}
-
\sqrt{v_\infty(\tau)}
=
\delta q.
\label{eq:critical-equality-expanded}
\end{equation}
Solving for the critical genome length \(L_c\) gives
\begin{equation}
L_c
=
\frac{v_1(\tau)}{\delta q^2+2\delta q\sqrt{v_\infty(\tau)}}.
\label{eq:Lc-exact-decomposition}
\end{equation}
Using Eq.~\eqref{eq:v1-definition} at the crossing time, where \(m_\tau=q_{\min}\), the same condition may be written entirely in terms of the infinite genome genealogical variance at crossing:
\begin{equation}
L_c
=
\frac{1-q_{\min}^2-v_\infty(\tau)}{\delta q^2+2\delta q\sqrt{v_\infty(\tau)}}.
\label{eq:Lc-vinfty-form}
\end{equation}
Equations~\eqref{eq:Lc-exact-decomposition} and \eqref{eq:Lc-vinfty-form} are exact consequences of the variance decomposition and the transient variance crossing criterion in Eq.~\eqref{eq:critical-equality}.  They do not yet provide a closed numerical prediction, because \(v_\infty(\tau)\) must still be evaluated.  The next section addresses this point by deriving an explicit solution of a reduced two dimensional matrix approximation for the infinite genome genealogical variance \(v_\infty(\tau)\); the following section then derives a simpler scalar closed form approximation, which facilitates the asymptotic analysis in Section~\ref{sec:asymptotic-regimes}.

\section{Closed form solution of the reduced matrix system}
\label{sec:matrix-closed-form}

The transient variance criterion in Section~\ref{sec:tv-criterion} reduces the finite genome length transition to the critical length formula in Eq.~\eqref{eq:Lc-vinfty-form}.  That formula is explicit except for the infinite genome genealogical variance at the crossing time, \(v_\infty(\tau)\).  Following the variance decomposition in Eq.~\eqref{eq:exact-variance-decomposition}, the purpose of this section is to evaluate this quantity in closed form.  We do this by analyzing the finite \(M\), infinite genome limit of the coupled variance and covariance moment recursions for the unrestricted homogeneous population model.  These coupled recursions are derived in the Supplementary Material; they revise the genealogical variance and covariance coefficients originally reported by Marquioni and de Aguiar~\citep{MarquioniAguiar2025}.  In their exact form these recursions are driven by two deterministic inputs, the mean overlap and the four individual overlap, and couple three second moment quantities to one another: the variance of an off diagonal overlap, the covariance between two overlaps that share one individual, and the covariance between two overlaps that share no individual.  This three variable second moment system can be solved in closed form using the spectral decomposition method utilized in Section~\ref{subsec:eigenvalue-solution}, but the resulting expression is not very informative.  For this reason, and in order to preserve a transparent closed form, we retain the first two and drop the third, whose feedback into the variance is of order \(M^{-2}\); the resulting reduced two variable system is then solved exactly in closed form.  The result gives a closed form matrix expression for \(v_\infty(\tau)\), and therefore a closed form formula for the critical genome length \(L_c\).  We verify in Section~\ref{sec:numerical-tests} that this two variable reduction does not degrade the accuracy of the resulting critical genome length relative to the full three variable system.

This section also prepares the asymptotic analysis that follows.  The matrix formula is the finite population expression that keeps the covariance between overlapping pairwise similarities.  The next section reduces this matrix formula to a simpler scalar approximation that ignores the covariance between two pairwise overlaps, which is easier to interpret and is the starting point for the asymptotic regimes in Section~\ref{sec:asymptotic-regimes}.  Thus the role of the present section is twofold: it supplies the closed form critical genome length implied by the transient variance criterion, and it shows precisely what is retained, and later simplified, in the simpler scalar approximation presented in Section~\ref{sec:scalar-reduction}.

The calculation proceeds in three steps.  First, we introduce the three second moment quantities \(V_L\), \(C_L\), and \(D_L\) at finite \(L\), defined explicitly in Eq.~\eqref{eq:matrix-V-C-def}, and recall their corrected finite genome variance and covariance recursions, separating the finite locus terms from the terms that survive as \(L\to\infty\).  Second, we take the infinite genome limit, thereby defining the genealogical second moments \(v_t\), \(c_t\), and \(d_t\), and obtaining the exact three variable genealogical system at finite \(M\).  Here \(v_t\) is the same quantity denoted \(v_\infty(t)\) earlier in the paper, while \(c_t\) and \(d_t\) are, respectively, the covariances between two pairwise overlaps that share one individual and between two pairwise overlaps that share no individual.  Third, we reduce this three variable system to a two variable system for \(v_t\) and \(c_t\) by dropping \(d_t\), whose feedback into the variance equation enters only at order \(M^{-2}\), and then solve the reduced system in closed form using the spectral decomposition of its \(2\times2\) propagation matrix.  Identifying the resulting solution \(v_t\) with the infinite genome genealogical variance \(v_\infty(t)\), we evaluate it at the crossing time \(\tau\), where the deterministic mean overlap crosses the mating threshold \(q_{\min}\), and substitute \(v_\infty(\tau)\) into the transient variance criterion.  This yields the closed form formula for the critical genome length \(L_c\).

Throughout this section we recall the notation
\begin{equation}
 a=\ee^{-4\mu},
 \qquad
 r=a\left(1-\frac1M\right),
 \qquad
 m_t=q_0+(1-q_0)r^t,
\label{eq:matrix-parameters}
\end{equation}
where \(m_t\) is the mean overlap trajectory of the unrestricted homogeneous population model and \(q_0\) is its fixed point, as defined in Eq.~\eqref{eq:hpm-stationary-overlap}.  We also write
\begin{equation}
q_t^{\alpha\beta\gamma\delta}
=
\frac{1}{L}\sum_{i=1}^L
s_{t,i}^{\alpha}
s_{t,i}^{\beta}
s_{t,i}^{\gamma}
s_{t,i}^{\delta},
\qquad
h_t=\E\!\left[q_t^{\alpha\beta\gamma\delta}\right]
\label{eq:matrix-h-def}
\end{equation}
for the expected four individual overlap with all four labels distinct.  For a clonal initial condition, \(m_0=1\) and \(h_0=1\).

\subsection{From the full moment equations to a reduced two dimensional system}
\label{subsec:full-to-reduced-system}

In this subsection we show how the two variables \(v_t\) and \(c_t\) in the reduced system arise from the full system of coupled moment equations.  This full system involves three second moment quantities,
\begin{equation}
 V_L(t)=\Var\!\left(q_t^{\alpha\beta}\right),
 \qquad
 C_L(t)=\Cov\!\left(q_t^{\alpha\beta},q_t^{\alpha\gamma}\right),
 \qquad
 D_L(t)=\Cov\!\left(q_t^{\alpha\beta},q_t^{\gamma\delta}\right),
\label{eq:matrix-V-C-def}
\end{equation}
where \(\alpha,\beta,\gamma,\delta\) are distinct individuals sampled from the unrestricted homogeneous population.  The first quantity is the variance of a randomly sampled off diagonal overlap.  The second is the covariance between two overlaps that share one individual.  The third is the covariance between two overlaps that share no individual.  Both covariances appear because overlaps that involve common ancestry are correlated, and at finite \(M\) the variance equation couples to both of them.

For the clonal initial condition, all genomes are identical at \(t=0\).  Hence every pairwise overlap and every four individual overlap is initially equal to one, and the second moment quantities have zero initial variance and covariance:
\begin{equation}
m_0=1,\qquad h_0=1,\qquad
V_L(0)=C_L(0)=D_L(0)=0 .
\label{eq:matrix-full-initial-conditions}
\end{equation}

The moment equations for the unrestricted model were first written down by Marquioni and de Aguiar~\citep{MarquioniAguiar2025}.  Their reported genealogical variance and covariance coefficients contain errors; the corrected hierarchy, rederived in the Supplementary Material, couples the three quantities \(V_L\), \(C_L\), and \(D_L\).  Rewritten with population size \(M\), genome length \(L\), and \(a=\ee^{-4\mu}\), the corrected finite genome variance recursion is
\begin{align}
V_L(t+1)
&=\frac1L
-\frac{a^2}{4L}
\left[
\left(1+\frac{2}{M(M-1)}\right)
+
\left(2+\frac{4(M-2)}{M(M-1)}\right)m_t
+
\frac{(M-2)(M-3)}{M(M-1)}h_t
\right]
\notag\\
&\quad+
\frac{a^2(M-2)^2}{4M^2(M-1)}(1-m_t)^2
+
\frac{a^2(M^2-2M+2)}{4M(M-1)}V_L(t)
\notag\\
&\quad+
\frac{a^2(M-2)}{2(M-1)}C_L(t)
+
\frac{a^2(M-2)(M-3)}{4M(M-1)}D_L(t).
\label{eq:matrix-full-variance-recursion}
\end{align}
The corrected covariance recursion for two overlaps sharing one individual is
\begin{align}
C_L(t+1)
&=
\frac{a}{L}
\left[
\frac1M+\left(1-\frac1M\right)m_t
\right]
\notag\\
&\quad-
\frac{a^2}{2L}
\left[
\frac{2}{M^2}+\frac1M
+
\left(1+\frac4M-\frac8{M^2}\right)m_t
+
\left(1-\frac2M\right)\left(1-\frac3M\right)h_t
\right]
\notag\\
&\quad+
\frac{a^2}{2M}V_L(t)
+
\frac{a^2(M^2-4)}{2M^2}C_L(t)
+
\frac{a^2(M-2)(M-3)}{2M^2}D_L(t).
\label{eq:matrix-full-covariance-recursion}
\end{align}
Finally, the covariance recursion for two overlaps sharing no individual is purely genealogical,
\begin{equation}
D_L(t+1)
=
\frac{2a^2(M-1)}{M^3}V_L(t)
+
\frac{4a^2(M-1)(M-2)}{M^3}C_L(t)
+
\frac{a^2(M-1)(M-2)(M-3)}{M^3}D_L(t).
\label{eq:matrix-full-disjoint-recursion}
\end{equation}
The terms proportional to \(1/L\) are finite locus terms.  They arise because an overlap is an average over only \(L\) loci.  The remaining terms are genealogical terms.  They come from the random choice of parents and survive even when \(L\to\infty\).  The disjoint pair covariance \(D_L\) has no finite locus term at this order, so Eq.~\eqref{eq:matrix-full-disjoint-recursion} is genealogical at every \(L\).

Equations~\eqref{eq:matrix-full-variance-recursion}--\eqref{eq:matrix-full-disjoint-recursion} do not form a closed system on their own: the variance recursion and the one common individual covariance recursion are forced by the mean overlap \(m_t\) and by the four individual overlap \(h_t\) of Eq.~\eqref{eq:matrix-h-def}.  The mean overlap is given in closed form by Eq.~\eqref{eq:matrix-parameters}, the solution of the homogeneous population recursion~\eqref{eq:hpm-mean-overlap-recursion-ar} derived earlier.  The four individual overlap obeys its own recursion, derived in the Supplementary Material,
\begin{equation}
h_{t+1}
=
\frac{a^2}{M^3}
\left[
(3M-2)+(M-1)(6M-8)m_t+(M-1)(M-2)(M-3)h_t
\right].
\label{eq:matrix-h-recursion}
\end{equation}
These two quantities are deterministic inputs: they evolve independently of the variance and covariance, and enter Eqs.~\eqref{eq:matrix-full-variance-recursion}--\eqref{eq:matrix-full-disjoint-recursion} only as known forcing.  The quantities that are coupled to one another are therefore just the three second moments \((V_L,C_L,D_L)\); these three recursions, together with the two deterministic inputs~\eqref{eq:matrix-parameters} and~\eqref{eq:matrix-h-recursion} and the clonal initial conditions in Eq.~\eqref{eq:matrix-full-initial-conditions}, form the complete moment system.  We retain \(h_t\) here because it enters the finite locus terms of Eqs.~\eqref{eq:matrix-full-variance-recursion} and \eqref{eq:matrix-full-covariance-recursion}; as shown below, it drops out once the infinite genome limit is taken, leaving the three genealogical second moments \((v_t,c_t,d_t)\).

To extract the infinite genome genealogical variance \(v_\infty(t)\), define
\begin{equation}
 v_t\equiv v_\infty(t)=\lim_{L\to\infty}V_L(t),
 \qquad
 c_t=\lim_{L\to\infty}C_L(t),
 \qquad
 d_t=\lim_{L\to\infty}D_L(t).
\label{eq:matrix-v-c-infinite-def}
\end{equation}
The notation \(v_t\) is used in this section only to keep the recursion compact; it is the same infinite genome variance \(v_\infty(t)\) that appears in the transient variance criterion.  The finite \(L\) clonal initial conditions in Eq.~\eqref{eq:matrix-full-initial-conditions} imply the corresponding initial conditions for the infinite genome second moments
\begin{equation}
v_0=c_0=d_0=0.
\label{eq:matrix-infinite-initial-conditions}
\end{equation}
Letting \(L\to\infty\) in Eqs.~\eqref{eq:matrix-full-variance-recursion}--\eqref{eq:matrix-full-disjoint-recursion} removes all explicit finite locus contributions and gives the three variable genealogical system
\begin{align}
 v_{t+1}
&=
\frac{a^2(M-2)^2}{4M^2(M-1)}(1-m_t)^2
+
\frac{a^2(M^2-2M+2)}{4M(M-1)}v_t
+
\frac{a^2(M-2)}{2(M-1)}c_t
+
\frac{a^2(M-2)(M-3)}{4M(M-1)}d_t,
\label{eq:matrix-infinite-L-three-variance}
\\
 c_{t+1}
&=
\frac{a^2}{2M}v_t
+
\frac{a^2(M^2-4)}{2M^2}c_t
+
\frac{a^2(M-2)(M-3)}{2M^2}d_t,
\label{eq:matrix-infinite-L-three-covariance}
\\
 d_{t+1}
&=
\frac{2a^2(M-1)}{M^3}v_t
+
\frac{4a^2(M-1)(M-2)}{M^3}c_t
+
\frac{a^2(M-1)(M-2)(M-3)}{M^3}d_t.
\label{eq:matrix-infinite-L-three-disjoint}
\end{align}
The four overlap \(h_t\), which is needed in the full finite \(L\) recursions, has disappeared because it enters only through finite locus terms.  Therefore the infinite genome genealogical second moments are governed exactly, at finite \(M\), by the triple \((v_t,c_t,d_t)\).

We now reduce this exact three variable system to a two variable system by dropping \(d_t\).  This reduction should not be confused with the two variable closure used by Marquioni and de Aguiar~\citep{MarquioniAguiar2025}.  Once the disjoint pair covariance \(d_t\) is included in the derivation, the finite \(M\) coefficients multiplying \(v_t\) and \(c_t\) are themselves changed.  The two variable system solved below is therefore the reduction of the full moment hierarchy, not a closure obtained by appending or removing a single term.  Starting from a clonal population, \(v_0=c_0=d_0=0\), and only the variance equation carries the source term \(\propto(1-m_t)^2\).  In the large population regimes used below this source is of order \(M^{-1}\), so \(v_t=O(M^{-1})\).  Equation~\eqref{eq:matrix-infinite-L-three-covariance} then generates \(c_t\) through the term \((a^2/2M)v_t=O(M^{-2})\), while its self coupling \(a^2(M^2-4)/(2M^2)\to a^2/2<1\) is a contraction, so \(c_t=O(M^{-2})\).  Equation~\eqref{eq:matrix-infinite-L-three-disjoint} generates \(d_t\) from \(v_t\) and \(c_t\) through coefficients of order \(M^{-2}\) and \(M^{-1}\), and its self coupling approaches \(a^2<1\) from below with a gap of order \(M^{-1}\); the resulting steady value is \(d_t=O(M^{-2})\).  The feedback of \(d_t\) into the variance equation through the coefficient \(a^2(M-2)(M-3)/(4M(M-1))=O(1)\) is therefore of order \(M^{-2}\), one order below the leading variance.  Dropping \(d_t\) thus changes \(v_t\), and hence the critical length, only at order \(M^{-2}\); numerically the resulting shift in \(L_c\) is below one percent for \(M\gtrsim50\) and negligible for larger populations.  Setting \(d_t=0\) in Eqs.~\eqref{eq:matrix-infinite-L-three-variance} and \eqref{eq:matrix-infinite-L-three-covariance} gives the reduced two variable system
\begin{align}
 v_{t+1}
&=
\frac{a^2(M-2)^2}{4M^2(M-1)}(1-m_t)^2
+
\frac{a^2(M^2-2M+2)}{4M(M-1)}v_t
+
\frac{a^2(M-2)}{2(M-1)}c_t,
\label{eq:matrix-infinite-L-exact-M-variance}
\\
 c_{t+1}
&=
\frac{a^2}{2M}v_t
+
\frac{a^2(M^2-4)}{2M^2}c_t.
\label{eq:matrix-infinite-L-exact-M-covariance}
\end{align}
These two equations are the finite \(M\), infinite genome genealogical recursions used in the remainder of this section.  They approximate the exact three variable system to order \(M^{-2}\) and are solved exactly in closed form below.

\subsection{The coupled variance and covariance recursion}
\label{subsec:coupled-variance-covariance-recursion}

Below we rewrite Eqs.~\eqref{eq:matrix-infinite-L-exact-M-variance} and \eqref{eq:matrix-infinite-L-exact-M-covariance} in a form that leads directly to the matrix solution.  It is convenient to name the source coefficient
\begin{equation}
A_M=
\frac{a^2(M-2)^2}{4M^2(M-1)},
\label{eq:matrix-AM-BM-def}
\end{equation}
which multiplies the additive variance source \((1-m_t)^2\).  With this notation the reduced genealogical system reads
\begin{align}
 v_{t+1}
&=A_M(1-m_t)^2
+
\frac{a^2(M^2-2M+2)}{4M(M-1)}v_t
+
\frac{a^2(M-2)}{2(M-1)}c_t,
\label{eq:matrix-coupled-v-recursion}
\\
 c_{t+1}
&=
\frac{a^2}{2M}v_t
+
\frac{a^2(M^2-4)}{2M^2}c_t.
\label{eq:matrix-coupled-c-recursion}
\end{align}
This is the coupled variance and covariance recursion underlying the matrix formula.  The additive term \(A_M(1-m_t)^2\) creates new genealogical variance in Eq.~\eqref{eq:matrix-coupled-v-recursion}.  The other terms carry forward the variance and covariance already present in the population.  Equation~\eqref{eq:matrix-coupled-c-recursion} shows why the covariance cannot be ignored: \(c_t\) is obtained from \(v_t\), and then feeds back into the next value of \(v_t\) through Eq.~\eqref{eq:matrix-coupled-v-recursion}.

The additive term is explicit because the mean overlap trajectory is explicit.  From Eq.~\eqref{eq:matrix-parameters},
\begin{equation}
1-m_t=(1-q_0)(1-r^t),
\end{equation}
and therefore
\begin{equation}
(1-m_t)^2=(1-q_0)^2(1-2r^t+r^{2t}).
\label{eq:matrix-source-shape}
\end{equation}
This identity is the key algebraic simplification: the additive part of the two dimensional recursion is a linear combination of the three scalar sequences \(1\), \(r^t\), and \(r^{2t}\).  The matrix solution below is obtained by summing the response of the coupled system to each of these three sequences.

\subsection{Matrix summation form}
\label{subsec:matrix-summation-form}

In this subsection we write the coupled recursion for the finite \(M\), infinite genome limit as a finite matrix sum.  Define
\begin{equation}
\bm z_t=\begin{pmatrix}v_t\\ c_t\end{pmatrix},
\qquad
\bm e_1=\begin{pmatrix}1\\0\end{pmatrix}.
\label{eq:matrix-z-e1-def}
\end{equation}
Equations~\eqref{eq:matrix-coupled-v-recursion} and \eqref{eq:matrix-coupled-c-recursion} can be written as
\begin{equation}
\bm z_{t+1}=K\bm z_t+D\bm e_1(1-2r^t+r^{2t}),
\label{eq:matrix-linear-recursion}
\end{equation}
where
\begin{equation}
K=
\begin{pmatrix}
\dfrac{a^2(M^2-2M+2)}{4M(M-1)}
&
\dfrac{a^2(M-2)}{2(M-1)}
\\[1em]
\dfrac{a^2}{2M}
&
\dfrac{a^2(M^2-4)}{2M^2}
\end{pmatrix}
\label{eq:matrix-K-def}
\end{equation}
and
\begin{equation}
D=A_M(1-q_0)^2
=
\frac{a^2(M-2)^2(1-q_0)^2}{4M^2(M-1)}.
\label{eq:matrix-D-def}
\end{equation}
For a clonal initial population, \(\bm z_0=\bm 0\).  Iterating Eq.~\eqref{eq:matrix-linear-recursion} shows how contributions from earlier generations accumulate.  A contribution created at generation \(s\) is multiplied by \(K\) once for each subsequent generation, so by the time it reaches generation \(t\) it has been propagated by \(K^{t-1-s}\).  Therefore
\begin{equation}
\bm z_t
=
D\sum_{s=0}^{t-1}K^{t-1-s}\bm e_1(1-2r^s+r^{2s}).
\label{eq:matrix-sum-solution}
\end{equation}
The first component of \(\bm z_t\) is the matrix expression for the infinite genome genealogical variance:
\begin{equation}
 v_\infty^{\mat}(t)=\bm e_1^{\top}\bm z_t.
\label{eq:matrix-vinfty-component}
\end{equation}

We can write Eq.~\eqref{eq:matrix-sum-solution} in a form that will simplify the analysis below.  For any scalar \(\eta\), define
\begin{equation}
\mathcal S_\eta(t)=\sum_{s=0}^{t-1}K^{t-1-s}\eta^s,
\label{eq:matrix-Seta-def}
\end{equation}
where the scalar \(\eta\) represents one of the three choices \(\eta=1\), \(\eta=r\), and \(\eta=r^2\), corresponding respectively to the three pieces \(1\), \(r^s\), and \(r^{2s}\) in Eq.~\eqref{eq:matrix-sum-solution}.  With this notation,
\begin{equation}
\bm z_t
=
D\left[\mathcal S_1(t)-2\mathcal S_r(t)+\mathcal S_{r^2}(t)\right]\bm e_1,
\label{eq:matrix-z-Seta}
\end{equation}
and hence, by Eq.~\eqref{eq:matrix-vinfty-component},
\begin{equation}
 v_\infty^{\mat}(t)
=
D\bm e_1^{\top}
\left[\mathcal S_1(t)-2\mathcal S_r(t)+\mathcal S_{r^2}(t)\right]
\bm e_1.
\label{eq:matrix-vinfty-Seta}
\end{equation}
The sum \(\mathcal S_\eta(t)\) in Eq.~\eqref{eq:matrix-Seta-def} is the matrix analogue of a finite geometric sum.  The next subsection evaluates it by diagonalizing \(K\).

\subsection{Eigenvalue solution of the matrix sum}
\label{subsec:eigenvalue-solution}

The remaining task is to evaluate the matrix powers in Eq.~\eqref{eq:matrix-sum-solution}.  We use spectral decomposition for this purpose.  Since \(K\) is a \(2\times2\) matrix with two distinct eigenvalues \(\lambda_+\) and \(\lambda_-\) for \(M\geq 3\), every power of \(K\) can be written as the sum of two contributions, one proportional to \(\lambda_+^n\) and the other proportional to \(\lambda_-^n\).  This converts the matrix expression in Eq.~\eqref{eq:matrix-vinfty-Seta} into two ordinary scalar sums.

Write
\begin{equation}
K=\begin{pmatrix}k_{11}&k_{12}\\ k_{21}&k_{22}\end{pmatrix},
\end{equation}
where
\begin{align}
k_{11}&=\frac{a^2(M^2-2M+2)}{4M(M-1)},
&
k_{12}&=\frac{a^2(M-2)}{2(M-1)},
\notag\\
k_{21}&=\frac{a^2}{2M},
&
k_{22}&=\frac{a^2(M^2-4)}{2M^2}.
\label{eq:matrix-K-entries}
\end{align}
The eigenvalues are the two roots of \(\lambda^2-(\operatorname{tr}K)\lambda+\det K=0\).  From the entries above,
\begin{align}
\operatorname{tr}K
&=
\frac{a^2(3M-4)(M^2-2)}{4M^2(M-1)},
\label{eq:matrix-K-trace}
\\
\det K
&=
\frac{a^4(M-2)^2(M^2-2)}{8M^3(M-1)}.
\label{eq:matrix-K-det}
\end{align}
The discriminant is
\begin{equation}
(\operatorname{tr}K)^2-4\det K
=
\frac{a^4(M^2-2)}{16M^4(M-1)^2}
\left(M^4+16M^3-66M^2+80M-32\right).
\label{eq:matrix-K-discriminant}
\end{equation}
Define
\begin{equation}
R_M=
\sqrt{(M^2-2)\left(M^4+16M^3-66M^2+80M-32\right)}.
\label{eq:matrix-RM-def}
\end{equation}
Then the two eigenvalues are
\begin{equation}
\lambda_\pm
=
\frac{a^2}{8M^2(M-1)}
\left[(3M-4)(M^2-2)\pm R_M\right].
\label{eq:matrix-eigenvalues}
\end{equation}
This eigenvalue formula follows directly by solving the quadratic equation determined by the trace and determinant of \(K\).  For \(M\geq 3\), the two eigenvalues are distinct because \(R_M>0\) and hence \(\lambda_+-\lambda_->0\).

Define the spectral projectors
\begin{equation}
P_+=\frac{K-\lambda_- I}{\lambda_+-\lambda_-},
\qquad
P_- =\frac{\lambda_+ I-K}{\lambda_+-\lambda_-}.
\label{eq:matrix-projectors}
\end{equation}
These projectors are an algebraic way to separate the two eigenvalue contributions without writing the eigenvectors explicitly.  They satisfy \(P_++P_-=I\), and they imply
\begin{equation}
K^n=\lambda_+^nP_+ + \lambda_-^nP_-.
\label{eq:matrix-power-spectral}
\end{equation}
This identity is the reason for introducing the projectors: it separates every matrix power in Eq.~\eqref{eq:matrix-sum-solution} into the two scalar powers \(\lambda_+^n\) and \(\lambda_-^n\).

We now apply Eq.~\eqref{eq:matrix-power-spectral} directly to the finite matrix sum.  Substituting it into Eq.~\eqref{eq:matrix-Seta-def} gives
\begin{align}
\mathcal S_\eta(t)
&=
\sum_{s=0}^{t-1}
\left(\lambda_+^{t-1-s}P_+ + \lambda_-^{t-1-s}P_-\right)\eta^s
\notag\\
&=
G_t(\lambda_+,\eta)P_+
+
G_t(\lambda_-,\eta)P_-,
\label{eq:matrix-Seta-spectral}
\end{align}
where
\begin{equation}
G_t(\lambda,\eta)=\sum_{s=0}^{t-1}\lambda^{t-1-s}\eta^s.
\label{eq:matrix-G-def}
\end{equation}
The scalar sum \(G_t\) is an ordinary finite geometric sum:
\[
G_t(\lambda,\eta)
=
\lambda^{t-1}+\lambda^{t-2}\eta+
\lambda^{t-3}\eta^2+
\cdots+\eta^{t-1}.
\]
Multiplying this expression by \(\lambda-\eta\) cancels all intermediate terms, leaving \(\lambda^t-\eta^t\).  Hence, for \(\lambda\neq\eta\),
\begin{equation}
G_t(\lambda,\eta)=\frac{\lambda^t-\eta^t}{\lambda-\eta}.
\label{eq:matrix-G-closed}
\end{equation}
If \(\lambda=\eta\), the finite sum in Eq.~\eqref{eq:matrix-G-def} gives the limiting value \(G_t(\lambda,\lambda)=t\lambda^{t-1}\).  Thus the quotient in Eq.~\eqref{eq:matrix-G-closed} is only a compact way of writing the finite sum; the original sum remains well defined.

Next combine the three pieces that appear in Eq.~\eqref{eq:matrix-vinfty-Seta}.  Using Eq.~\eqref{eq:matrix-Seta-spectral}, the coefficient of \(P_+\) is
\[
G_t(\lambda_+,1)-2G_t(\lambda_+,r)+G_t(\lambda_+,r^2),
\]
and the coefficient of \(P_-\) is the same expression with \(\lambda_-\) in place of \(\lambda_+\).  Define the scalar response function
\begin{equation}
F_t(\lambda)
=
G_t(\lambda,1)-2G_t(\lambda,r)+G_t(\lambda,r^2).
\label{eq:matrix-F-def}
\end{equation}
Using Eq.~\eqref{eq:matrix-G-closed}, this can be written as
\begin{equation}
F_t(\lambda)
=
\frac{\lambda^t-1}{\lambda-1}
-
2\frac{\lambda^t-r^t}{\lambda-r}
+
\frac{\lambda^t-r^{2t}}{\lambda-r^2},
\label{eq:matrix-F-closed}
\end{equation}
with the corresponding finite sum interpretation whenever a denominator vanishes.  Therefore
\begin{equation}
\mathcal S_1(t)-2\mathcal S_r(t)+\mathcal S_{r^2}(t)
=
F_t(\lambda_+)P_+ + F_t(\lambda_-)P_-.
\label{eq:matrix-spectral-response}
\end{equation}

Finally, substitute Eq.~\eqref{eq:matrix-spectral-response} into Eq.~\eqref{eq:matrix-vinfty-Seta} to extract the variance component:
\begin{align}
 v_\infty^{\mat}(t)
&=
D\bm e_1^\top
\left[F_t(\lambda_+)P_+ + F_t(\lambda_-)P_-\right]
\bm e_1
\notag\\
&=
D\left[W_+F_t(\lambda_+)+W_-F_t(\lambda_-)\right],
\label{eq:matrix-vinfty-closed-t}
\end{align}
where \(F_t(\lambda_+)\) and \(F_t(\lambda_-)\) are defined by Eq.~\eqref{eq:matrix-F-def}, and
\begin{equation}
W_+=\bm e_1^\top P_+\bm e_1,
\qquad
W_- =\bm e_1^\top P_-\bm e_1.
\label{eq:matrix-weights-def}
\end{equation}
Using the projector formulas in Eq.~\eqref{eq:matrix-projectors},
\begin{equation}
W_+=\frac{k_{11}-\lambda_-}{\lambda_+-\lambda_-},
\qquad
W_- =\frac{\lambda_+-k_{11}}{\lambda_+-\lambda_-}.
\label{eq:matrix-weights-k11}
\end{equation}
Substituting \(k_{11}=a^2(M^2-2M+2)/(4M(M-1))\) and the eigenvalues in Eq.~\eqref{eq:matrix-eigenvalues} gives
\begin{equation}
W_+
=
\frac{R_M-(M^3-10M+8)}{2R_M},
\qquad
W_-
=
\frac{R_M+(M^3-10M+8)}{2R_M}.
\label{eq:matrix-projector-weights}
\end{equation}
These weights add to one because the projectors sum to the identity.

\subsection{Closed form matrix expression for \texorpdfstring{\(v_\infty(t)\)}{v infinity(t)} and \texorpdfstring{\(v_1(t)\)}{v1(t)}}
\label{subsec:matrix-vinfty}

The calculation above yields the closed form
\begin{equation}
 v_\infty^{\mat}(t)
=
D\left[
W_+F_t(\lambda_+)
+
W_-F_t(\lambda_-)
\right].
\label{eq:matrix-vinfty-closed-summary}
\end{equation}
Equivalently, substituting \(D\) and the weights explicitly,
\begin{equation}
\begin{split}
 v_\infty^{\mat}(t)
=
\frac{a^2(M-2)^2(1-q_0)^2}{4M^2(M-1)}
\Bigg[
&\frac{R_M-(M^3-10M+8)}{2R_M}F_t(\lambda_+)
\\
&+
\frac{R_M+(M^3-10M+8)}{2R_M}F_t(\lambda_-)
\Bigg].
\end{split}
\label{eq:matrix-vinfty-closed-expanded}
\end{equation}
This is the closed form evaluation of Eq.~\eqref{eq:matrix-vinfty-Seta}; the sums contained in \(\mathcal S_\eta(t)\) have been evaluated through the two eigenvalues of \(K\).  No expansion in \(1/M\) has been made in deriving it.  It is the exact closed form solution of the reduced two variable variance and covariance system in Eqs.~\eqref{eq:matrix-infinite-L-exact-M-variance} and \eqref{eq:matrix-infinite-L-exact-M-covariance}, which itself approximates the full finite \(M\), infinite genome three variable genealogical system in Eqs.~\eqref{eq:matrix-infinite-L-three-variance}--\eqref{eq:matrix-infinite-L-three-disjoint} to order \(M^{-2}\).  Retaining \(c_t\) changes the propagation matrix \(K\), and therefore changes both the eigenvalues \(\lambda_\pm\) and the weights \(W_\pm\) that determine how genealogical variance is carried through time.

The corresponding finite genome correction coefficient \(v_1^{\mat}(t)\) follows from the exact variance decomposition in Section~\ref{subsec:exact-variance-decomposition}:
\begin{equation}
\sigma_L^2(t)=v_\infty(t)+\frac{v_1(t)}{L},
\qquad
v_1(t)=1-m_t^2-v_\infty(t).
\label{eq:matrix-v1-from-decomposition}
\end{equation}
That is, once the matrix formula supplies \(v_\infty^{\mat}(t)\), the associated finite genome correction coefficient is fixed:
\begin{equation}
 v_1^{\mat}(t)
=
1-\left[q_0+(1-q_0)r^t\right]^2-v_\infty^{\mat}(t).
\label{eq:matrix-v1-t}
\end{equation}

\subsection{Evaluation at the crossing time}
\label{subsec:matrix-crossing-time}

We next specialize the closed form expressions in Section~\ref{subsec:matrix-vinfty} to the threshold crossing time \(\tau\).  At \(t=\tau\), Eq.~\eqref{eq:r-tau} gives
\begin{equation}
r^\tau=\frac{q_{\min}-q_0}{1-q_0}.
\label{eq:matrix-r-tau-crossing}
\end{equation}
Thus the scalar response function becomes
\begin{equation}
F_\tau(\lambda)
=
\frac{\lambda^\tau-1}{\lambda-1}
-
2\frac{\lambda^\tau-r^\tau}{\lambda-r}
+
\frac{\lambda^\tau-r^{2\tau}}{\lambda-r^2}.
\label{eq:matrix-F-tau}
\end{equation}
The matrix expression at the crossing time is therefore
\begin{equation}
 v_\infty^{\mat}(\tau)
=
D\left[
W_+F_\tau(\lambda_+)
+
W_-F_\tau(\lambda_-)
\right],
\label{eq:matrix-vinfty-tau}
\end{equation}
with \(D\), \(\lambda_\pm\), \(W_\pm\), and \(F_\tau(\lambda_\pm)\) defined in Eqs.~\eqref{eq:matrix-D-def}, \eqref{eq:matrix-eigenvalues}, \eqref{eq:matrix-projector-weights}, and \eqref{eq:matrix-F-tau}, respectively.

Because the stochastic model evolves in discrete generations, one may alternatively evaluate the finite matrix sum in Eq.~\eqref{eq:matrix-sum-solution} at \(\lfloor\tau\rfloor\) or \(\lceil\tau\rceil\).  Equation~\eqref{eq:matrix-vinfty-tau} uses the standard continuous interpolation obtained by allowing powers such as \(\lambda^\tau\) and \(r^\tau\) in the closed form.  This is the same interpolation already used in defining the deterministic crossing time.

At crossing, \(m_\tau=q_{\min}\).  Hence Eq.~\eqref{eq:matrix-v1-t} becomes
\begin{equation}
 v_1^{\mat}(\tau)
=
1-q_{\min}^2-v_\infty^{\mat}(\tau).
\label{eq:matrix-v1-tau}
\end{equation}
Equations~\eqref{eq:matrix-vinfty-tau} and \eqref{eq:matrix-v1-tau} are the two variance quantities needed, by the transient variance criterion, to evaluate the critical genome length \(L_c\), as shown below.

\subsection{Matrix formula for the critical genome length}
\label{subsec:matrix-critical-length}

We can now insert the closed form expressions in Eqs.~\eqref{eq:matrix-vinfty-tau} and \eqref{eq:matrix-v1-tau} into the critical length formula derived in Section~\ref{subsec:critical-length-condition}.  Combining Eqs.~\eqref{eq:Lc-exact-decomposition}, \eqref{eq:matrix-vinfty-tau}, and \eqref{eq:matrix-v1-tau} gives
\begin{equation}
 L_c^{\mat}
=
\frac{1-q_{\min}^2-v_\infty^{\mat}(\tau)}
{\delta q^2+2\delta q\sqrt{v_\infty^{\mat}(\tau)}}.
\label{eq:matrix-Lc}
\end{equation}
Here
\begin{equation}
\delta q=\frac{q_{\min}/q_0-1}{M-1},
\qquad
\tau=\frac{\log\!\left[(q_{\min}-q_0)/(1-q_0)\right]}{\log r},
\qquad
r=a\left(1-\frac1M\right),
\label{eq:matrix-Lc-ingredients}
\end{equation}
and \(v_\infty^{\mat}(\tau)\) is given by Eq.~\eqref{eq:matrix-vinfty-tau}.  Equation~\eqref{eq:matrix-Lc} is the closed form matrix version of the critical genome length prediction.  Its input \(v_\infty^{\mat}(\tau)\) comes from the closed form solution of the reduced finite \(M\), infinite genome variance and covariance system, and its numerator uses the finite genome correction coefficient fixed by the exact variance decomposition of Eq.~\eqref{eq:exact-variance-decomposition}.

The next section derives a scalar reduction of this expression.  The scalar formula for the critical genome length is less accurate than the closed form expression in Eq.~\eqref{eq:matrix-Lc} for small \(M\), because it simplifies the coupled variance and covariance propagation.  It becomes rapidly accurate as \(M\) grows.  Thus the matrix formula in Eq.~\eqref{eq:matrix-Lc} should be regarded as the natural finite \(M\) closed form version of the critical genome length prediction, while the scalar formula should be regarded as its leading large \(M\) reduction, which is more transparent and useful for the asymptotic analysis in Section~\ref{sec:asymptotic-regimes}.

\section{Scalar reduction of the matrix formula}
\label{sec:scalar-reduction}

The matrix formula for the critical genome length in Eq.~\eqref{eq:matrix-Lc} is the finite population closed form associated with the transient variance criterion.  In deriving this formula in Section~\ref{sec:matrix-closed-form}, we kept both the infinite genome genealogical variance of the off diagonal overlap and the covariance between two overlaps that share one individual.  This gives the most faithful expression derived here for the infinite genome genealogical variance at the crossing time, \(v_\infty^{\mat}(\tau)\), and therefore for the critical genome length \(L_c^{\mat}\).  For the asymptotic analysis in Section~\ref{sec:asymptotic-regimes}, however, it is useful to have a simpler expression for \(v_\infty(\tau)\) that retains the leading large \(M\) contribution while suppressing lower order covariance feedback.  This section derives that scalar reduction.  As shown in Section~\ref{sec:numerical-tests}, the critical genome length produced by the scalar reduction is in excellent agreement with the matrix formula \(L_c^{\mat}\) and with the full hierarchy numerical benchmark across a broad parameter range.

Below we first derive the scalar recursion equation from the reduced finite \(M\), infinite genome variance and covariance coupled system in Eqs.~\eqref{eq:matrix-infinite-L-exact-M-variance} and \eqref{eq:matrix-infinite-L-exact-M-covariance}.  Solving this scalar recursion gives a closed form approximation \(v_\infty^{\scal}(\tau)\) to the infinite genome genealogical variance \(v_\infty(\tau)\).  We then insert \(v_\infty^{\scal}(\tau)\) into the critical length formula, giving a closed form scalar expression \(L_c^{\scal}\).  Finally, we explain how \(v_\infty^{\scal}(\tau)\) is related to the matrix expression \(v_\infty^{\mat}(\tau)\), why it captures the leading large \(M\) behavior, and when the scalar formula is sufficient.

\subsection{The scalar recursion equation}
\label{subsec:scalar-tv-equation}

The starting point for the derivation of the scalar recursion equation is the reduced finite \(M\), infinite genome coupled system in Eqs.~\eqref{eq:matrix-infinite-L-exact-M-variance} and \eqref{eq:matrix-infinite-L-exact-M-covariance}.  With the source coefficient \(A_M\) of Eq.~\eqref{eq:matrix-AM-BM-def}, this system is
\begin{align}
 v_{t+1}
&=A_M(1-m_t)^2
+
\frac{a^2(M^2-2M+2)}{4M(M-1)}v_t
+
\frac{a^2(M-2)}{2(M-1)}c_t,
\label{eq:scalar-start-v}
\\
 c_{t+1}
&=
\frac{a^2}{2M}v_t
+
\frac{a^2(M^2-4)}{2M^2}c_t.
\label{eq:scalar-start-c}
\end{align}
These equations describe two quantities: \(v_t=v_\infty(t)\), the infinite genome genealogical variance of an off diagonal overlap, and \(c_t\), the infinite genome covariance between two pairwise overlaps that share one individual.  The scalar reduction below keeps the leading large \(M\) contribution to the variance equation and suppresses the covariance feedback, that is, it sets \(c_t\) to zero at leading order.

To see precisely what is retained, first expand the finite population coefficients.  The source coefficient is
\begin{align}
A_M
&=
\frac{a^2(M-2)^2}{4M^2(M-1)}
=
\frac{a^2}{4M}\frac{(1-2/M)^2}{1-1/M}
\notag\\
&=
\frac{a^2}{4}
\left(
\frac1M-
\frac3{M^2}
+
O(M^{-3})
\right).
\label{eq:scalar-AM-two-term}
\end{align}
Only its leading \(1/M\) piece contributes to the leading scalar variance, because \(A_M\) multiplies the order one source \((1-m_t)^2\).  The two propagation coefficients of the variance equation expand as
\begin{align}
k_{11}=\frac{a^2(M^2-2M+2)}{4M(M-1)}
&=
\frac{a^2}{4}
\left(1-\frac1M\right)+O(M^{-2}),
\label{eq:scalar-v-coefficient-expansion}
\\
k_{12}=\frac{a^2(M-2)}{2(M-1)}
&=
\frac{a^2}{2}
\left(1-\frac1M\right)+O(M^{-2}).
\label{eq:scalar-c-feedback-expansion}
\end{align}
The covariance equation has the exact source coupling \(k_{21}=a^2/(2M)\), while its self coupling expands as
\begin{equation}
k_{22}=\frac{a^2(M^2-4)}{2M^2}=
\frac{a^2}{2}
\left(1-\frac0M\right)+O(M^{-2}),
\label{eq:scalar-c-self-expansion}
\end{equation}
that is, its first order \(1/M\) correction vanishes.  Thus, retaining terms through first order in \(1/M\) in the coefficients, the reduced finite \(M\), infinite genome system becomes the large population system
\begin{align}
 v_{t+1}
&=
\frac{a^2}{4}
\left[
\frac1M(1-m_t)^2
+
\left(1-\frac1M\right)v_t
+
2\left(1-\frac1M\right)c_t
\right],
\label{eq:scalar-largeM-v-recursion}
\\
 c_{t+1}
&=
\frac{a^2}{2}
\left[
\frac1M v_t+
c_t
\right].
\label{eq:scalar-largeM-c-recursion}
\end{align}
The omitted terms are coefficient corrections of order \(M^{-2}\).  This is the infinite genome, large population reduction of the coupled variance and covariance system.

We now use Eq.~\eqref{eq:scalar-largeM-c-recursion} to justify the scalar approximation.  We start from a clonal population, so \(v_0=c_0=0\).  The source term \(a^2(1-m_t)^2/(4M)\) in Eq.~\eqref{eq:scalar-largeM-v-recursion} is of order \(M^{-1}\), and hence the leading variance obtained from the recursion is \(v_t=O(M^{-1})\) in the large population regimes considered below.  Equation~\eqref{eq:scalar-largeM-c-recursion} then generates \(c_t\) through the term \(M^{-1}v_t\).  Thus \(c_t=O(M^{-2})\), as long as the linear response remains bounded.  Its feedback into the variance equation in Eq.~\eqref{eq:scalar-largeM-v-recursion} is therefore also of order \(M^{-2}\), below the leading scalar contribution.  This scale separation motivates the scalar step: while the reduced coupled system in Eqs.~\eqref{eq:scalar-start-v} and \eqref{eq:scalar-start-c} keeps this covariance feedback, the scalar reduction omits it.

Setting \(c_t=0\) in Eq.~\eqref{eq:scalar-largeM-v-recursion} gives
\begin{equation}
 v_{\infty}^{\scal}(t+1)
=
\frac{a^2}{4}\left(1-\frac1M\right)v_{\infty}^{\scal}(t)
+
\frac{a^2}{4M}(1-m_t)^2,
\qquad
v_{\infty}^{\scal}(0)=0.
\label{eq:scalar-recursion-pre-source}
\end{equation}
Using the explicit mean trajectory in Eq.~\eqref{eq:matrix-parameters},
\begin{equation}
(1-m_t)^2=(1-q_0)^2(1-2r^t+r^{2t}),
\label{eq:scalar-source-shape}
\end{equation}
we define
\begin{equation}
 b=
\frac{a^2}{4}\left(1-\frac1M\right),
\qquad
 D^{\scal}=
\frac{a^2(1-q_0)^2}{4M}.
\label{eq:scalar-b-D-def}
\end{equation}
The scalar recursion equation is therefore
\begin{equation}
 v_{\infty}^{\scal}(t+1)=
 b\,v_{\infty}^{\scal}(t)
 +D^{\scal}(1-2r^t+r^{2t}),
\qquad
v_{\infty}^{\scal}(0)=0.
\label{eq:scalar-tv-recursion}
\end{equation}
This equation is a one variable approximation to the reduced two variable system in Eqs.~\eqref{eq:matrix-infinite-L-exact-M-variance} and \eqref{eq:matrix-infinite-L-exact-M-covariance}.  It propagates only the infinite genome genealogical variance and does not keep a separate covariance variable.

\subsection{Closed form solution of the scalar recursion and relation to the matrix formula}
\label{subsec:scalar-matrix-relation}

In this subsection we solve Eq.~\eqref{eq:scalar-tv-recursion} and show that the scalar closed form for \(v_\infty^{\scal}(t)\) is the one dimensional analogue of the finite matrix sum in Eq.~\eqref{eq:matrix-sum-solution}.

Iterating Eq.~\eqref{eq:scalar-tv-recursion} from \(v_\infty^{\scal}(0)=0\) gives
\begin{equation}
 v_\infty^{\scal}(t)
=
D^{\scal}
\sum_{s=0}^{t-1}
 b^{t-1-s}(1-2r^s+r^{2s}).
\label{eq:scalar-finite-sum}
\end{equation}
This is the scalar counterpart of Eq.~\eqref{eq:matrix-sum-solution}: the matrix power \(K^{t-1-s}\) has been replaced by the scalar power \(b^{t-1-s}\), and the matrix source coefficient \(D\) has been replaced by \(D^{\scal}\).

Splitting the sum into the three sequences \(1\), \(r^s\), and \(r^{2s}\), we obtain
\begin{equation}
 v_\infty^{\scal}(t)
=
D^{\scal}
\left[
\sum_{s=0}^{t-1}b^{t-1-s}
-2\sum_{s=0}^{t-1}b^{t-1-s}r^s
+\sum_{s=0}^{t-1}b^{t-1-s}r^{2s}
\right].
\label{eq:scalar-three-sums}
\end{equation}
For a scalar \(\eta\), define
\begin{equation}
G_t(b,\eta)=\sum_{s=0}^{t-1}b^{t-1-s}\eta^s.
\label{eq:scalar-G-def}
\end{equation}
This is the same scalar geometric sum used in Eq.~\eqref{eq:matrix-G-def}, with the eigenvalue \(\lambda\) replaced by the damping coefficient \(b\).  Hence
\begin{equation}
G_t(b,\eta)=\frac{b^t-\eta^t}{b-\eta},
\label{eq:scalar-G-closed}
\end{equation}
with the finite sum interpretation if \(b=\eta\).  It follows that
\begin{equation}
 v_\infty^{\scal}(t)
=
D^{\scal}F_t(b),
\label{eq:scalar-v-F}
\end{equation}
where the same response function introduced in Eq.~\eqref{eq:matrix-F-def} is evaluated at \(\lambda=b\):
\begin{equation}
F_t(b)
=
\frac{b^t-1}{b-1}
-
2\frac{b^t-r^t}{b-r}
+
\frac{b^t-r^{2t}}{b-r^2}.
\label{eq:scalar-F-closed}
\end{equation}
Thus the scalar expression is obtained from the matrix expression by replacing the two eigenvalue response
\[
D\left[W_+F_t(\lambda_+)+W_-F_t(\lambda_-)\right]
\]
with the single response \(D^{\scal}F_t(b)\).  This replacement is the precise sense in which the scalar formula reduces the matrix formula.  Section~\ref{subsec:scalar-leading-largeM} makes this connection more explicit.

\subsection{The scalar critical genome length formula}
\label{subsec:scalar-critical-length}

In this subsection we insert the scalar expression for \(v_\infty(\tau)\) at the crossing time \(\tau\) into the critical genome length formula in Eq.~\eqref{eq:Lc-exact-decomposition}, as implied by the transient variance criterion.  Recall that at the crossing time, Eq.~\eqref{eq:r-tau} gives
\begin{equation}
r^\tau=\frac{q_{\min}-q_0}{1-q_0}.
\label{eq:scalar-r-tau}
\end{equation}
Therefore Eq.~\eqref{eq:scalar-F-closed} gives
\begin{equation}
F_\tau(b)
=
\frac{b^\tau-1}{b-1}
-
2\frac{b^\tau-r^\tau}{b-r}
+
\frac{b^\tau-r^{2\tau}}{b-r^2}.
\label{eq:scalar-F-tau}
\end{equation}
By Eq.~\eqref{eq:scalar-v-F}, the scalar approximation to the infinite genome genealogical variance at crossing is therefore
\begin{equation}
 v_\infty^{\scal}(\tau)
=
D^{\scal}F_\tau(b)
=
\frac{a^2(1-q_0)^2}{4M}
\left[
\frac{b^\tau-1}{b-1}
-
2\frac{b^\tau-r^\tau}{b-r}
+
\frac{b^\tau-r^{2\tau}}{b-r^2}
\right].
\label{eq:scalar-vinfty-tau}
\end{equation}
The finite genome correction coefficient \(v_1^{\scal}(\tau)\) associated with the scalar approximation follows from the exact variance decomposition:
\begin{equation}
 v_1^{\scal}(\tau)=1-q_{\min}^2-v_\infty^{\scal}(\tau).
\label{eq:scalar-v1-tau}
\end{equation}
Substitution into Eq.~\eqref{eq:Lc-exact-decomposition} gives the scalar critical genome length formula
\begin{equation}
 L_c^{\scal}
=
\frac{1-q_{\min}^2-v_\infty^{\scal}(\tau)}
{\delta q^2+2\delta q\sqrt{v_\infty^{\scal}(\tau)}}.
\label{eq:scalar-Lc}
\end{equation}
Here \(\delta q\) is the deterministic crossing scale in Eq.~\eqref{eq:delta-q-definition}, \(\tau\) is given by Eq.~\eqref{eq:tau-formula}, and \(v_\infty^{\scal}(\tau)\) is given by Eq.~\eqref{eq:scalar-vinfty-tau}.

For leading large \(M\) asymptotic scaling, we can further simplify Eq.~\eqref{eq:scalar-Lc}.  Since by Eq.~\eqref{eq:scalar-vinfty-tau} \(v_\infty^{\scal}(\tau)=O(M^{-1})\) in the regimes considered below, the numerator in Eq.~\eqref{eq:scalar-Lc} may be replaced at leading order by \(1-q_{\min}^2\), provided \(q_{\min}\) is not so close to one that this numerator is itself small.  This gives the leading scalar critical length
\begin{equation}
 \widetilde L_c^{\scal}
=
\frac{1-q_{\min}^2}
{\delta q^2+2\delta q\sqrt{v_\infty^{\scal}(\tau)}}.
\label{eq:scalar-Lc-leading}
\end{equation}
Equation~\eqref{eq:scalar-Lc} is the more accurate scalar reduction of the matrix formula.  The leading scalar critical length \(\widetilde L_c^{\scal}\) in Eq.~\eqref{eq:scalar-Lc-leading} is the simpler form used to make the asymptotic scaling laws of the critical genome length explicit, and it will be the basis for the asymptotic analysis in Section~\ref{sec:asymptotic-regimes}.

\subsection{Leading large \texorpdfstring{\(M\)}{M} behavior and range of validity}
\label{subsec:scalar-leading-largeM}
\label{subsec:when-scalar-sufficient}

In this subsection we explain why the scalar equation captures the leading large \(M\) behavior of the matrix formula and summarize when the scalar formula is an appropriate replacement for the full matrix expression.  Recall that the matrix expression in Eq.~\eqref{eq:matrix-vinfty-closed-summary} can be written as
\begin{equation}
 v_\infty^{\mat}(t)
=
D\left[W_+F_t(\lambda_+)+W_-F_t(\lambda_-)\right].
\label{eq:scalar-start-from-matrix}
\end{equation}
The large \(M\) expansions of the quantities in this expression are
\begin{align}
D
&=
\frac{a^2(1-q_0)^2}{4M}
\left[1+O\!\left(M^{-1}\right)\right],
\label{eq:scalar-D-largeM}
\\
\lambda_+
&=
\frac{a^2}{2}+\frac{a^2}{M}+O\!\left(M^{-2}\right),
&
W_+
&=
\frac4M+O\!\left(M^{-2}\right),
\label{eq:scalar-plus-largeM}
\\
\lambda_-
&=
\frac{a^2}{4}-\frac{5a^2}{4M}+O\!\left(M^{-2}\right),
&
W_-
&=
1-\frac4M+O\!\left(M^{-2}\right).
\label{eq:scalar-minus-largeM}
\end{align}
The scalar damping factor satisfies
\begin{equation}
 b=\frac{a^2}{4}-\frac{a^2}{4M}.
\label{eq:scalar-b-largeM}
\end{equation}
Thus \(\lambda_-\), the eigenvalue carrying the \(O(1)\) weight in Eq.~\eqref{eq:scalar-start-from-matrix}, differs from \(b\) only at order \(M^{-1}\), while the weight of the \(\lambda_-\) contribution in Eq.~\eqref{eq:scalar-start-from-matrix} is one up to order \(M^{-1}\).  By contrast, the \(\lambda_+\) contribution has weight \(O(M^{-1})\).  Since the common prefactor \(D\) is itself \(O(M^{-1})\), the \(\lambda_+\) contribution to \(v_\infty^{\mat}(t)\) is lower order whenever the response functions remain bounded on the scale being considered.

Consequently, for the large population regimes used in the asymptotic analysis,
\begin{equation}
 v_\infty^{\mat}(t)
=
D^{\scal}F_t(b)+O\!\left(M^{-2}\right)
=
v_\infty^{\scal}(t)+O\!\left(M^{-2}\right).
\label{eq:scalar-leading-relation}
\end{equation}
This is the main justification for the scalar reduction.  The scalar equation does not reproduce the full finite \(M\) variance and covariance dynamics.  Rather, it captures the dominant contribution to \(v_\infty(t)\) in the large population limit.  The omitted covariance feedback, the difference between \(D\) and \(D^{\scal}\), the difference between \(\lambda_-\) and \(b\), and the contribution proportional to \(W_+\) all enter as finite \(M\) corrections.

It is useful to distinguish the two scalar critical length formulas introduced above.  The full scalar critical length \(L_c^{\scal}\) in Eq.~\eqref{eq:scalar-Lc} keeps the numerator correction \(-v_\infty^{\scal}(\tau)\) within the scalar approximation.  The leading scalar critical length \(\widetilde L_c^{\scal}\) in Eq.~\eqref{eq:scalar-Lc-leading} drops that numerator correction and is intended for leading large \(M\) scaling.  If \(q_{\min}\) is fixed away from one, then \(1-q_{\min}^2\) is order one, while \(v_\infty^{\scal}(\tau)=O(M^{-1})\).  In that case, the subtraction by \(v_\infty^{\scal}(\tau)\) changes the finite \(M\) value but not the leading scaling, so \(\widetilde L_c^{\scal}\) is appropriate for the asymptotic regimes below.  If one wants a more accurate scalar estimate while still neglecting covariance feedback, then \(L_c^{\scal}\) should be used instead.

The full matrix formula should be used when the aim is quantitative finite \(M\) prediction, especially for small or moderate populations.  It should also be preferred when covariance feedback is not expected to be negligible, or when the threshold is close enough to one that the numerator correction may matter.  In these cases the appropriate formula is Eq.~\eqref{eq:matrix-Lc}, which retains the covariance feedback present in the reduced finite \(M\), infinite genome variance and covariance system.  The scalar formula in Eq.~\eqref{eq:scalar-Lc} is best viewed as the leading large population reduction of the matrix formula, and Eq.~\eqref{eq:scalar-Lc-leading} is the further simplification used for transparent asymptotic estimates in Section~\ref{sec:asymptotic-regimes}.

\section{Asymptotic analysis}
\label{sec:asymptotic-regimes}

The closed form matrix expression in Eq.~\eqref{eq:matrix-Lc} is the appropriate formula for quantitative finite population estimates.  The purpose of this section is different: we use the scalar reduction of Section~\ref{sec:scalar-reduction} to isolate the dominant scaling mechanisms that control the critical genome length.  The results are meaningful only when the characteristic overlap actually falls below the mating threshold, that is, when \(q_{\min}>q_0\).

Two quantities organize the asymptotics.  The first is the compound mutation parameter
\begin{equation}
 \rho=M\left(\ee^{4\mu}-1\right),
\label{eq:asymp-rho-def}
\end{equation}
which controls the equilibrium mean overlap between two genomes,
\(q_0 = 1/(1+\rho)\). For small mutation rate,
\(\rho=4M\mu+O(M\mu^2)\). Thus the product \(M\mu\), rather than
\(M\) or \(\mu\) alone, controls \(q_0\). Since the transient variance criterion is meaningful only when the characteristic overlap can cross the threshold, we require \(q_{\min}>q_0\); the product \(M\mu\) therefore determines how far the chosen mating threshold lies above this boundary. The second organizing quantity is
\begin{equation}
 \Gamma=q_{\min}(1+\rho)-1
 =
 \frac{q_{\min}}{q_0}-1,
\label{eq:asymp-Gamma-def}
\end{equation}
which is the scaled distance of the mating threshold from the boundary value \(q_0\).  The crossing condition \(q_{\min}>q_0\) is therefore equivalent to \(\Gamma>0\).  By Eq.~\eqref{eq:delta-q-definition}, the deterministic crossing scale is
\begin{equation}
 \delta q=\frac{\Gamma}{M-1}.
\label{eq:asymp-deltaq-Gamma}
\end{equation}

As discussed in Section~\ref{subsec:scalar-leading-largeM}, the asymptotic analysis below starts from the scalar critical length in Eq.~\eqref{eq:scalar-Lc}.  For leading large \(M\) scaling, the numerator correction in that formula is lower order, so we also use the leading scalar critical length
\begin{equation}
 \widetilde L_c^{\scal}
 =
 \frac{1-q_{\min}^2}{\delta q^2+2\delta q\sqrt{v_\infty^{\scal}(\tau)}}.
\label{eq:asymp-leading-scalar-Lc}
\end{equation}
Thus the formulas below may be read at two scalar levels: \(L_c^{\scal}\) keeps the numerator correction \(-v_\infty^{\scal}(\tau)\), while \(\widetilde L_c^{\scal}\) drops it to expose the leading scaling.  Relative to \(L_c^{\scal}\), the matrix formula in Eq.~\eqref{eq:matrix-Lc} replaces \(v_\infty^{\scal}\) by \(v_\infty^{\mat}\).  These refinements matter for finite \(M\), but, in the large population regimes considered here, they do not change the leading powers of \(M\), \(\mu\), or \(q_{\min}-q_0\).  The scaling is determined by which denominator term is dominant: the deterministic crossing term \(\delta q^2\), or the transient genealogical width term \(2\delta q\sqrt{v_\infty^{\scal}(\tau)}\).  The subsections below track this competition in the main asymptotic regimes.

\subsection{Large \texorpdfstring{\(M\)}{M}, small \texorpdfstring{\(\mu\)}{mu} approximation for \texorpdfstring{\(v_\infty(\tau)\)}{v infinity(tau)}}
\label{subsec:largeM-smallmu-vinfty}

We begin with a variance estimate that will be used repeatedly below. In the small mutation, large population regimes where the crossing time \(\tau\) is asymptotically long, the infinite genome genealogical variance at the crossing time \(\tau\) has the simple leading form
\begin{equation}
 v_\infty^{\scal}(\tau)\sim \frac{(1-q_{\min})^2}{3M}.
\label{eq:asymp-vinfty-simple}
\end{equation}
This estimate is used in the regimes below where \(\mu\to0\).  It is not needed in the fixed mutation rate limit of Section~\ref{subsec:largeM-fixed-mu}, where only the order estimate \(v_\infty^{\scal}(\tau)=O(M^{-1})\) is required.

To derive Eq.~\eqref{eq:asymp-vinfty-simple}, we start from the scalar expression \(v_\infty^{\scal}(\tau)\) in Eq.~\eqref{eq:scalar-vinfty-tau}.  By Eq.~\eqref{eq:r-tau},
\begin{equation}
 x=r^\tau=\frac{q_{\min}-q_0}{1-q_0}.
\label{eq:asymp-x-def}
\end{equation}
Then
\begin{equation}
 v_\infty^{\scal}(\tau)
 =
 D^{\scal}
 \left[
 \frac{b^\tau-1}{b-1}
 -2\frac{b^\tau-x}{b-r}
 +\frac{b^\tau-x^2}{b-r^2}
 \right],
\label{eq:asymp-vinfty-start}
\end{equation}
where
\begin{equation}
 D^{\scal}=\frac{a^2(1-q_0)^2}{4M},
 \qquad
 b=\frac{a^2}{4}\left(1-\frac1M\right),
 \qquad
 a=\ee^{-4\mu}.
\label{eq:asymp-D-b-a}
\end{equation}
For small \(\mu\) and large \(M\), \(a\approx1\), \(b\approx1/4\), and \(r\approx1\).  In the regimes where Eq.~\eqref{eq:asymp-vinfty-simple} is used, the crossing time \(\tau\) is large.  More importantly, the two decay rates are separated: \(b\) is bounded well below one, whereas \(r\) is close to one.  Thus \(b^\tau\) decays rapidly, while \(x=r^\tau\) must be retained.  Substitution of \(b^\tau\approx0\), \(b\approx1/4\), and \(r\approx1\) into the bracket in Eq.~\eqref{eq:asymp-vinfty-start} gives
\begin{align}
 \frac{b^\tau-1}{b-1}
 -2\frac{b^\tau-x}{b-r}
 +\frac{b^\tau-x^2}{b-r^2}
 &\approx
 \frac43-\frac{8x}{3}+\frac{4x^2}{3} 
 \notag\\
 &=
 \frac43(1-x)^2.
\label{eq:asymp-bracket-simple}
\end{align}
Therefore
\begin{equation}
 v_\infty^{\scal}(\tau)
 \approx
 \frac{(1-q_0)^2}{4M}\,\frac43(1-x)^2.
\label{eq:asymp-vinfty-x-form}
\end{equation}
Finally, Eq.~\eqref{eq:asymp-x-def} implies
\begin{equation}
 1-q_{\min}=(1-q_0)(1-x).
\label{eq:asymp-one-minus-q}
\end{equation}
Combining Eqs.~\eqref{eq:asymp-vinfty-x-form} and \eqref{eq:asymp-one-minus-q} gives Eq.~\eqref{eq:asymp-vinfty-simple}.

We note that Eq.~\eqref{eq:scalar-leading-relation} implies that, in the same small mutation, large population range in which Eq.~\eqref{eq:asymp-vinfty-simple} was derived, the matrix expression has the same leading behavior:
\begin{equation}
 v_\infty^{\mat}(\tau)
 =
 \frac{(1-q_{\min})^2}{3M}+O(M^{-2}).
\label{eq:asymp-matrix-vinfty-leading}
\end{equation}

As \(q_{\min}\downarrow q_0\), the crossing time diverges, and the scalar variance approaches the fixed point of the scalar recursion in Eq.~\eqref{eq:scalar-tv-recursion}.  Setting \(v_{\infty}^{\scal}(t+1)=v_{\infty}^{\scal}(t)=v_*^{\scal}\) and \(r^t\to0\) in that equation gives
\begin{equation}
 v_*^{\scal}=b\,v_*^{\scal}+D^{\scal},
 \qquad
 v_*^{\scal}=\frac{D^{\scal}}{1-b}.
\label{eq:asymp-vstar-consistency}
\end{equation}
When \(a\approx1\), \(D^{\scal}\approx(1-q_0)^2/(4M)\) and \(1-b\approx3/4\), so \(v_*^{\scal}\approx(1-q_0)^2/(3M)\).  This is precisely the value obtained by evaluating the leading variance estimate in Eq.~\eqref{eq:asymp-vinfty-simple} at \(q_{\min}=q_0\).  This agreement shows that the variance approximation in Eq.~\eqref{eq:asymp-vinfty-simple} is consistent with the fixed point behavior of the scalar recursion at the boundary \(q_{\min}=q_0\).

\subsection{Regime I: large \texorpdfstring{\(M\)}{M} with fixed \texorpdfstring{\(\mu>0\)}{mu > 0}}
\label{subsec:largeM-fixed-mu}

In this subsection we show that when \(M\to\infty\) at fixed \(\mu>0\), the critical genome length approaches a finite limit controlled by the mutation rate.  Define
\begin{equation}
 s_\mu=\ee^{4\mu}-1.
\label{eq:asymp-smu-def}
\end{equation}
Then \(\rho=M s_\mu\), and Eq.~\eqref{eq:asymp-rho-def} gives
\begin{equation}
 q_0=\frac{1}{1+M s_\mu}\sim\frac{1}{M s_\mu}\to0.
\label{eq:asymp-q0-fixed-mu}
\end{equation}
Since \(\rho=M s_\mu\), Eq.~\eqref{eq:asymp-deltaq-Gamma} gives the deterministic crossing scale:
\begin{equation}
 \delta q
 =
 \frac{q_{\min}(1+M s_\mu)-1}{M-1}
 \sim
 q_{\min} s_\mu
 =
 q_{\min}\left(\ee^{4\mu}-1\right).
\label{eq:asymp-delta-fixed-mu}
\end{equation}
Thus \(\delta q\) has a nonzero limit as \(M\to\infty\).

To determine the asymptotic order of \(v_\infty^{\scal}(\tau)\), we first note that, in this fixed mutation rate regime, the crossing time \(\tau\) remains finite rather than diverging.  Indeed, since \(q_0\to0\), Eq.~\eqref{eq:r-tau} gives \(r^\tau=(q_{\min}-q_0)/(1-q_0)\to q_{\min}\), while \(r=a(1-1/M)\to a\); hence \(\tau\to\log(q_{\min})/\log(a)\), a finite positive limit.  Because \(\tau\) stays finite while \(b\to a^2/4\in(0,1)\), the power \(b^\tau\) tends to a nonzero constant rather than to zero.  The simplified variance in Eq.~\eqref{eq:asymp-vinfty-simple} was derived under the opposite assumption \(b^\tau\approx0\), so we do \emph{not} use it here; we need only the order of \(v_\infty^{\scal}(\tau)\), which is insensitive to the precise limiting value of \(b^\tau\).  Since \(D^{\scal}\) in Eq.~\eqref{eq:scalar-b-D-def} is \(O(M^{-1})\), while the bracket in Eq.~\eqref{eq:scalar-vinfty-tau} tends to a finite limit and hence remains \(O(1)\), we conclude that \(v_\infty^{\scal}(\tau)=O(M^{-1})\) and \(\sqrt{v_\infty^{\scal}(\tau)}=O(M^{-1/2})\).  Recalling from Eq.~\eqref{eq:asymp-delta-fixed-mu} that \(\delta q\) has a nonzero limit, the cross term \(2\delta q\sqrt{v_\infty^{\scal}(\tau)}=O(M^{-1/2})\) in the denominator of Eq.~\eqref{eq:asymp-leading-scalar-Lc} is negligible compared with \(\delta q^2=O(1)\).  The denominator is therefore dominated by \(\delta q^2\). Hence
\begin{equation}
 L_c^{\mat}
 \sim
 L_c^{\scal}
 \sim
 \widetilde L_c^{\scal}
 \sim
 \frac{1-q_{\min}^2}{q_{\min}^2\left(\ee^{4\mu}-1\right)^2}.
\label{eq:asymp-Lc-fixed-mu}
\end{equation}
If in addition the fixed mutation rate \(\mu\) is itself small, then \(\ee^{4\mu}-1\sim4\mu\), and Eq.~\eqref{eq:asymp-Lc-fixed-mu} becomes
\begin{equation}
 L_c
 \sim
 \frac{1-q_{\min}^2}{16q_{\min}^2\mu^2}.
\label{eq:asymp-Lc-fixed-small-mu}
\end{equation}
This regime shows that once \(M\) is sufficiently large at fixed \(\mu\), increasing the population size further does not change the leading critical genome length, whose leading scale is set by mutation through \(\mu^{-2}\).

\subsection{Regime II: \texorpdfstring{\(M\mu=O(1)\)}{M mu = O(1)}}
\label{subsec:Mmu-order-one}

In this subsection we take \(M\to\infty\) while holding \(M\mu=O(1)\).  Let
\begin{equation}
 \kappa=4M\mu=O(1).
\label{eq:asymp-kappa-def}
\end{equation}
Since \(\mu=\kappa/(4M)\) is small, we have, to leading order,
\begin{equation}
 \rho=M\left(\ee^{4\mu}-1\right)\approx 4M\mu=\kappa,
 \qquad
 q_0\approx\frac{1}{1+\kappa}.
\label{eq:asymp-rho-kappa}
\end{equation}
The crossing condition \(q_{\min}>q_0\) becomes
\begin{equation}
 q_{\min}>\frac{1}{1+\kappa}.
\label{eq:asymp-crossing-kappa}
\end{equation}
Define the corresponding positive rescaled distance from the limiting threshold
\(q_0\approx 1/(1+\kappa)\) by
\begin{equation}
 \gamma=q_{\min}(1+\kappa)-1>0.
\label{eq:asymp-gamma-def}
\end{equation}
Replacing \(\rho\) by \(\kappa\) in Eq.~\eqref{eq:asymp-deltaq-Gamma} gives
\begin{equation}
\delta q\sim\frac{q_{\min}(1+\kappa)-1}{M}=\frac{\gamma}{M}.
\label{eq:dq-II}
\end{equation}
Here \(\mu\) is small and the transient is long, so the simplified variance in
Eq.~\eqref{eq:asymp-vinfty-simple} applies, giving
\begin{equation}
 \sqrt{v_\infty^{\scal}(\tau)}
 \sim
 \frac{1-q_{\min}}{\sqrt{3M}}.
\label{eq:asymp-sqrt-v-kappa}
\end{equation}
The terms in the denominator of the scalar critical genome length formula in
Eq.~\eqref{eq:scalar-Lc} are
\begin{align}
 \delta q^2
 &\sim
 \frac{\gamma^2}{M^2},
 \\
 2\delta q\sqrt{v_\infty^{\scal}(\tau)}
 &\sim
 \frac{2\gamma(1-q_{\min})}{\sqrt3\,M^{3/2}}.
\label{eq:asymp-kappa-denominator-terms}
\end{align}
These estimates lead to the asymptotic scalar critical genome length formula
\begin{equation}
 L_c^{\scal}
 \sim
 \frac{\left(1-q_{\min}^2-\dfrac{(1-q_{\min})^2}{3M}\right)M^2}
 {\gamma^2+\dfrac{2\gamma(1-q_{\min})}{\sqrt3}\sqrt M}.
\label{eq:asymp-Lc-kappa-uniform}
\end{equation}
For fixed \(q_{\min}<1\) and fixed \(\gamma>0\), the term proportional to \(M^{1/2}\) in the denominator of Eq.~\eqref{eq:asymp-Lc-kappa-uniform} dominates the constant term \(\gamma^2\).  Thus
\begin{equation}
 L_c^{\mat}
 \sim
 L_c^{\scal}
 \sim
 \widetilde L_c^{\scal}
 \sim
 \frac{\sqrt3(1+q_{\min})}{2\,[q_{\min}(1+\kappa)-1]}
 M^{3/2}.
\label{eq:asymp-Lc-kappa-M32}
\end{equation}
This is one of the main scaling predictions: in the regime \(M\mu=O(1)\), the critical length scales as \(M^{3/2}\).  If the genealogical variance \(v_\infty^{\scal}(\tau)\) were ignored, the denominator would be only \(\delta q^2\), and the same calculation would give
\begin{equation}
 L_c^{\scal}\sim \frac{(1-q_{\min}^2)M^2}{\gamma^2}=O(M^2).
\label{eq:asymp-Lc-kappa-without-width}
\end{equation}
The genealogical width \(2\delta q\sqrt{v_\infty^{\scal}(\tau)}\) therefore changes the leading scaling from \(M^2\) to \(M^{3/2}\).

\subsection{Regime III: \texorpdfstring{\(M\mu\gg 1\)}{M mu >> 1} with small \texorpdfstring{\(\mu\)}{mu}}
\label{subsec:Mmu-large-small-mu}

Assume \(M\mu\gg1\) and \(\mu\ll1\).  The smallness of \(\mu\) gives \(\ee^{4\mu}-1\approx4\mu\), so
\begin{equation}
 \rho=M\left(\ee^{4\mu}-1\right)\approx4M\mu\gg1,
 \qquad
 q_0=\frac{1}{1+\rho}\approx\frac{1}{\rho}\approx\frac{1}{4M\mu}\ll1.
\label{eq:asymp-rho-large-Mmu}
\end{equation}
For a threshold \(q_{\min}\) bounded away from zero, the crossing condition \(q_{\min}>q_0\) therefore holds automatically once \(M\mu\) is large enough.  Because \(\rho\gg1\), the crossing scale becomes
\begin{equation}
 \delta q
 =
 \frac{q_{\min}(1+\rho)-1}{M-1}
 \sim
 4q_{\min}\mu.
\label{eq:asymp-delta-large-Mmu}
\end{equation}
Combining this with the simplified variance in Eq.~\eqref{eq:asymp-vinfty-simple} gives the two denominator terms
\begin{align}
 \delta q^2
 &\sim
 16q_{\min}^2\mu^2,
 \\
 2\delta q\sqrt{v_\infty^{\scal}(\tau)}
 &\sim
 \frac{8q_{\min}\mu(1-q_{\min})}{\sqrt{3M}}.
\label{eq:asymp-large-Mmu-denominator-terms}
\end{align}
Thus the corresponding asymptotic scalar critical genome length formula is
\begin{equation}
 L_c^{\scal}
 \sim
 \frac{1-q_{\min}^2-\dfrac{(1-q_{\min})^2}{3M}}
 {16q_{\min}^2\mu^2+\dfrac{8q_{\min}\mu(1-q_{\min})}{\sqrt{3M}}}.
\label{eq:asymp-Lc-large-Mmu-two-term}
\end{equation}
Because the two denominator terms in the asymptotic formula for \(L_c^{\scal}\) compete, the condition \(M\mu\gg1\) alone does not determine the leading scaling; the scaling depends on whether the mutation term or the genealogical width term dominates.  Equating these two terms gives the leading order crossover scale
\begin{equation}
 \mu\sqrt M\approx\frac{1-q_{\min}}{2q_{\min}\sqrt3},
\label{eq:asymp-mu-sqrtM-crossover}
\end{equation}
so, at the level of scaling, the transition occurs when \(\mu\sqrt M\) is of order one.  This crossover separates the following two subregimes according to which denominator term is dominant.

\paragraph{Subregime III(a): \texorpdfstring{\(\mu\sqrt{M}\gg1\)}{mu sqrtM large}.}
In this case, the first denominator term \(\delta q^2\) dominates the denominator and
\begin{equation}
 L_c^{\mat}
 \sim
 L_c^{\scal}
 \sim
 \widetilde L_c^{\scal}
 \sim
 \frac{1-q_{\min}^2}{16q_{\min}^2\mu^2}.
\label{eq:asymp-Lc-large-Mmu-mutation-dominated}
\end{equation}
This coincides with the small \(\mu\) limit in Eq.~\eqref{eq:asymp-Lc-fixed-small-mu} of Regime~I, as it should.  In both limits, \(q_0\ll1\), the deterministic crossing scale satisfies \(\delta q\sim4q_{\min}\mu\), and the genealogical width term is lower order than \(\delta q^2\).  The leading critical length is therefore set by the same deterministic mutation scale, and hence has the same \(\mu^{-2}\) scaling.

\paragraph{Subregime III(b): \texorpdfstring{\(M^{-1}\ll\mu\ll M^{-1/2}\)}{1/M << mu << 1/sqrtM}.}
Here \(M\mu\gg1\) but \(\mu\sqrt{M}\ll1\).  In this window, the genealogical width term dominates the denominator, so
\begin{equation}
 L_c^{\mat}
 \sim
 L_c^{\scal}
 \sim
 \widetilde L_c^{\scal}
 \sim
 \frac{\sqrt3(1+q_{\min})}{8q_{\min}}\,\frac{\sqrt M}{\mu}.
\label{eq:asymp-Lc-large-Mmu-width-dominated}
\end{equation}
Together, Subregimes III(a) and III(b) show that the large \(M\mu\) regime contains two distinct leading balances.  When \(\mu\sqrt M\gg1\), the common leading critical length is mutation dominated and scales as \(\mu^{-2}\).  When \(M^{-1}\ll\mu\ll M^{-1/2}\), it is controlled by the genealogical width and scales as \(\sqrt M/\mu\).

The formula for the case controlled by the transient width in Eq.~\eqref{eq:asymp-Lc-large-Mmu-width-dominated} also has an overlap with Regime~II.  This matching is verified explicitly in Subsection~\ref{subsec:asymptotic-consistency}, where the large \(\kappa\) limit of the Regime~II formula is shown to reproduce Eq.~\eqref{eq:asymp-Lc-large-Mmu-width-dominated}.

\subsection{Regime IV: \texorpdfstring{\(M\mu\ll 1\)}{M mu << 1}}
\label{subsec:Mmu-small}

Assume \(M\mu\ll1\), and define
\begin{equation}
 \kappa=4M\mu,
 \qquad
 \kappa\ll1.
\label{eq:asymp-kappa-small-def}
\end{equation}
In this weak mutation limit,
\begin{equation}
 \rho=M\left(\ee^{4\mu}-1\right)\approx4M\mu=\kappa\ll1,
 \qquad
 q_0=\frac{1}{1+\rho}\approx\frac{1}{1+\kappa}=1-\kappa+O(\kappa^2).
\label{eq:asymp-q0-small-Mmu}
\end{equation}
Because \(q_0\) is close to one, the crossing condition \(q_{\min}>q_0\) can hold only if \(q_{\min}\) is also close to one.  Write
\begin{equation}
 q_{\min}=1-s,
 \qquad
 0<s\ll1.
\label{eq:asymp-q-one-minus-s}
\end{equation}
Then the crossing condition gives, to first order,
\begin{equation}
 s<\kappa\ll1.
\label{eq:asymp-small-Mmu-crossing-condition}
\end{equation}
Equivalently, at this order the existence condition for this regime is
\begin{equation}
 4M\mu>1-q_{\min}.
\label{eq:asymp-small-Mmu-existence-condition}
\end{equation}
More exactly, by Eq.~\eqref{eq:asymp-Gamma-def}, the crossing condition is \(q_{\min}(1+\rho)>1\); Eq.~\eqref{eq:asymp-small-Mmu-existence-condition} is its first order form for small \(\kappa\).  For a fixed threshold \(q_{\min}<1\), this condition eventually fails as \(M\mu\to0\).  Thus the only nontrivial crossings in Regime~IV occur in a limit with a very high similarity threshold, in which the mating threshold itself approaches perfect similarity, \(q_{\min}\uparrow1\), as \(M\mu\downarrow0\).

The asymptotic form of the crossing scale \(\delta q\) is obtained by expanding \(q_{\min}/q_0\).  Using \(q_{\min}=1-s\), \(q_0\approx1-\kappa\), and \((1-\kappa)^{-1}=1+\kappa+O(\kappa^2)\),
\begin{equation}
 \frac{q_{\min}}{q_0}
 \approx
 (1-s)(1+\kappa)
 =
 1+\kappa-s-s\kappa
 =
 1+\kappa-s+O(\kappa^2).
\label{eq:asymp-qmin-over-q0-small-Mmu}
\end{equation}
Therefore \(q_{\min}/q_0-1\sim\kappa-s\), and, dividing by \(M\),
\begin{equation}
 \delta q
 =
 \frac{q_{\min}/q_0-1}{M-1}
 \sim
 \frac{\kappa-s}{M}.
\label{eq:asymp-delta-small-Mmu}
\end{equation}
This approximation keeps only the first order contribution to the crossing scale \(\delta q\); the exact boundary limit \(q_{\min}\downarrow q_0\) is treated directly in Regime~V below.

By Eq.~\eqref{eq:asymp-vinfty-simple},
\begin{equation}
 \sqrt{v_\infty^{\scal}(\tau)}
 \sim
 \frac{1-q_{\min}}{\sqrt{3M}}
 =
 \frac{s}{\sqrt{3M}},
 \qquad
 1-q_{\min}^2=1-(1-s)^2=2s-s^2.
\label{eq:asymp-sqrt-v-small-Mmu}
\end{equation}
The denominator terms in Eq.~\eqref{eq:scalar-Lc} are therefore
\begin{align}
 \delta q^2
 &\sim
 \frac{(\kappa-s)^2}{M^2},
 \\
 2\delta q\sqrt{v_\infty^{\scal}(\tau)}
 &\sim
 \frac{2s(\kappa-s)}{\sqrt3\,M^{3/2}}.
\label{eq:asymp-small-Mmu-denominator-terms}
\end{align}
Substituting these estimates into Eq.~\eqref{eq:scalar-Lc} gives the asymptotic scalar critical genome length formula
\begin{equation}
\begin{gathered}
 L_c^{\scal}
 \sim
 \frac{\left(2s-s^2-\dfrac{s^2}{3M}\right)M^2}
 {(\kappa-s)^2+\dfrac{2s(\kappa-s)}{\sqrt3}\sqrt M},
 \\
 s=1-q_{\min},
 \qquad
 \kappa=4M\mu,
 \qquad
 0<s<\kappa\ll1.
\end{gathered}
\label{eq:asymp-Lc-small-Mmu-exact-numerator}
\end{equation}
Dropping the lower order numerator terms gives the simpler leading expression
\begin{equation}
 \widetilde L_c^{\scal}
 \sim
 \frac{2sM^2}
 {(\kappa-s)^2+\dfrac{2s(\kappa-s)}{\sqrt3}\sqrt M}.
\label{eq:asymp-Lc-small-Mmu-leading}
\end{equation}
This leading form also displays the approach toward the boundary \(q_{\min}\downarrow q_0\) in this first order limit.  As \(s\uparrow\kappa\), the threshold \(q_{\min}=1-s\) approaches \(q_0\approx1-\kappa\) from above, the factor \(\kappa-s\) tends to zero, and \(\widetilde L_c^{\scal}\) diverges.  This is the small \(M\mu\) version of the boundary divergence analyzed next, where the exact boundary distance \(q_{\min}-q_0\) is kept explicitly.

\subsection{Regime V: near the infinite genome boundary \texorpdfstring{\(q_{\min}\downarrow q_0\)}{qmin down to q0}}
\label{subsec:near-boundary}

Finally, we consider the regime in which \(q_{\min}\) approaches the boundary value \(q_0\) from above, without assuming \(M\mu\ll1\).  We show that the scalar critical genome length has a linear boundary divergence, proportional to \((q_{\min}-q_0)^{-1}\), rather than the quadratic divergence predicted if the genealogical variance were ignored, as shown below.

Let
\begin{equation}
 q_{\min}=q_0+\varepsilon,
 \qquad
 0<\varepsilon\ll1.
\label{eq:asymp-q-near-boundary}
\end{equation}
Then Eq.~\eqref{eq:delta-q-definition} gives
\begin{equation}
 \delta q
 =
 \frac{q_{\min}/q_0-1}{M-1}
 \sim
 \frac{\varepsilon}{q_0(M-1)}.
\label{eq:asymp-delta-near-boundary}
\end{equation}
Also,
\begin{equation}
 r^\tau
 =
 \frac{q_{\min}-q_0}{1-q_0}
 =
 \frac{\varepsilon}{1-q_0}
 \longrightarrow 0
 \qquad
 (\varepsilon\downarrow0).
\label{eq:asymp-r-tau-boundary}
\end{equation}
Since \(0<r<1\), Eq.~\eqref{eq:asymp-r-tau-boundary} implies that the crossing time diverges, \(\tau\to\infty\), as \(\varepsilon\downarrow0\).  The infinite genome genealogical variance \(v_\infty^{\scal}\) therefore tends to the fixed point of the scalar recursion in Eq.~\eqref{eq:scalar-tv-recursion}, as in Eq.~\eqref{eq:asymp-vstar-consistency}:
\begin{equation}
 v_*^{\scal}
 =
 \lim_{t\to\infty}v_\infty^{\scal}(t)
 =
 \frac{D^{\scal}}{1-b}
 =
 \frac{a^2(1-q_0)^2}{4M\left[1-\dfrac{a^2}{4}\left(1-\dfrac1M\right)\right]}.
\label{eq:asymp-vstar-scalar}
\end{equation}
At the crossing time \(\tau\), \(v_\infty^{\scal}(\tau)\approx v_*^{\scal}\), which is the appropriate limit because \(\tau\to\infty\).  In the denominator of Eq.~\eqref{eq:scalar-Lc}, the term \(\delta q^2\) is \(O(\varepsilon^2)\), whereas \(2\delta q\sqrt{v_*^{\scal}}\) is \(O(\varepsilon)\).  Hence, for sufficiently small \(\varepsilon\), the linear term dominates.  Replacing \(1-q_{\min}^2\) by its boundary value \(1-q_0^2\), which is correct to leading order in \(\varepsilon\), gives
\begin{equation}
 L_c^{\scal}
 \sim
 \frac{\left(1-q_0^2-v_*^{\scal}\right)q_0(M-1)}
 {2\varepsilon\sqrt{v_*^{\scal}}},
 \qquad
 \varepsilon=q_{\min}-q_0.
\label{eq:asymp-Lc-boundary-linear}
\end{equation}
Using the leading numerator approximation,
\begin{equation}
\begin{aligned}
 \widetilde L_c^{\scal}
 &\sim
 \frac{1-q_0^2}{2\delta q\sqrt{v_*^{\scal}}}
 =
 \frac{(1-q_0^2)q_0(M-1)}
 {2(q_{\min}-q_0)\sqrt{v_*^{\scal}}},
 \\
 v_*^{\scal}
 &=
 \frac{D^{\scal}}{1-b}.
\end{aligned}
\label{eq:asymp-Lc-boundary-leading-numerator}
\end{equation}
Thus, near the boundary value \(q_0\), both scalar formulas have the same linear boundary exponent,
\begin{equation}
 L_c^{\scal}\propto \frac{1}{q_{\min}-q_0},
 \qquad
 \widetilde L_c^{\scal}\propto \frac{1}{q_{\min}-q_0},
 \qquad
 \text{as }q_{\min}\downarrow q_0.
\label{eq:asymp-Lc-boundary-proportional}
\end{equation}
If the genealogical variance were ignored, the denominator would scale as \(\delta q^2\propto\varepsilon^2\), giving a quadratic divergence proportional to \((q_{\min}-q_0)^{-2}\).  The transient genealogical width therefore softens the boundary divergence from quadratic to linear.

\subsection{Asymptotic matching between overlapping regimes}
\label{subsec:asymptotic-consistency}

All five regimes were obtained by taking different limits of the same scalar transient variance formula.  They should therefore agree wherever their domains of validity overlap.  The matching calculations below are not extra assumptions; they show how the asymptotic reductions fit together into a single coherent picture of the underlying expression.  Because the same expression is being viewed from several limiting directions, the analysis below is organized by the limiting assumptions being matched and by the term in the denominator of Eq.~\eqref{eq:scalar-Lc} that controls the leading behavior.

\paragraph{Regime~I and Regime~III(a): the weak mutation overlap of the fixed \texorpdfstring{\(\mu\)}{mu} formula.}
Regime~I takes \(M\to\infty\) at fixed \(\mu>0\), whereas Regime~III takes \(M\mu\gg1\) with \(\mu\ll1\).  These limits overlap when \(M\) is large, \(\mu\) is small, and still \(M\mu\gg1\) and \(\mu\sqrt M\gg1\).  In this overlap the deterministic crossing contribution \(\delta q^2\) in the denominator of Eq.~\eqref{eq:scalar-Lc} is larger than the genealogical width contribution \(2\delta q\sqrt{v_\infty^{\scal}(\tau)}\).  Thus the leading expression is obtained by taking the small \(\mu\) limit of the fixed \(\mu\) Regime~I formula.  From Eq.~\eqref{eq:asymp-Lc-fixed-mu},
\begin{equation}
 \frac{1-q_{\min}^2}{q_{\min}^2\left(\ee^{4\mu}-1\right)^2}
 \sim
 \frac{1-q_{\min}^2}{16q_{\min}^2\mu^2},
\label{eq:asymp-consistency-I-IIIa}
\end{equation}
which is exactly the Regime~III(a) formula in Eq.~\eqref{eq:asymp-Lc-large-Mmu-mutation-dominated}.  Thus Regime~III(a) is the weak mutation, large population limit of Regime~I in the part of parameter space where \(\delta q^2\) controls the denominator.

\paragraph{Regime~II and Regime~III(b): the large \texorpdfstring{\(\kappa\)}{kappa} overlap \texorpdfstring{\(1\ll\kappa\ll\sqrt M\)}{1 << kappa << sqrt(M)}.}
Regime~II was derived for \(\kappa=4M\mu=O(1)\), but the expression containing both denominator terms in Eq.~\eqref{eq:asymp-Lc-kappa-uniform} can also be read in an overlap window where \(\kappa\) grows with \(M\).  For fixed \(q_{\min}\) and \(\kappa\gg1\),
\begin{equation}
 \gamma=q_{\min}(1+\kappa)-1\sim q_{\min}\kappa.
\label{eq:asymp-consistency-gamma-large-kappa}
\end{equation}
The Regime~II formula in Eq.~\eqref{eq:asymp-Lc-kappa-M32} is obtained from Eq.~\eqref{eq:asymp-Lc-kappa-uniform} when the genealogical width term in Eq.~\eqref{eq:asymp-kappa-denominator-terms} dominates \(\delta q^2\).  This dominance requires \(\gamma\ll\sqrt M\).  Using Eq.~\eqref{eq:asymp-consistency-gamma-large-kappa}, this becomes
\begin{equation}
 1\ll\kappa\ll\sqrt M.
\label{eq:asymp-consistency-kappa-window}
\end{equation}
Since \(\kappa=4M\mu\), the same window is, up to constants,
\begin{equation}
 M^{-1}\ll\mu\ll M^{-1/2},
\label{eq:asymp-consistency-mu-window}
\end{equation}
which is precisely Regime~III(b).  Substituting \(\kappa=4M\mu\) and Eq.~\eqref{eq:asymp-consistency-gamma-large-kappa} into Eq.~\eqref{eq:asymp-Lc-kappa-M32} gives
\begin{equation}
 \frac{\sqrt3(1+q_{\min})}{2\,[q_{\min}(1+\kappa)-1]}M^{3/2}
 \sim
 \frac{\sqrt3(1+q_{\min})}{2q_{\min}\kappa}M^{3/2}
 =
 \frac{\sqrt3(1+q_{\min})}{8q_{\min}}\,\frac{\sqrt M}{\mu},
\label{eq:asymp-consistency-II-IIIb}
\end{equation}
which is the Regime~III(b) result in Eq.~\eqref{eq:asymp-Lc-large-Mmu-width-dominated}.  Thus Regime~II and Regime~III(b) meet smoothly in the window \(1\ll\kappa\ll\sqrt M\).

\paragraph{Regime~II and Regime~V: the threshold limit \texorpdfstring{\(q_{\min}\downarrow q_0\)}{qmin -> q0}.}
Now let \(q_{\min}\) approach \(q_0\) from above while remaining within Regime~II.  Equivalently, we approach the boundary \(q_{\min}=q_0\) of the crossing condition \(q_{\min}>q_0\) in Eq.~\eqref{eq:tv-regime-qmin}.  Keep \(\kappa=O(1)\) and write
\begin{equation}
 q_{\min}=q_0+\varepsilon,
 \qquad
 0<\varepsilon\ll1,
 \qquad
 q_0\approx\frac{1}{1+\kappa},
\label{eq:asymp-consistency-II-boundary-def}
\end{equation}
where the last approximation is the Regime~II relation in Eq.~\eqref{eq:asymp-rho-kappa}.  The parameter \(\gamma\) can then be expanded directly:
\begin{align}
 \gamma
 &=q_{\min}(1+\kappa)-1 \nonumber\\
 &=(q_0+\varepsilon)(1+\kappa)-1 \nonumber\\
 &=\bigl[q_0(1+\kappa)-1\bigr]+(1+\kappa)\varepsilon \nonumber\\
 &\sim (1+\kappa)\varepsilon
 \sim \frac{\varepsilon}{q_0}.
\label{eq:asymp-consistency-gamma-epsilon}
\end{align}
The bracketed term vanishes to leading order because \(q_0\approx1/(1+\kappa)\).  The last step uses the same relation, \(1+\kappa\approx1/q_0\).

At this stage it is simplest to apply the leading Regime~II formula in Eq.~\eqref{eq:asymp-Lc-kappa-M32}.  Since \(\gamma=q_{\min}(1+\kappa)-1\),
\begin{align}
 L_c^{\scal}
 &\sim
 \frac{\sqrt3(1+q_{\min})}{2\,[q_{\min}(1+\kappa)-1]}M^{3/2} \nonumber\\
 &=
 \frac{\sqrt3(1+q_{\min})}{2\gamma}M^{3/2} \nonumber\\
 &\sim
 \frac{\sqrt3\,q_0(1+q_0)}{2\varepsilon}M^{3/2}.
\label{eq:asymp-consistency-II-V-from-II}
\end{align}
The final line uses \(q_{\min}\sim q_0\) together with Eq.~\eqref{eq:asymp-consistency-gamma-epsilon}.

The same expression is obtained from the Regime~V leading scalar formula in Eq.~\eqref{eq:asymp-Lc-boundary-leading-numerator}.  Indeed, in the Regime~II scaling,
\begin{equation}
 M\to\infty,
 \qquad
 \kappa=O(1),
 \qquad
 \mu=\frac{\kappa}{4M}.
\label{eq:asymp-consistency-II-scaling-restated}
\end{equation}
At the threshold limit \(q_{\min}=q_0+\varepsilon\) with \(\varepsilon\downarrow0\), Regime~V has \(\tau\to\infty\), so \(v_\infty^{\scal}(\tau)\approx v_*^{\scal}\).  At the order used in Regime~II, this fixed point value is equivalently obtained from Eq.~\eqref{eq:asymp-sqrt-v-kappa}:
\begin{equation}
 \sqrt{v_*^{\scal}}
 \sim
 \sqrt{v_\infty^{\scal}(\tau)}
 \sim
 \frac{1-q_{\min}}{\sqrt{3M}}
 \sim
 \frac{1-q_0}{\sqrt{3M}}.
\label{eq:asymp-consistency-sqrt-vstar-kappa}
\end{equation}
Now apply Eq.~\eqref{eq:asymp-Lc-boundary-leading-numerator}.  Using \(q_{\min}-q_0=\varepsilon\), \(M-1\sim M\), and Eq.~\eqref{eq:asymp-consistency-sqrt-vstar-kappa},
\begin{equation}
 2(q_{\min}-q_0)\sqrt{v_*^{\scal}}
 \sim
 2\varepsilon\,\frac{1-q_0}{\sqrt{3M}}.
\label{eq:asymp-consistency-boundary-denominator-expanded}
\end{equation}
Hence
\begin{equation}
\begin{aligned}
 \widetilde L_c^{\scal}
 &\sim
 \frac{(1-q_0^2)q_0M}
 {2\varepsilon(1-q_0)/\sqrt{3M}} \\
 &=
 \frac{(1-q_0)(1+q_0)q_0M\sqrt{3M}}
 {2\varepsilon(1-q_0)} \\
 &=
 \frac{\sqrt3\,q_0(1+q_0)}{2\varepsilon}M^{3/2}.
\end{aligned}
\label{eq:asymp-consistency-II-V-from-V}
\end{equation}
Equation~\eqref{eq:asymp-consistency-II-V-from-V} is the same leading expression as Eq.~\eqref{eq:asymp-consistency-II-V-from-II}.  Thus the Regime~II law contains the Regime~V linear divergence in the explicit threshold distance \(q_{\min}-q_0=\varepsilon\).

\paragraph{Regime~IV and Regime~V: the small \texorpdfstring{\(M\mu\)}{M mu} threshold limit \texorpdfstring{\(s\uparrow\kappa\)}{s -> kappa}.}
Now consider Regime~IV, for which \(\kappa=4M\mu\ll1\), \(q_{\min}=1-s\), \(q_0\approx1-\kappa\), and \(0<s<\kappa\ll1\).  In this regime, we also obtain
\begin{equation}
 q_{\min}-q_0=(1-s)-(1-\kappa)+O(\kappa^2)
 =\kappa-s+O(\kappa^2).
\label{eq:asymp-consistency-kappa-minus-s-distance}
\end{equation}
Thus, at the level of approximation used in Regime~IV, approaching the threshold boundary \(q_{\min}=q_0\) is the same as taking \(s\uparrow\kappa\), or \(\kappa-s\downarrow0\).

Starting from the leading Regime~IV formula in Eq.~\eqref{eq:asymp-Lc-small-Mmu-leading},
\begin{equation}
 \widetilde L_c^{\scal}
 \sim
 \frac{2sM^2}
 {(\kappa-s)^2+\dfrac{2s(\kappa-s)}{\sqrt3}\sqrt M}.
\label{eq:asymp-consistency-IV-leading-restated}
\end{equation}
Near \(s=\kappa\), the denominator has a quadratic deterministic term and a linear genealogical width term.  Their ratio is
\begin{equation}
 \frac{(\kappa-s)^2}
 {\dfrac{2s(\kappa-s)}{\sqrt3}\sqrt M}
 =
 \frac{\sqrt3(\kappa-s)}{2s\sqrt M}.
\label{eq:asymp-consistency-IV-denominator-ratio}
\end{equation}
Therefore, in the boundary layer where \(\kappa-s\ll s\sqrt M\), the linear genealogical width term dominates, and Eq.~\eqref{eq:asymp-consistency-IV-leading-restated} reduces to
\begin{equation}
 \widetilde L_c^{\scal}
 \sim
 \frac{2sM^2}{2s(\kappa-s)\sqrt M/\sqrt3}
 =
 \frac{\sqrt3 M^{3/2}}{\kappa-s}.
\label{eq:asymp-consistency-IV-V-from-IV}
\end{equation}

Regime~V gives the same small \(\kappa\) limit from Eq.~\eqref{eq:asymp-Lc-boundary-leading-numerator}.  Indeed, in the small \(\kappa\) scaling,
\begin{equation}
 q_0\sim1,
 \qquad
 1-q_0\sim\kappa,
 \qquad
 1-q_0^2\sim2\kappa,
 \qquad
 \sqrt{v_*^{\scal}}\sim\frac{\kappa}{\sqrt{3M}}.
\label{eq:asymp-consistency-small-kappa-boundary}
\end{equation}
Together with \(q_{\min}-q_0\sim\kappa-s\) and \(M-1\sim M\), Eq.~\eqref{eq:asymp-Lc-boundary-leading-numerator} gives
\begin{equation}
 \widetilde L_c^{\scal}
 \sim
 \frac{2\kappa M}{2(\kappa-s)\kappa/\sqrt{3M}}
 =
 \frac{\sqrt3 M^{3/2}}{\kappa-s}.
\label{eq:asymp-consistency-IV-V-from-V}
\end{equation}
This matches Eq.~\eqref{eq:asymp-consistency-IV-V-from-IV}.  Thus the Regime~IV formula reduces to the Regime~V threshold law when \(s\uparrow\kappa\), i.e., when \(q_{\min}=1-s\) approaches \(q_0\approx1-\kappa\) from above.

Taken together, the above analysis shows that the five regimes are not separate, ad hoc approximations but different projections of a single transient variance formula.  Regime~I and Regime~III(a) match in the weak mutation overlap where \(\delta q^2\) dominates; Regime~II and Regime~III(b) match in the large \(\kappa\) overlap where \(2\delta q\sqrt{v_\infty^{\scal}(\tau)}\) dominates; and both Regime~II and Regime~IV reduce to the same linear Regime~V threshold divergence as \(q_{\min}\downarrow q_0\).

\begin{figure}[p]
\centering
\begin{tikzpicture}[
  x=1.20cm,y=1.20cm,
  every node/.style={font=\scriptsize},
  regionbox/.style={align=center, rounded corners=5pt, inner sep=2.6pt, fill=white, fill opacity=0.80, text opacity=1, draw=black!16},
  linebox/.style={align=center, rounded corners=5pt, inner sep=2.4pt, fill=white, fill opacity=0.84, text opacity=1, draw=black!16},
  note/.style={align=center, rounded corners=5pt, inner sep=3pt, fill=blue!4, fill opacity=0.94, draw=blue!35!black, text opacity=1}
]
  \def\xmax{11.2}
  \def\ymax{7.35}
  \def\c{6.2}
  \def\d{1.25}

  \fill[green!16] (0,\c) -- (0,\ymax) -- (\xmax,\ymax) -- (\xmax,\d) -- cycle;
  \fill[orange!20] (0,\c) -- (\c,0) -- (\xmax,0) -- (\xmax,\d) -- cycle;
  \fill[cyan!17] (0,0) -- (0,\c) -- (\c,0) -- cycle;

  \draw[line width=17pt, draw=violet!27, opacity=0.80] (0,\c) -- (\c,0);

  \draw[dashed, very thick, violet!70!black] (0,\c) -- (\c,0);
  \draw[dotted, very thick, blue!70!black] (0,\c) -- (\xmax,\d);

  \draw[->, thick] (0,0) -- (\xmax+0.35,0) node[right] {\(\log_{10}M\)};
  \draw[->, thick] (0,0) -- (0,\ymax+0.35) node[above] {\(\log_{10}\mu\)};
  \node[anchor=north east, font=\scriptsize] at (\xmax,0) {larger \(M\)};

  \node[linebox, text width=2.95cm] at ($(0,\c)!0.25!(\c,0)$)
    {dashed: \(M\mu=O(1)\)};
  \node[linebox, text width=3.05cm] at ($(0,\c)!0.67!(\xmax,\d)$)
    {dotted: \(\mu\sqrt M=O(1)\)};

  \node[regionbox, text width=3.85cm] at (7.85,5.62)
  {\textcolor{green!45!black}{\textbf{Regime I and III(a)}}\\[1.5pt]
   mutation term dominates\\[1.5pt]
   \(\displaystyle
   \widetilde L_c^{\scal}\sim
   \frac{1-q_{\min}^2}{16q_{\min}^2\mu^2}
   \)};

  \node[regionbox, text width=3.70cm] at (8.20,0.95)
  {\textcolor{orange!75!black}{\textbf{Regime III(b)}}\\[1.5pt]
   genealogical width dominates\\[1.5pt]
   \(\displaystyle
   \widetilde L_c^{\scal}\sim
   \frac{\sqrt3(1+q_{\min})}{8q_{\min}}\frac{\sqrt M}{\mu}
   \)};

  \node[regionbox, text width=3.80cm] at (2.05,1.05)
  {\textcolor{cyan!60!black}{\textbf{Regime IV}}\\[1.5pt]
   \(M\mu\ll1\), \(q_{\min}=1-s\)\\[1.5pt]
   \(\displaystyle
   \widetilde L_c^{\scal}\sim
   \frac{2sM^2}{(\kappa-s)^2+2s(\kappa-s)\sqrt M/\sqrt3}
   \)};

  \node[regionbox, text width=3.55cm] at (2.95,3.25)
  {\textcolor{violet!40}{\textbf{Regime II}}\\[1.5pt]
   \(\kappa=4M\mu=O(1)\), \\
   \(\gamma=q_{\min}(1+\kappa)-1\)\\[1.5pt]
   \(\displaystyle
   \widetilde L_c^{\scal}\sim
   \frac{\sqrt3(1+q_{\min})}{2\gamma}M^{3/2}
   \)};

  \node[note, text width=8.5cm, anchor=north] at (5.60,-0.72)
  {\textbf{Regime V is a boundary in threshold space:}
   \(q_{\min}\downarrow q_0\), with
   \(\widetilde L_c^{\scal}\propto(q_{\min}-q_0)^{-1}\).
   It is not a separate region in the \((M,\mu)\) plane.};
\end{tikzpicture}
\caption{Schematic map of the asymptotic regimes in the \((\log_{10}M,\log_{10}\mu)\) plane.  The dashed curve marks \(M\mu=O(1)\), and the dotted curve marks \(\mu\sqrt M=O(1)\).  The colors identify the regimes: \textcolor{green!45!black}{green} marks Regime~I and Regime~III(a); \textcolor{orange!75!black}{orange} marks Regime~III(b); \textcolor{cyan!60!black}{cyan} marks Regime~IV; and the \textcolor{violet!40}{violet} band marks Regime~II.  The corresponding leading critical genome length formula is displayed inside each region.  The map is schematic, not a sharp phase diagram; constants multiplying the crossover curves depend on \(q_{\min}\).}
\label{fig:asymp-regime-map}
\end{figure}
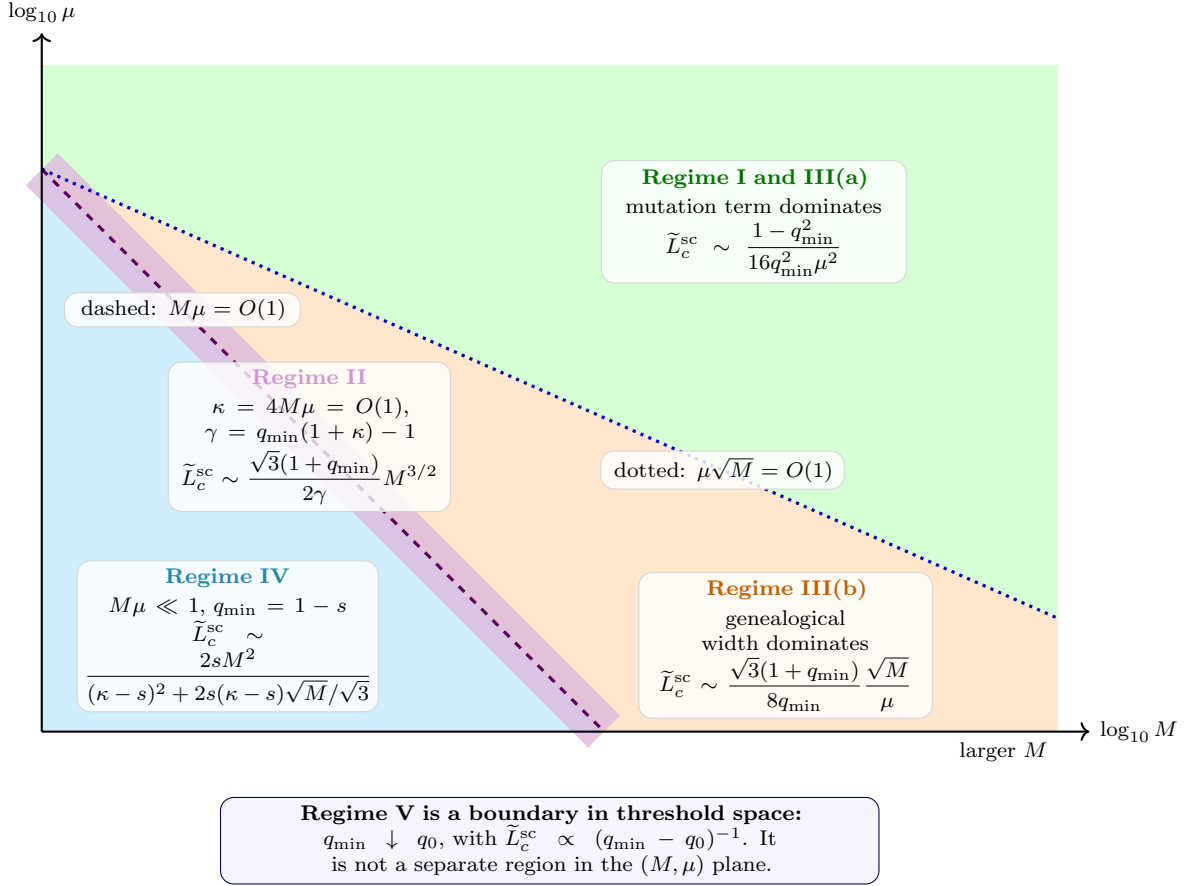

Figure~\ref{fig:asymp-regime-map} illustrates these regimes schematically, with the colored regions serving as visual guides to the parameter ranges of each leading formula.  Table~\ref{tab:asymp-summary} provides a more detailed summary, collecting the limiting laws developed in this section together with their assumptions.

\clearpage
\clearpage

\begin{table}[p]
\centering
\caption{Summary of the main asymptotic regimes for the critical genome length.  The first column gives the parameter assumptions defining each regime.  The second column reports the leading scalar approximation \(\widetilde L_c^{\scal}\) and briefly indicates how the deterministic crossing term \(\delta q^2\) and the genealogical width term \(2\delta q\sqrt{v_\infty^{\scal}(\tau)}\) enter the scaling.  Only \(\widetilde L_c^{\scal}\) is displayed.  The full scalar formula \(L_c^{\scal}\) retains the scalar numerator correction, while the matrix formula replaces \(v_\infty^{\scal}\) by the closed form \(v_\infty^{\mat}\).  These refinements do not change the leading powers shown here.}
\label{tab:asymp-summary}
\vspace{0.8em}
\small
\renewcommand{\arraystretch}{1.23}
\begin{tabular}{>{\raggedright\arraybackslash}p{0.27\textwidth}>{\raggedright\arraybackslash}p{0.65\textwidth}}
\toprule
\rowcolor{blue!8}
\textcolor{blue!60!black}{\large\bfseries Regime} &
\textcolor{blue!60!black}{\large\bfseries Leading behavior of \(\widetilde L_c^{\scal}\)} \\
\midrule
\textbf{Regime I:} large \(M\), fixed \(\mu>0\) &
\(\displaystyle
\widetilde L_c^{\scal}
\sim
\frac{1-q_{\min}^2}{q_{\min}^2(\ee^{4\mu}-1)^2}
\).
For small fixed \(\mu\), this reduces to
\(\displaystyle \widetilde L_c^{\scal}\sim\frac{1-q_{\min}^2}{16q_{\min}^2\mu^2}\).
The deterministic crossing term \(\delta q^2\) sets the scale. \\
\addlinespace[0.7em]
\textbf{Regime II:} \(\kappa=4M\mu=O(1)\) &
Crossing requires \(q_{\min}>1/(1+\kappa)\).  With
\(\gamma=q_{\min}(1+\kappa)-1>0\),
\(\displaystyle
\widetilde L_c^{\scal}
\sim
\frac{\sqrt3(1+q_{\min})}{2\gamma}M^{3/2}
\).
The genealogical width term changes the scaling from the deterministic \(M^2\) estimate to \(M^{3/2}\). \\
\addlinespace[0.7em]
\textbf{Regime III(a):} \(M\mu\gg1\), \(\mu\ll1\), \(\mu\sqrt M\gg1\) &
\(\displaystyle
\widetilde L_c^{\scal}
\sim
\frac{1-q_{\min}^2}{16q_{\min}^2\mu^2}
\).
The mutation term \(\delta q^2\) dominates, so the leading scale is \(\mu^{-2}\). \\
\addlinespace[0.7em]
\textbf{Regime III(b):} \(M\mu\gg1\), \(\mu\ll1\), \(\mu\sqrt M\ll1\) &
Equivalently, \(M^{-1}\ll\mu\ll M^{-1/2}\).  The genealogical width term dominates and
\(\displaystyle
\widetilde L_c^{\scal}
\sim
\frac{\sqrt3(1+q_{\min})}{8q_{\min}}\frac{\sqrt M}{\mu}
\).
This is the overlap window with Regime~II when \(1\ll\kappa\ll\sqrt M\). \\
\addlinespace[0.7em]
\textbf{Regime IV:} \(M\mu\ll1\) &
A crossing is possible only for a threshold close to perfect similarity.  Write
\(q_{\min}=1-s\), \(\kappa=4M\mu\), and \(0<s<\kappa\ll1\).  Then
\(\displaystyle
\widetilde L_c^{\scal}
\sim
\frac{2sM^2}{(\kappa-s)^2+2s(\kappa-s)\sqrt M/\sqrt3}
\).
This first order formula displays the approach to the boundary as \(s\uparrow\kappa\). \\
\addlinespace[0.7em]
\textbf{Regime V:} \(q_{\min}\downarrow q_0\) &
No assumption \(M\mu\ll1\) is needed.  With \(v_*^{\scal}=D^{\scal}/(1-b)\),
\(\displaystyle
\widetilde L_c^{\scal}
\sim
\frac{(1-q_0^2)q_0(M-1)}{2(q_{\min}-q_0)\sqrt{v_*^{\scal}}}
\),
so \(\displaystyle \widetilde L_c^{\scal}\propto(q_{\min}-q_0)^{-1}\).  The genealogical width term softens the boundary divergence from quadratic to linear. \\
\bottomrule
\end{tabular}
\end{table}
\clearpage

\section{Numerical evaluation and simulation tests}
\label{sec:numerical-tests}

This section evaluates the theory at three progressively more demanding levels.  The first subsection asks a quantitative finite \(M\) question: how closely do the closed form matrix and scalar predictions reproduce the benchmark obtained from the full system of coupled moment equations.  The second asks a scaling question: once the scalar formula is reduced further to its limiting asymptotic forms, how much accuracy is retained, and in which regions of parameter space.  The third moves outside the unrestricted moment hierarchy and compares the theory directly with stochastic simulations of the nonlinear species formation model, both at finite parameter values and through finite range tests of the predicted scaling laws.

\subsection{Accuracy of the closed form formulas}
\label{subsec:comparison-closed-form-critical-lengths}

The purpose of this subsection is to test whether the closed form formulas reproduce the critical genome length obtained from the most detailed system of coupled moment equations, presented in Section~\ref{subsec:full-to-reduced-system}.  We use this full hierarchy of moment equations as the benchmark.  This hierarchy evolves the deterministic inputs \(m_t\) and \(h_t\), together with the three genealogical second moments \(v_t\), \(c_t\), and \(d_t\).  We denote by \(L_c^{\full}\) the corresponding critical genome length, obtained by iterating the full system of coupled moment equations (see the Supplementary Material for details).  In the comparison below, the two approximations are the reduced matrix value \(L_c^{\mat}\) from Eq.~\eqref{eq:matrix-Lc}, which keeps the covariance \(c_t\) but drops the disjoint pair covariance \(d_t\), and the scalar value \(L_c^{\scal}\) from Eq.~\eqref{eq:scalar-Lc}, which keeps only the leading scalar propagation of the genealogical variance.

The results are presented in Tables~\ref{tab:closed-form-accuracy-M1000} and \ref{tab:closed-form-accuracy-M5000}.  All quantities are evaluated at the continuous crossing time \(\tau\), where \(m_\tau=q_{\min}\).  To compare the formulas on the same scale, we report in these tables the signed relative error of each approximation against the full system, in percent: for any quantity \(X\in\{L_c,v_\infty,v_1\}\) and label \(\alpha\in\{\mat,\scal\}\), \(\mathcal{E}_X^{\alpha}=X^{\alpha}/X^{\full}-1\), where \(X^{\full}\) is the full hierarchy value obtained by iterating the full system of coupled moment equations. A positive value means the approximation overestimates the benchmark and a negative value that it underestimates it.  The error in \(v_\infty\) measures how well the approximations capture the infinite genome genealogical width; the error in \(v_1\) measures the associated finite locus correction coefficient; and the error in \(L_c\) measures the combined effect on the critical genome length.  Thus the entry \(+0.0028/-0.0883\) in the \(\mathcal{E}_{L_c}^{\mat}/\mathcal{E}_{L_c}^{\scal}\) column of Table~\ref{tab:closed-form-accuracy-M1000} means that the matrix formula overestimates \(L_c^{\full}\) by \(0.0028\%\), whereas the scalar formula underestimates it by \(0.0883\%\).

\begin{figure}[p]
\centering
\includegraphics[width=0.82\textwidth]{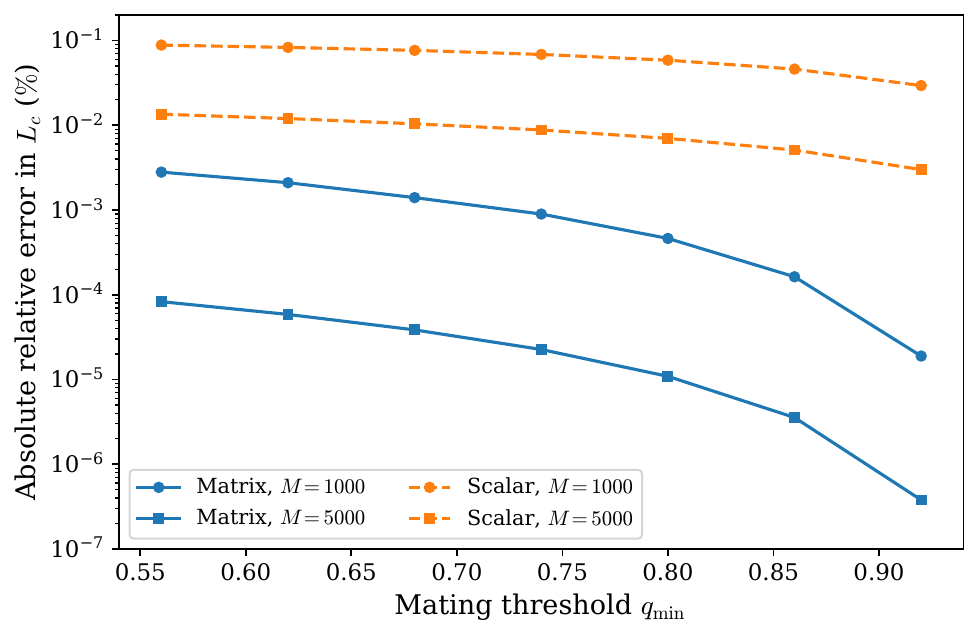}
\caption{Absolute relative error in the critical genome length against the full hierarchy benchmark for \(\mu=0.0025\).  Solid blue curves show the reduced matrix formula; dashed orange curves show the scalar formula.  Circles correspond to \(M=1000\), and squares to \(M=5000\).  The matrix formula is already a very close finite \(M\) reduction, while the scalar formula retains excellent accuracy, which improves rapidly with population size.}
\label{fig:closed-form-accuracy-profile}
\end{figure}

As seen in Tables~\ref{tab:closed-form-accuracy-M1000} and \ref{tab:closed-form-accuracy-M5000}, the matrix formula is almost indistinguishable from the full hierarchy even at \(M=1000\), with errors below \(3\times10^{-3}\%\) throughout the entire range of \(q_{\min}\) values.  The scalar formula is deliberately simpler, but its critical genome length error is still below \(0.09\%\) for \(M=1000\) and below \(0.014\%\) for \(M=5000\).  The scalar error in \(v_\infty\) is larger than its error in \(L_c\), because \(v_\infty\) enters the critical length only through the combination \(2\delta q\sqrt{v_\infty}\) in the denominator and because \(v_1=1-q_{\min}^2-v_\infty\) is numerically dominated by \(1-q_{\min}^2\) over this range.  Thus the scalar formula captures the final critical genome length more accurately than it captures the genealogical variance by itself.  Figure~\ref{fig:closed-form-accuracy-profile} summarizes the same comparison by plotting the absolute relative error in \(L_c\) on a logarithmic scale.

The conclusion from this comparison is that the reduced hierarchy is not merely qualitatively correct.  The two variable matrix formula reproduces the full moment hierarchy at numerical precision relevant for the present theory, and the scalar formula supplies a highly accurate leading reduction.  This justifies using \(L_c^{\mat}\) for finite \(M\) numerical prediction and \(L_c^{\scal}\), or its leading version \(\widetilde L_c^{\scal}\), for the asymptotic analysis.

\begin{table}[p]
\centering

\caption{Closed form accuracy for \(M=1000\) and \(\mu=0.0025\) for various \(q_{\min}\) values.  The full hierarchy value \(L_c^{\full}\) is computed from the coupled five variable moment equations \((m_t,h_t,v_t,c_t,d_t)\).  The matrix value \(L_c^{\mat}\) is the reduced two variable closed form, and \(L_c^{\scal}\) is the scalar closed form.  The paired error columns report matrix/scalar relative errors, in percent, against the full hierarchy values.}
\label{tab:closed-form-accuracy-M1000}
\vspace{8pt}
{\scriptsize
\setlength{\tabcolsep}{3.0pt}
\renewcommand{\arraystretch}{1.18}
\resizebox{\textwidth}{!}{%
\begin{tabular}{@{}c c r r r c c c@{}}
\toprule
\rowcolor{blue!8}
\textcolor{blue!60!black}{\bfseries \(q_{\min}\)} &
\textcolor{blue!60!black}{\bfseries \(\tau\)} &
\textcolor{blue!60!black}{\bfseries \(L_c^{\full}\)} &
\textcolor{violet!70!black}{\bfseries \(L_c^{\mat}\)} &
\textcolor{orange!75!black}{\bfseries \(L_c^{\scal}\)} &
\textcolor{cyan!60!black}{\bfseries \(\mathcal{E}_{L_c}^{\mat}/\mathcal{E}_{L_c}^{\scal}\) (\%)} &
\textcolor{cyan!60!black}{\bfseries \(\mathcal{E}_{v_\infty}^{\mat}/\mathcal{E}_{v_\infty}^{\scal}\) (\%)} &
\textcolor{cyan!60!black}{\bfseries \(\mathcal{E}_{v_1}^{\mat}/\mathcal{E}_{v_1}^{\scal}\) (\%)} \\
\midrule
0.56 & 60.11 & 6360.32 & 6360.50 & 6354.71 & \(+0.0028/-0.0883\) & \(-0.0075/+0.24\) & \(6.67\times10^{-7}/-2.09\times10^{-5}\) \\
0.62 & 49.18 & 5459.20 & 5459.31 & 5454.67 & \(+0.0021/-0.0829\) & \(-0.0060/+0.24\) & \(4.37\times10^{-7}/-1.74\times10^{-5}\) \\
0.68 & 39.42 & 4651.23 & 4651.29 & 4647.67 & \(+0.0014/-0.0765\) & \(-0.0046/+0.24\) & \(2.66\times10^{-7}/-1.41\times10^{-5}\) \\
0.74 & 30.61 & 3887.40 & 3887.44 & 3884.74 & \(8.96\times10^{-4}/-0.0685\) & \(-0.0032/+0.25\) & \(1.45\times10^{-7}/-1.11\times10^{-5}\) \\
0.80 & 22.57 & 3126.45 & 3126.47 & 3124.62 & \(4.62\times10^{-4}/-0.0586\) & \(-0.0020/+0.25\) & \(6.43\times10^{-8}/-8.16\times10^{-6}\) \\
0.86 & 15.20 & 2327.08 & 2327.09 & 2326.01 & \(1.63\times10^{-4}/-0.0460\) & \(-9.23\times10^{-4}/+0.26\) & \(1.90\times10^{-8}/-5.39\times10^{-6}\) \\
0.92 & 8.37  & 1440.07 & 1440.07 & 1439.64 & \(1.89\times10^{-5}/-0.0294\) & \(-1.80\times10^{-4}/+0.28\) & \(1.75\times10^{-9}/-2.73\times10^{-6}\) \\
\bottomrule
\end{tabular}%
}}

\vspace{3.2em}

\caption{Closed form accuracy for \(M=5000\) and \(\mu=0.0025\) for various values of \(q_{\min}\), with the same notation and error convention as in Table~\ref{tab:closed-form-accuracy-M1000}.  Increasing \(M\) reduces the relative errors of both the matrix and scalar formulas, as expected from the reduction hierarchy.}
\label{tab:closed-form-accuracy-M5000}
\vspace{8pt}
{\scriptsize
\setlength{\tabcolsep}{3.0pt}
\renewcommand{\arraystretch}{1.18}
\resizebox{\textwidth}{!}{%
\begin{tabular}{@{}c c r r r c c c@{}}
\toprule
\rowcolor{blue!8}
\textcolor{blue!60!black}{\bfseries \(q_{\min}\)} &
\textcolor{blue!60!black}{\bfseries \(\tau\)} &
\textcolor{blue!60!black}{\bfseries \(L_c^{\full}\)} &
\textcolor{violet!70!black}{\bfseries \(L_c^{\mat}\)} &
\textcolor{orange!75!black}{\bfseries \(L_c^{\scal}\)} &
\textcolor{cyan!60!black}{\bfseries \(\mathcal{E}_{L_c}^{\mat}/\mathcal{E}_{L_c}^{\scal}\) (\%)} &
\textcolor{cyan!60!black}{\bfseries \(\mathcal{E}_{v_\infty}^{\mat}/\mathcal{E}_{v_\infty}^{\scal}\) (\%)} &
\textcolor{cyan!60!black}{\bfseries \(\mathcal{E}_{v_1}^{\mat}/\mathcal{E}_{v_1}^{\scal}\) (\%)} \\
\midrule
0.56 & 58.39 & 9900.72 & 9900.73 & 9899.39 & \(8.29\times10^{-5}/-0.0135\) & \(-2.98\times10^{-4}/+0.0483\) & \(5.27\times10^{-9}/-8.55\times10^{-7}\) \\
0.62 & 48.07 & 8232.14 & 8232.15 & 8231.16 & \(5.86\times10^{-5}/-0.0120\) & \(-2.38\times10^{-4}/+0.0487\) & \(3.46\times10^{-9}/-7.09\times10^{-7}\) \\
0.68 & 38.73 & 6738.22 & 6738.23 & 6737.52 & \(3.85\times10^{-5}/-0.0104\) & \(-1.81\times10^{-4}/+0.0491\) & \(2.11\times10^{-9}/-5.74\times10^{-7}\) \\
0.74 & 30.21 & 5364.04 & 5364.04 & 5363.57 & \(2.26\times10^{-5}/-0.0088\) & \(-1.28\times10^{-4}/+0.0498\) & \(1.15\times10^{-9}/-4.47\times10^{-7}\) \\
0.80 & 22.37 & 4070.17 & 4070.17 & 4069.88 & \(1.09\times10^{-5}/-0.0070\) & \(-7.91\times10^{-5}/+0.0507\) & \(5.12\times10^{-10}/-3.28\times10^{-7}\) \\
0.86 & 15.10 & 2826.96 & 2826.96 & 2826.82 & \(3.56\times10^{-6}/-0.0051\) & \(-3.67\times10^{-5}/+0.0524\) & \(1.52\times10^{-10}/-2.16\times10^{-7}\) \\
0.92 & 8.34  & 1611.00 & 1611.00 & 1610.95 & \(3.79\times10^{-7}/-0.0030\) & \(-7.14\times10^{-6}/+0.0558\) & \(1.39\times10^{-11}/-1.09\times10^{-7}\) \\
\bottomrule
\end{tabular}%
}}
\end{table}

\subsection{Accuracy of the limiting asymptotic formulas}
\label{subsec:asymptotic-accuracy}

The purpose of this subsection is different from that of the preceding finite \(M\) comparison.  The closed form formulas in Eqs.~\eqref{eq:matrix-Lc} and \eqref{eq:scalar-Lc} are intended as numerical predictions of the critical genome length.  The limiting formulas in Section~\ref{sec:asymptotic-regimes} have a different role: they explain scaling by identifying which term in the transient variance denominator controls \(L_c\).  This is why the same closed form expression can reduce, in different limits, to powers such as \(\mu^{-2}\), \(M^{3/2}\), \(\sqrt M/\mu\), or \((q_{\min}-q_0)^{-1}\).

\begin{figure}[p]
\centering
\IfFileExists{asymptotic_ratios.pdf}{%
\includegraphics[width=\textwidth]{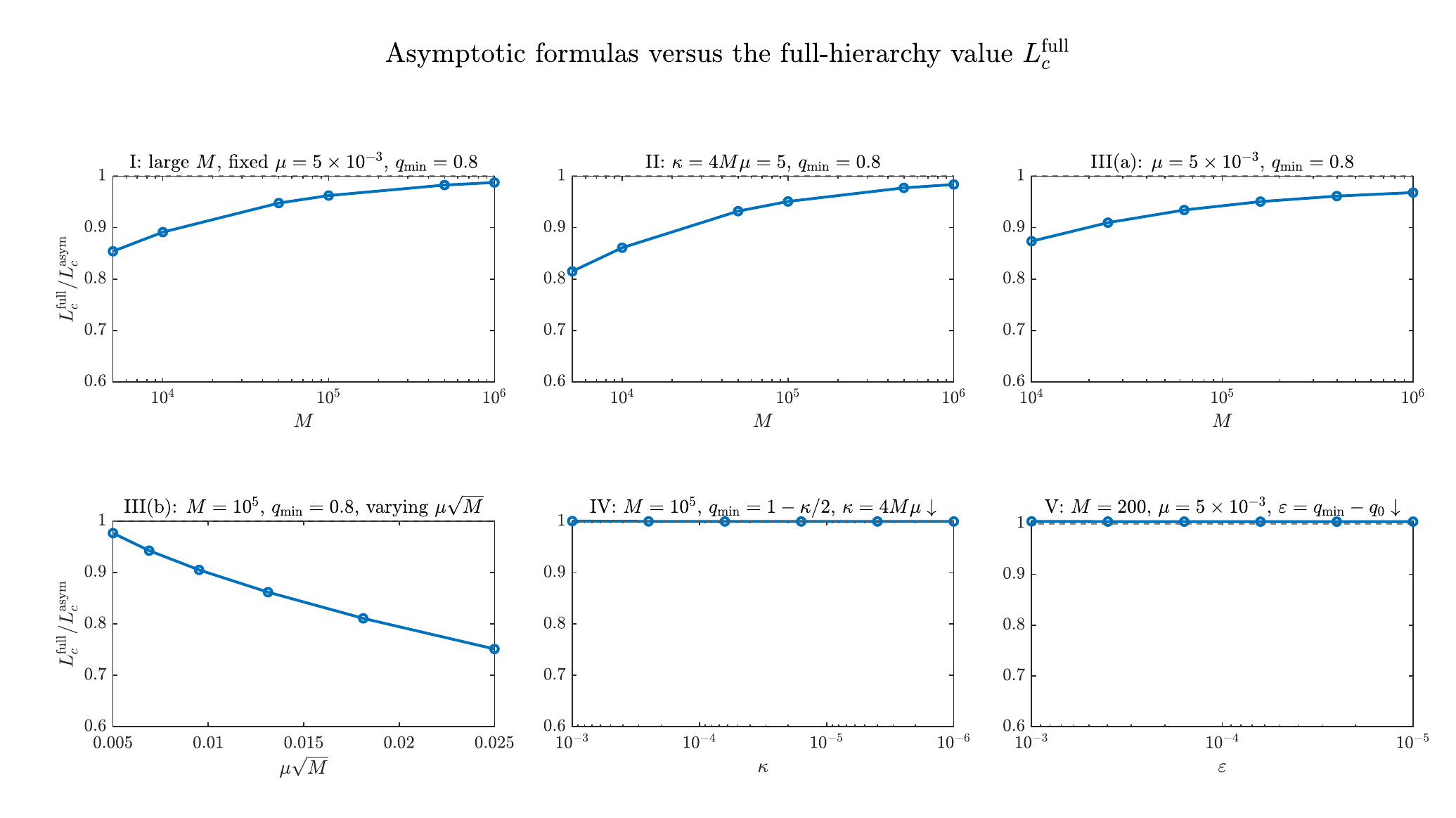}%
}{%
\missingfigure{asymptotic_ratios.pdf}%
}
\caption{Convergence of the full hierarchy critical genome length \(L_c^{\full}\) to the asymptotic formulas of Section~\ref{sec:asymptotic-regimes}.  Each panel plots the ratio \(L_c^{\full}/L_c^{\mathrm{asym}}\), and the dashed horizontal line marks perfect agreement.  The six panels test Regime~I, Regime~II, Regime~III(a), Regime~III(b), Regime~IV, and the boundary Regime~V.  The fixed population panels III(b) and IV use \(M=10^5\).}
\label{fig:asymptotic-ratio-tests}
\end{figure}

We first test the accuracy of these asymptotic formulas by approaching the limits in several distinct ways: fixed \(\mu\) with \(M\to\infty\), fixed \(\kappa=4M\mu\), the two subregimes of \(M\mu\gg1\), the regime with small \(M\mu\) near the threshold, and the boundary limit \(q_{\min}\downarrow q_0\).  Figure~\ref{fig:asymptotic-ratio-tests} compares each asymptotic expression directly with the corresponding full hierarchy value \(L_c^{\full}\), evaluated from the full system of coupled moment equations at the deterministic crossing time.  The ratios approach one rapidly in Regimes~I, II, IV, and V over the displayed numerical ranges.  Regimes~III(a) and III(b) converge more slowly, because their assumptions require two scale separations simultaneously: \(M\mu\gg1\) and either \(\mu\sqrt M\gg1\) or \(\mu\sqrt M\ll1\).  The remaining discrepancies therefore reflect the fact that the asymptotic limits have not yet been reached fully; the limiting expressions nevertheless become accurate over reasonable parameter ranges.

\begin{figure}[p]
\centering
\includegraphics[width=\textwidth]{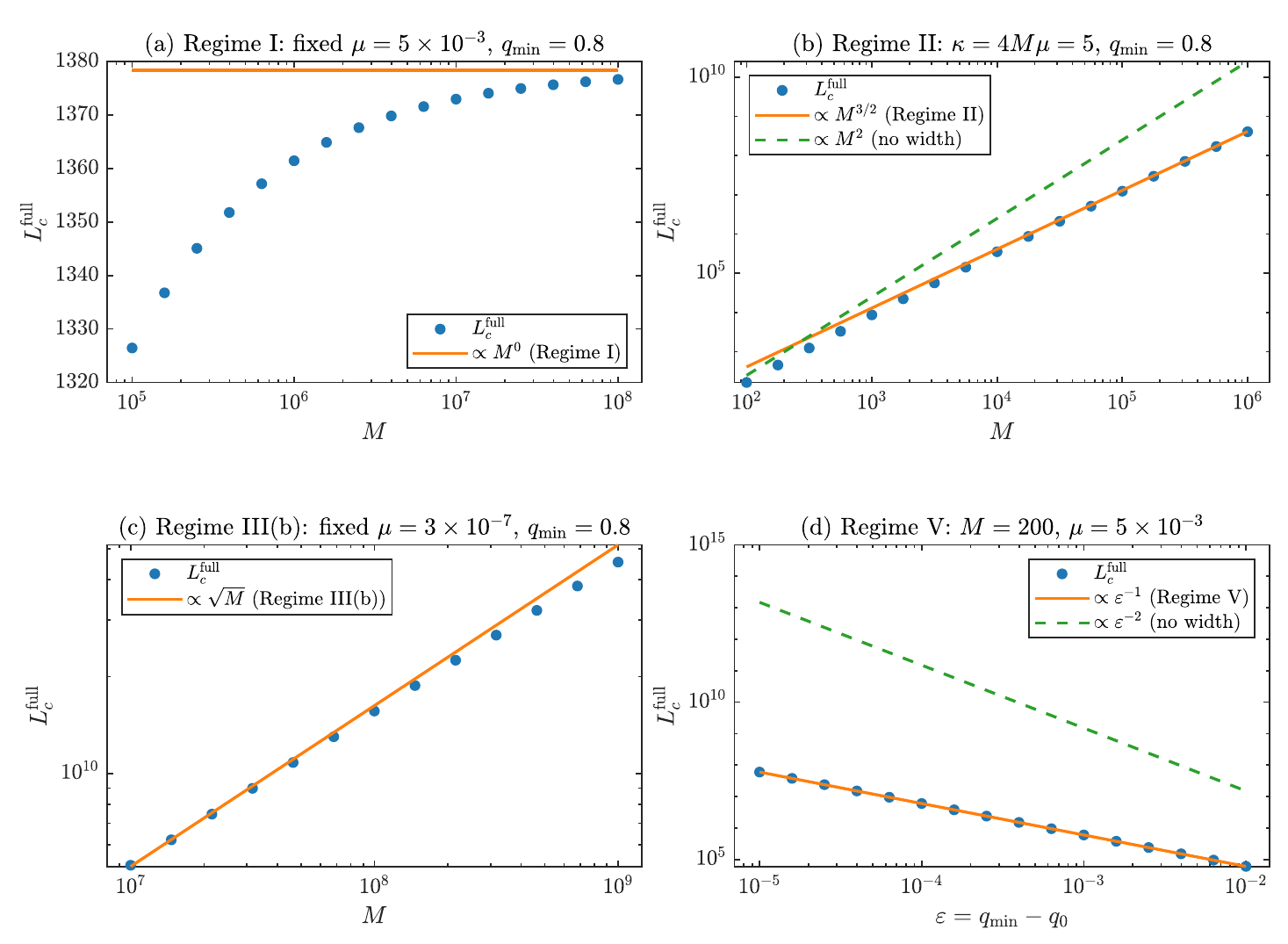}
\caption{Scaling tests using the full moment hierarchy.  Each panel plots the full hierarchy critical genome length \(L_c^{\full}\) against the corresponding asymptotic scaling law.  Top left: Regime~I with fixed \(\mu=0.005\) and \(q_{\min}=0.8\), showing the approach to the \(M^0\) fixed mutation asymptote, whose small \(\mu\) reduction is Eq.~\eqref{eq:asymp-Lc-fixed-small-mu}.  Top right: Regime~II with \(\kappa=4M\mu=5\) and \(q_{\min}=0.8\), where the full hierarchy values follow the \(M^{3/2}\) law rather than the \(M^2\) law obtained when the width is omitted.  Bottom left: Regime~III(b) with fixed \(\mu=3\times10^{-7}\) and \(q_{\min}=0.8\), showing the predicted \(\sqrt M\) scaling.  Bottom right: Regime~V with \(M=200\), \(\mu=0.005\), and \(\varepsilon=q_{\min}-q_0\), showing the linear boundary divergence \(\varepsilon^{-1}\) rather than the quadratic divergence \(\varepsilon^{-2}\) obtained when the width is omitted.  In all four panels the vertical axis is \(L_c^{\full}\).}
\label{fig:asymptotic-scaling-tests}
\end{figure}

We next use the same full hierarchy benchmark to illustrate the scaling consequences of the transient genealogical width.  Figure~\ref{fig:asymptotic-scaling-tests} compares \(L_c^{\full}\) with the leading powers predicted by the asymptotic analysis, again evaluated at the deterministic crossing time \(\tau\).  In Regime~I, at fixed small mutation rate, \(L_c^{\full}\) approaches an \(M^0\) law, so increasing the population size no longer changes the leading critical genome length.  In Regime~II, retaining the transient width term \(2\delta q\sqrt{v_\infty}\) changes the population size scaling from \(M^2\), which is obtained when the width is omitted, to \(M^{3/2}\).  In Regime~III(b), along a fixed small \(\mu\) path in the window \(M^{-1}\ll\mu\ll M^{-1/2}\), the full hierarchy values follow the predicted \(\sqrt M\) dependence.  Finally, near the infinite genome boundary, the same width term changes the divergence from \((q_{\min}-q_0)^{-2}\) to \((q_{\min}-q_0)^{-1}\).  These are not small numerical corrections; they are changes in exponent, and they are already visible over the numerical ranges shown.

Overall, the asymptotic tests show that the limiting formulas do what they are intended to do.  They are not a replacement for the matrix closed form when high finite \(M\) accuracy is needed.  Rather, they identify which part of the transient variance denominator controls the critical genome length and thereby explain why different powers of \(M\), \(\mu\), and \(q_{\min}-q_0\) appear in different limits.  The finite \(M\) benchmark in the previous subsection shows that these asymptotic explanations remain close to the full hierarchy of coupled moment equations.

\clearpage
\subsection{Direct simulations of the species formation model}
\label{subsec:direct-simulations}

Sections~\ref{subsec:comparison-closed-form-critical-lengths} and \ref{subsec:asymptotic-accuracy} tested the theory internally: first against the full hierarchy of the unrestricted coupled moment equations and then against the asymptotic reductions derived from the closed form expression.  We now make the more demanding comparison with the nonlinear species formation model restricted by the mating threshold.  The analysis has two parts.  We first ask whether the matrix prediction \(L_c^{\mat}\) in Eq.~\eqref{eq:matrix-Lc}, evaluated at finite population size, predicts the critical genome length \(L_c^{\mathrm{sim}}\) obtained from simulations as the mating threshold \(q_{\min}\), mutation rate \(\mu\), and population size \(M\) are varied.  We then ask whether direct simulations display the scaling predicted in Regimes~I, II, and V.  Because the asymptotic limits require either very large populations or extremely small distances from the boundary, these comparisons use finite parameter ranges in which the limiting behavior has not yet been fully reached.

\subsubsection{Tests of the matrix formula over finite parameter ranges}
\label{subsubsec:simulation-matrix-tests}

The stochastic simulations implement the species formation dynamics described in Section~\ref{subsec:species-formation-model}.  Each replicate begins with a clonal population of \(M\) haploid binary genomes.  At every generation, the compatibility graph is constructed from the condition \(q_t^{\alpha\beta}\ge q_{\min}\); first parents are sampled from individuals having at least one compatible partner, second parents are sampled uniformly from the corresponding compatible set, and offspring are produced by free recombination followed by mutation.  Species are identified with connected components of the compatibility graph.

For a given parameter set and trial genome length \(L\), each simulation is run for \(\lceil 3\tau\rceil\) generations, where \(\tau\) is the deterministic threshold crossing time in
Eq.~\eqref{eq:tau-formula}. The first \(\lceil 2\tau\rceil\) generations are treated as a transient period, and \(\overline S_r(L)\) denotes the number of connected components in replicate \(r\), averaged over the final \(\lceil\tau\rceil\) generations. This observation window samples the late dynamics on the same characteristic time scale used by the analytical criterion and determines whether reproductive fragmentation persists over that time scale, without requiring the full species abundance distribution to have reached stationarity.

Because the raw number of connected components depends strongly on
\(M\), \(\mu\), and \(q_{\min}\), it cannot be compared directly across parameter sets. Following the operational normalization introduced by Marquioni and de Aguiar~\citep{MarquioniAguiar2025}, we normalize the late time component count by the following reference richness scale:
\begin{equation}
S_e(M,\mu,q_{\min})
=
M\,\frac{\ee^{4\mu}-1}{q_{\min}^{-1}-1}.
\label{eq:simulation-expected-species}
\end{equation}
To interpret this normalization, let \(M^*\) denote the population size whose unrestricted equilibrium overlap \(q_0\) is exactly the mating threshold \(q_{\min}\). Using Eq.~\eqref{eq:hpm-stationary-overlap}, the condition \(q_0(M^*,\mu)=q_{\min}\) gives
\[
M^*=\frac{q_{\min}^{-1}-1}{\ee^{4\mu}-1},
\]
and therefore \(S_e=M/M^*\). Thus \(M^*\) can be interpreted as the
characteristic component size for which the unrestricted mean pairwise overlap \(q_0\) equals the mating threshold \(q_{\min}\), while \(S_e\) is the corresponding reference number of components in a population of size \(M\). It is a parameter dependent benchmark for the development of a multispecies configuration, rather than an exact prediction of the stationary number of species.

We therefore define
\begin{equation}
R(L)=\frac{\E_{\mathrm{rep}}[\overline S_r(L)]}{S_e},
\qquad
R\!\left(L_c^{\mathrm{sim}}\right)=1,
\label{eq:simulation-critical-crossing}
\end{equation}
where \(\E_{\mathrm{rep}}[\cdot]\) is the arithmetic mean over
independent simulation replicates. Thus \(R(L)<1\) means that the mean component count remains below the reference richness, whereas
\(R(L)\geq1\) means that the simulations sustain at least that
parameter adjusted level of reproductive fragmentation over the
observation window. The simulation estimate \(L_c^{\mathrm{sim}}\) is taken as the genome length at which this dimensionless ratio crosses one. This provides a reproducible operational marker of the onset of fragmentation that can be applied consistently across parameter sets. Because \(S_e\) is a reference scale rather than the exact stationary richness, \(L_c^{\mathrm{sim}}\) should be interpreted as a simulation estimate of the fragmentation threshold, not as a complete characterization of the eventual species abundance distribution. This crossing provides the simulation counterpart of the analytical transient variance threshold. The transient variance criterion predicts when finite genome broadening becomes too small to maintain sufficient compatibility above the mating threshold, whereas Eq.~\eqref{eq:simulation-critical-crossing} records the resulting fragmentation of the compatibility graph in the restricted model. Although the two criteria use different observables, they are intended to locate the same finite genome transition.

The crossing is located by local linear interpolation in \(\log L\) between the neighboring tested lengths for which the mean ratio across replicates lies below and above one.  The initial length grid is centered on \(L_c^{\mat}\) and is expanded automatically when it does not bracket the crossing.  The reported runs use ten pilot replicates per tested length and twenty additional replicates on a refined grid near the crossing.  Confidence intervals of 95\% are obtained from 1000 nonparametric bootstrap samples of the replicate values at each \(L\), with the crossing recomputed in every bootstrap sample.

\begin{figure}[p]
\centering
\includegraphics[width=\textwidth]{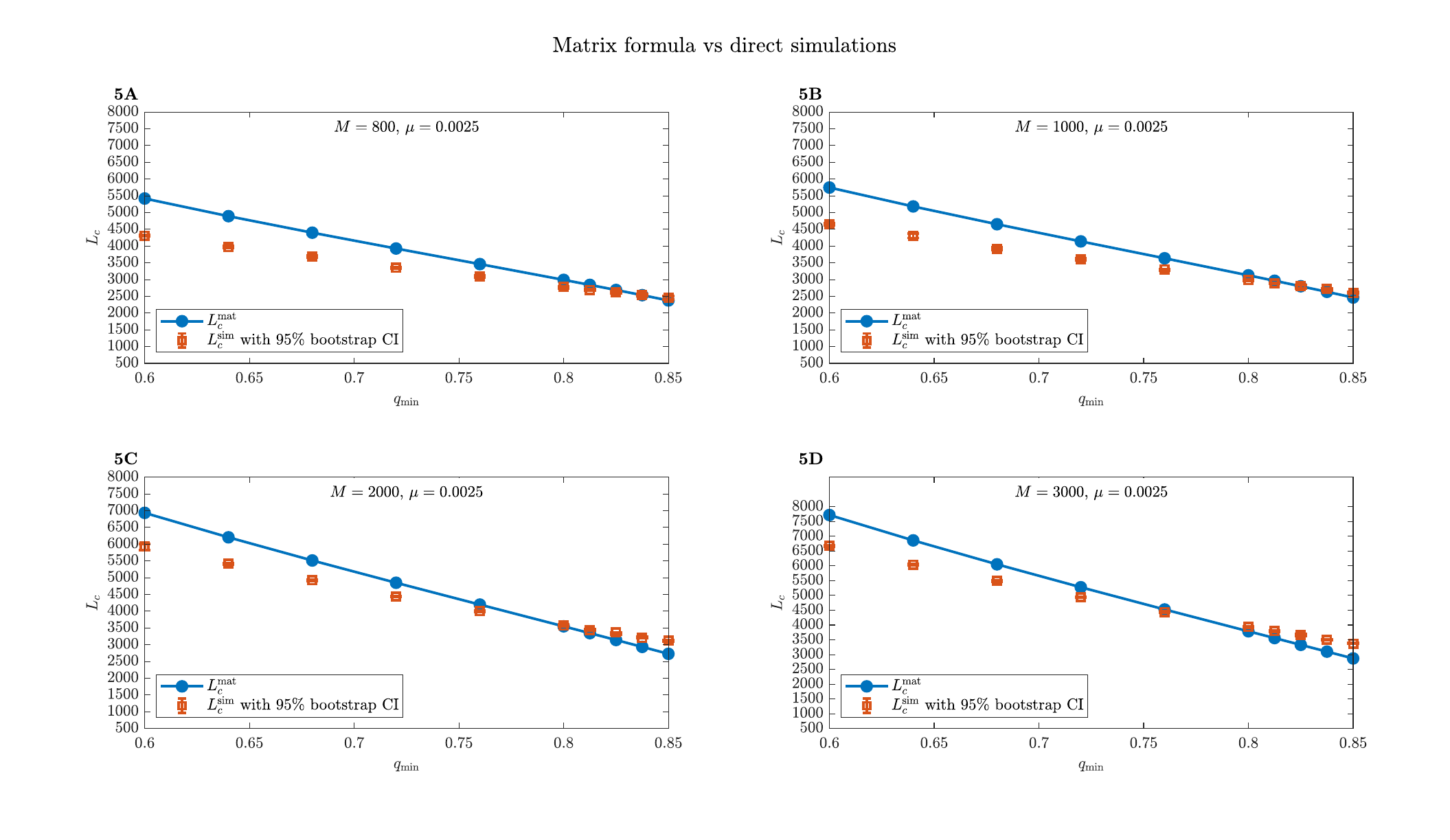}
\caption{Matrix prediction and direct simulation estimates of the critical genome length as functions of the mating threshold.  The mutation rate is fixed at \(\mu=0.0025\), and the panels show (5A) \(M=800\), (5B) \(M=1000\), (5C) \(M=2000\), and (5D) \(M=3000\).  Blue circles connected by lines show the closed form matrix prediction \(L_c^{\mat}\).  Orange squares show the simulation estimate \(L_c^{\mathrm{sim}}\), with vertical bars denoting 95\% bootstrap confidence intervals.  Across the four population sizes, the matrix formula reproduces the scale and the systematic decline of the critical genome length as the mating threshold becomes more restrictive.}
\label{fig:simulation-qmin-sweep}
\end{figure}

Figure~\ref{fig:simulation-qmin-sweep} compares \(L_c^{\mat}\) with \(L_c^{\mathrm{sim}}\) as the mating threshold is varied at fixed \(\mu=0.0025\), for four population sizes.  The matrix formula captures both the decrease of the critical length with increasing \(q_{\min}\) and its change with \(M\).  At the lower thresholds the matrix prediction tends to lie above the simulation estimate.  The discrepancy generally narrows toward intermediate and high thresholds, although the improvement is not monotone: at the upper end of the range the difference can change sign, especially for the larger populations.  Thus the figure does not support a universal claim that accuracy must increase with \(q_{\min}\).  It does show that a formula derived from the transient moments of the unrestricted model tracks the critical genome length of the nonlinear restricted model over a substantial threshold interval.

\begin{figure}[!htbp]
\centering
\includegraphics[width=0.76\textwidth]{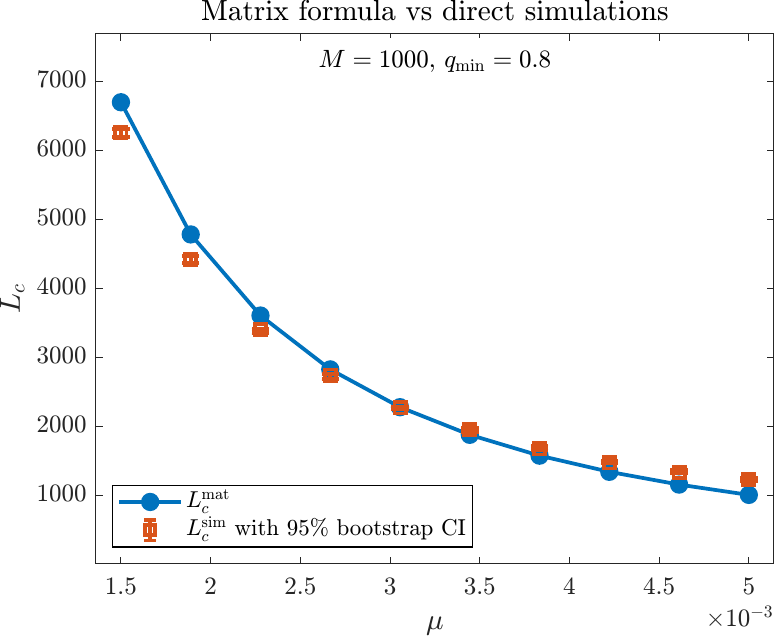}
\caption{Matrix prediction and direct simulation estimates of the critical genome length as the mutation rate varies.  The population size and mating threshold are fixed at \(M=1000\) and \(q_{\min}=0.8\), and \(\mu\) ranges from \(0.0015\) to \(0.005\).  Blue circles and the connecting line show \(L_c^{\mat}\); orange squares and 95\% bootstrap confidence intervals show \(L_c^{\mathrm{sim}}\).  The matrix formula captures the pronounced decrease in the critical genome length across the full range of mutation rates.}
\label{fig:simulation-mu-sweep}
\end{figure}

Figure~\ref{fig:simulation-mu-sweep} provides a complementary test in which \(M=1000\) and \(q_{\min}=0.8\) are fixed while \(\mu\) varies from \(0.0015\) to \(0.005\).  Over this more than threefold range of mutation rates, the critical length changes by more than a factor of five.  The matrix expression follows this strong variation closely.  Moreover, the simulation points pass from slightly below to slightly above the matrix curve, so the remaining difference is not well described by a single multiplicative bias.  Together, Figures~\ref{fig:simulation-qmin-sweep} and \ref{fig:simulation-mu-sweep} indicate that the transient variance construction captures the leading dependence of \(L_c^{\mathrm{sim}}\) on the three model parameters \(M\), \(\mu\), and \(q_{\min}\), even though the restricted species formation model is much more complicated because the evolving compatibility graph continuously feeds back into the choice of parents.

The values of \(\mu\) in Figure~\ref{fig:simulation-mu-sweep} are small on the dimensionless scale of the model, which is the range in which the small mutation rate approximations developed in Section~\ref{sec:asymptotic-regimes} become relevant.  We do not, however, attach a universal biological calibration to either \(\mu\) or \(q_{\min}\): their empirical interpretation depends on the organism, the genomic representation, and the effective time unit.

\FloatBarrier

\subsubsection{Finite size tests of the asymptotic scaling laws}
\label{subsubsec:simulation-asymptotic-tests}

Testing an asymptotic exponent directly is more computationally demanding than the comparisons above.  Increasing \(M\) increases both the number of genomes that must be evolved and the number of pairwise compatibility relations that must be evaluated, while approaching the Regime~V boundary makes the critical genome length itself diverge.  Consequently, the simulations cannot reach arbitrarily large \(M\) or arbitrarily small \(\varepsilon=q_{\min}-q_0\).  We therefore compare three slopes obtained in logarithmic coordinates: the slope obtained by linearly regressing \(\log L_c^{\mathrm{sim}}\) on \(\log M\), or on \(\log\varepsilon\) in Regime~V, over the tested range; the corresponding slope obtained by applying the same regression to \(L_c^{\mat}\) at the same parameter values; and the limiting exponent predicted by the asymptotic analysis.  Agreement between the first two indicates whether the closed form captures the trend of the nonlinear model over the tested finite range.  Proximity to the third indicates how closely the tested parameters approach the predicted asymptotic limit.

\begin{figure}[p]
\centering
\includegraphics[width=0.86\textwidth]{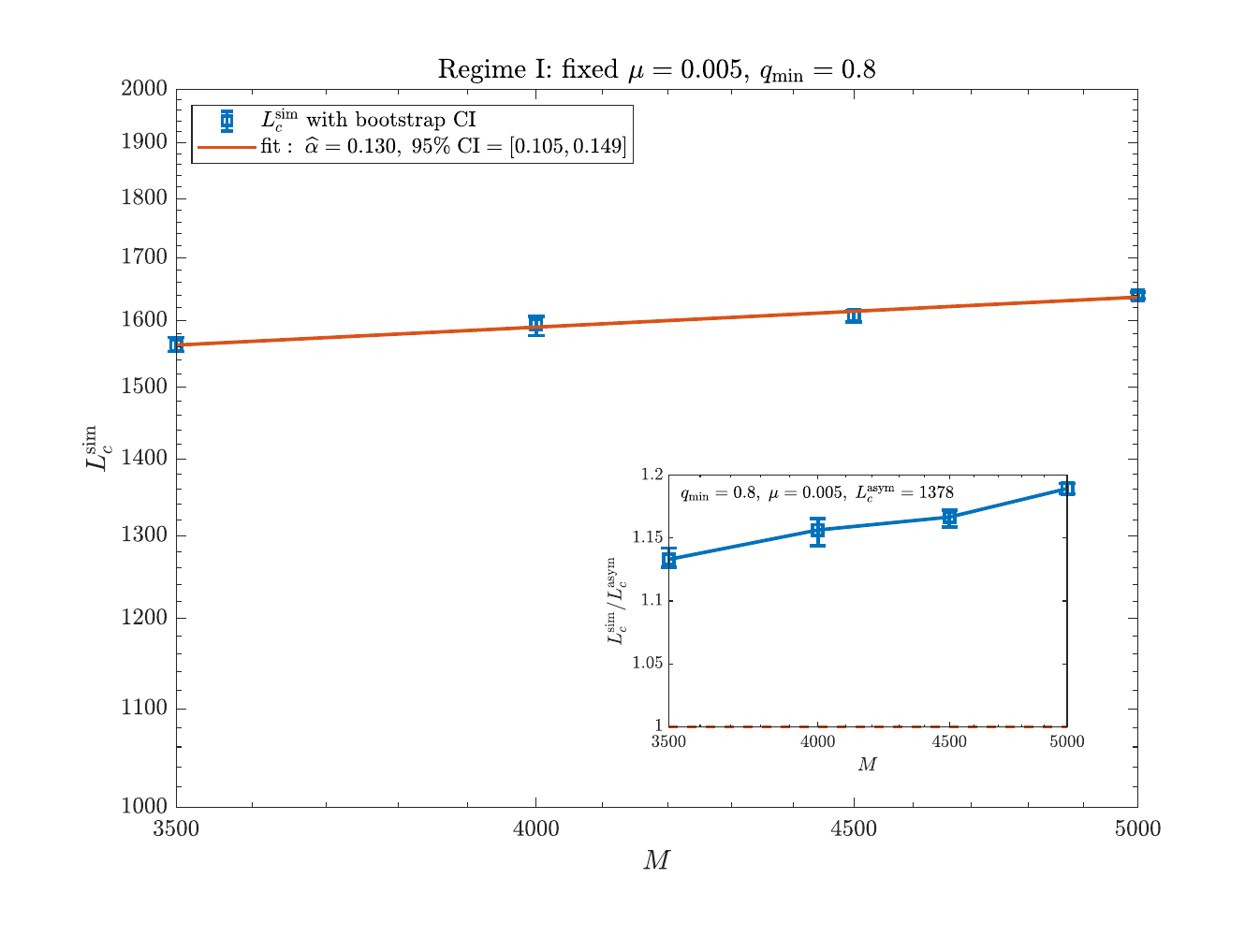}
\caption{Test of Regime~I over a finite population range at fixed \(\mu=0.005\) and \(q_{\min}=0.8\).  Squares show \(L_c^{\mathrm{sim}}\) with 95\% bootstrap confidence intervals for \(M=3500,4000,4500,5000\); the solid line is a power law fit to \(\log L_c^{\mathrm{sim}}\) versus \(\log M\), with \(\widehat\alpha=0.130\) and bootstrap interval \([0.105,0.149]\).  Regime~I predicts the limiting exponent \(\alpha=0\).  The inset shows \(L_c^{\mathrm{sim}}/L_c^{\mathrm{asym}}\), where \(L_c^{\mathrm{asym}}=1378\) is the fixed mutation rate asymptote from Eq.~\eqref{eq:asymp-Lc-fixed-mu}; the dashed line marks unity.  The weak positive slope and ratios above one show that the limiting behavior has not yet been fully reached, while also showing that the dependence on population size is already small.}
\label{fig:simulation-regime-I}
\end{figure}

\paragraph{Regime I: fixed mutation rate.}
Regime~I predicts that at fixed \(\mu>0\), the critical length approaches the constant in Eq.~\eqref{eq:asymp-Lc-fixed-mu}; equivalently, the asymptotic slope of \(\log L_c\) versus \(\log M\) is zero.  Figure~\ref{fig:simulation-regime-I} tests this prediction at \(\mu=0.005\), \(q_{\min}=0.8\), and \(M=3500,4000,4500,5000\).  A linear fit of \(\log L_c^{\mathrm{sim}}\) against \(\log M\) gives
\[
\widehat\alpha_{\mathrm{sim}}=0.130,
\qquad
95\%\ \mathrm{CI}=[0.105,0.149].
\]
Over the same four population sizes, the matrix prediction \(L_c^{\mat}\) has the slope \(\alpha_{\mat}=0.076\) in logarithmic coordinates.  The fitted simulation slope is therefore positive and statistically distinguishable from the limiting value zero over this finite range, but it is small and shows a similarly weak dependence on \(M\) as the matrix prediction over the same range.  The inset gives a second comparison: \(L_c^{\mathrm{sim}}/L_c^{\mathrm{asym}}\) lies approximately between \(1.13\) and \(1.19\), above rather than centered on one, and increases slowly across the displayed range.  These observations show that the tested values of \(M\) have not yet reached either the limiting slope or the asymptotic value.  Nevertheless, the dependence on \(M\) is already weak and close to the predicted limit that is independent of \(M\).  The remaining increase in \(L_c^{\mathrm{sim}}\) over the tested range suggests that still larger populations would be needed for the slope to approach zero.

\begin{figure}[p]
\centering
\includegraphics[width=\textwidth]{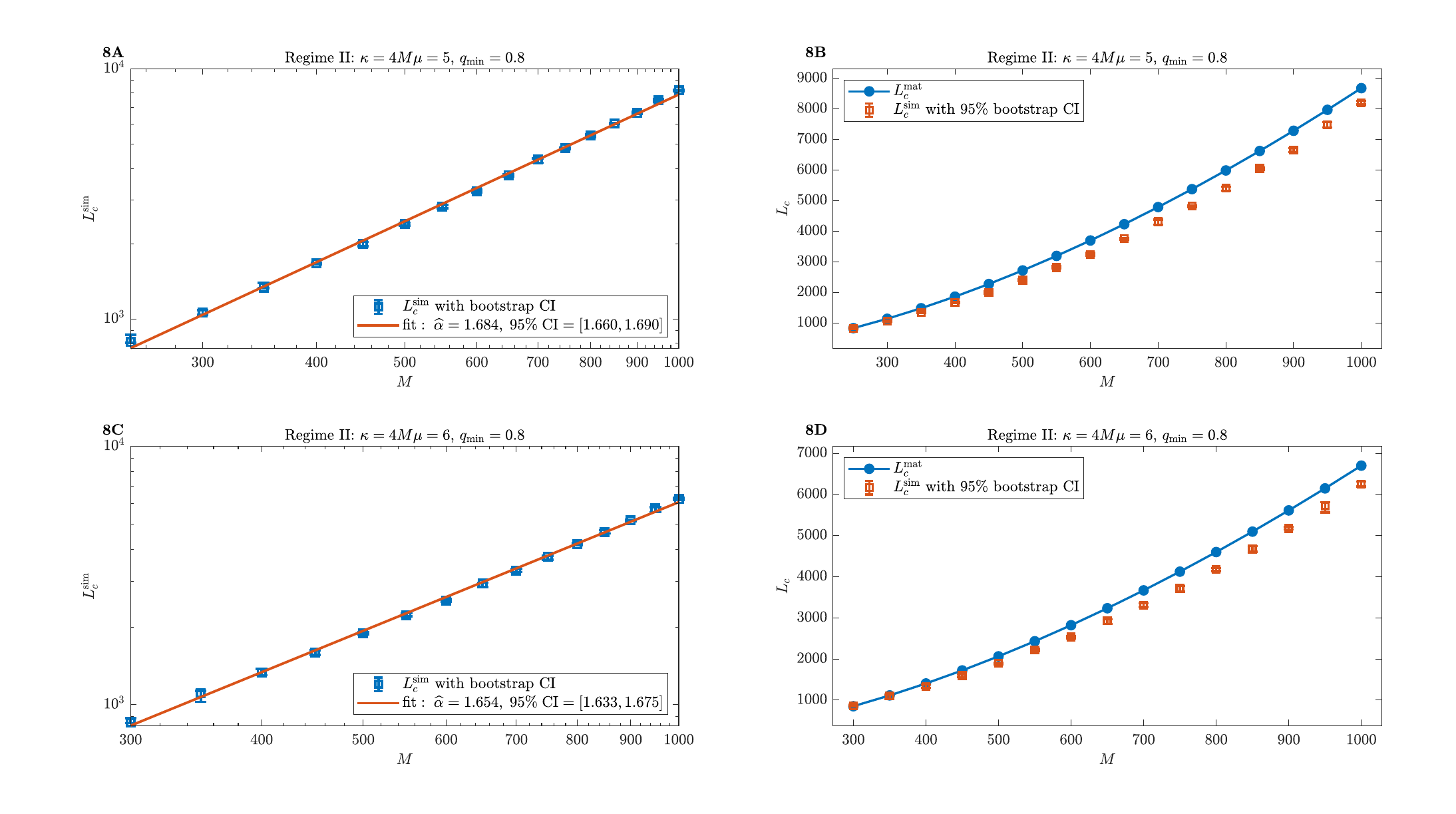}
\caption{Tests of Regime~II over finite population ranges at \(q_{\min}=0.8\).  Panels~8A and 8B use \(\kappa=4M\mu=5\) and \(250\le M\le1000\); panels~8C and 8D use \(\kappa=6\) and \(300\le M\le1000\).  Panels~8A and 8C show \(L_c^{\mathrm{sim}}\) with 95\% bootstrap confidence intervals and power law fits to \(\log L_c^{\mathrm{sim}}\) versus \(\log M\).  The fitted exponents are \(1.684\) with interval \([1.660,1.690]\) for \(\kappa=5\), and \(1.654\) with interval \([1.633,1.675]\) for \(\kappa=6\).  Panels~8B and 8D compare the matrix prediction \(L_c^{\mat}\) at finite \(M\) with \(L_c^{\mathrm{sim}}\) over the same ranges.  The observed exponents are close to the slopes from the matrix expression over the same ranges, \(1.691\) and \(1.710\), and are closer to the \(M^{3/2}\) prediction obtained when the transient genealogical width is included than to the \(M^2\) law obtained when it is omitted.}
\label{fig:simulation-regime-II}
\end{figure}

\paragraph{Regime II: fixed \(\kappa=4M\mu\).}
Regime~II predicts \(L_c\propto M^{3/2}\), in contrast with the \(M^2\) law obtained when the transient genealogical width is omitted.  Figure~\ref{fig:simulation-regime-II} examines two fixed values of \(\kappa\) at \(q_{\min}=0.8\).  For \(\kappa=5\), the range is \(250\le M\le1000\); for \(\kappa=6\), it is \(300\le M\le1000\).  In both cases the lower endpoint corresponds to \(\mu=0.005\), and \(\mu\) decreases as \(M^{-1}\) along the tested path.

The comparison is especially informative because these population ranges are not yet deep in the asymptotic regime.  Regressing \(\log L_c^{\mat}\) on \(\log M\) over exactly these ranges gives slopes \(1.691\) for \(\kappa=5\) and \(1.710\) for \(\kappa=6\).  The simulations give, respectively,
\[
\widehat\alpha_{\mathrm{sim}}=1.684,
\qquad 95\%\ \mathrm{CI}=[1.660,1.690],
\]
and
\[
\widehat\alpha_{\mathrm{sim}}=1.654,
\qquad 95\%\ \mathrm{CI}=[1.633,1.675].
\]
Thus the simulation exponent is nearly identical to the matrix exponent over the same range for \(\kappa=5\), and remains close for \(\kappa=6\).  Both are already much nearer to the prediction \(3/2\), obtained when the transient genealogical width is included, than to the exponent \(2\), obtained when it is omitted.  Panels~8B and 8D provide the corresponding comparison of the numerical values.  Across the two data sets, \(L_c^{\mathrm{sim}}\) is up to about \(12\%\) below \(L_c^{\mat}\), while following the same curvature with \(M\).  Regime~II therefore provides the clearest combined evidence that the transient genealogical width controls both the exponent and much of the numerical value of the critical genome length in the restricted model.

\begin{figure}[p]
\centering
\includegraphics[width=0.86\textwidth]{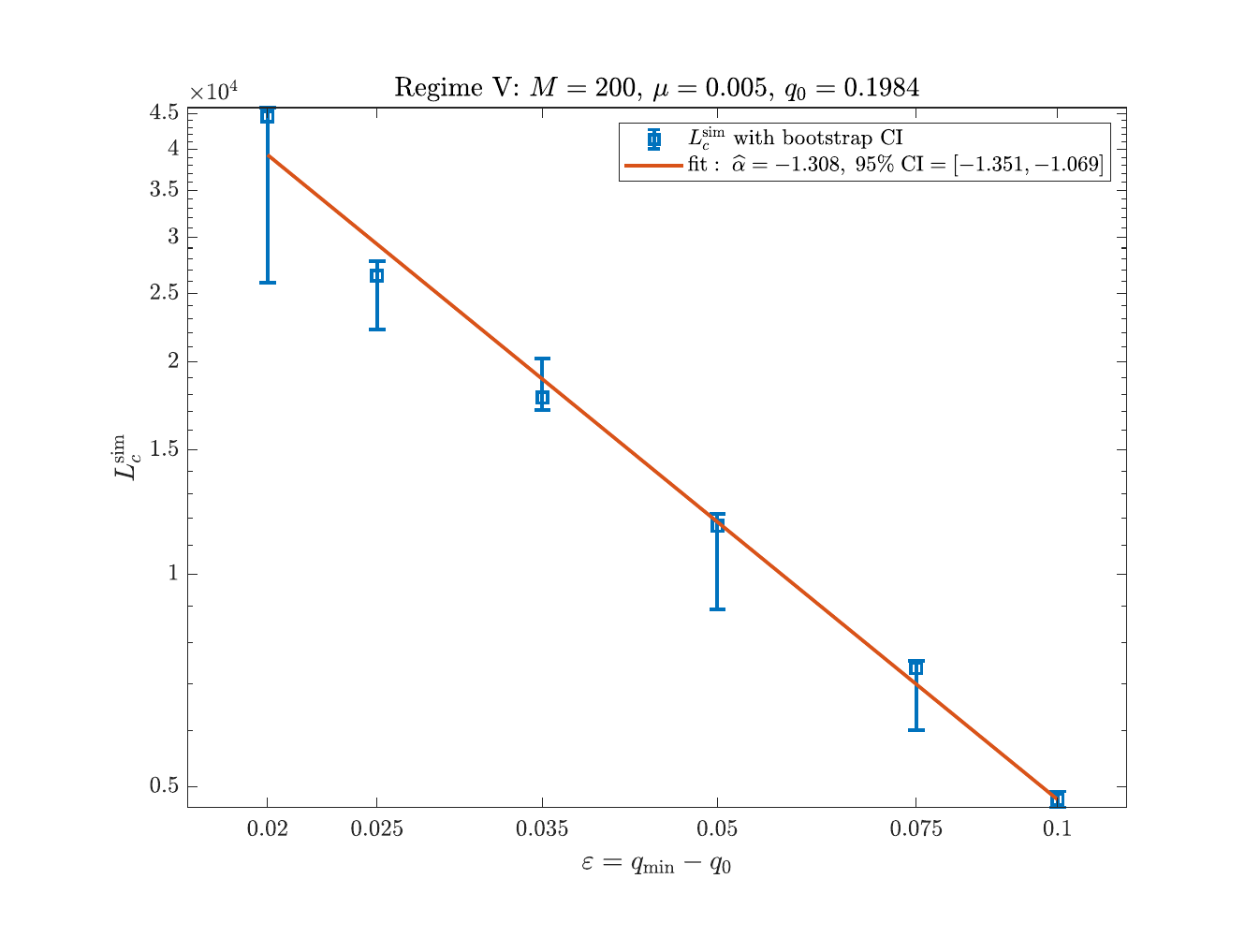}
\caption{Test of Regime~V over a finite range of distances from the boundary.  The population size and mutation rate are fixed at \(M=200\) and \(\mu=0.005\), giving \(q_0=0.1984\), and the threshold is varied through \(\varepsilon=q_{\min}-q_0\).  Squares show \(L_c^{\mathrm{sim}}\) with 95\% bootstrap confidence intervals.  The solid line is a power law fit to \(\log L_c^{\mathrm{sim}}\) versus \(\log\varepsilon\), with exponent \(-1.308\) and bootstrap interval \([-1.351,-1.069]\).  Over the same interval, the slope from the matrix expression is \(-0.981\).  Although the simulation point estimate is steeper than the Regime~V value \(-1\), the upper end of its confidence interval lies close to \(-1\), and the result remains clearly separated from the prediction \(-2\) obtained when the genealogical width is omitted.}
\label{fig:simulation-regime-V}
\end{figure}

\paragraph{Regime V: approach to the infinite genome boundary.}
For the revised boundary test we fix \(M=200\) and \(\mu=0.005\), for which \(q_0=0.198402\), and set \(q_{\min}=q_0+\varepsilon\) using
\[
\varepsilon\in\{0.02,0.025,0.035,0.05,0.075,0.10\}.
\]
Regime~V predicts the linear divergence \(L_c\propto\varepsilon^{-1}\) in Eq.~\eqref{eq:asymp-Lc-boundary-proportional}; omitting the genealogical width would instead give \(\varepsilon^{-2}\).  Regressing \(\log L_c^{\mat}\) on \(\log\varepsilon\) over this interval gives the slope \(-0.981\), already very close to the limiting value \(-1\).  The direct simulations shown in Figure~\ref{fig:simulation-regime-V} give
\[
\widehat\alpha_{\mathrm{sim}}=-1.308,
\qquad 95\%\ \mathrm{CI}=[-1.351,-1.069].
\]
The fitted simulation slope is steeper than both the matrix slope and the asymptotic prediction.  Its confidence interval does not contain \(-1\), but its upper endpoint, \(-1.069\), lies close to that value, and the entire interval remains far from the exponent \(-2\) obtained when the genealogical width is omitted.  The result therefore supports the prediction that the critical genome length depends approximately inversely on the distance from the boundary when the transient genealogical width is included, while also showing that the tested range has not yet produced precise agreement with the limiting exponent.  The comparison of numerical values is more favorable: \(L_c^{\mathrm{sim}}/L_c^{\mat}\) ranges from approximately \(0.77\) to \(1.47\), and the matrix prediction lies within the bootstrap interval at the three smallest tested values of \(\varepsilon\).  The largest relative difference occurs at \(\varepsilon=0.02\), where the estimated critical length is largest and its uncertainty is correspondingly broad.  Smaller values of \(\varepsilon\) would provide a more decisive exponent test, but the divergence of \(L_c\) makes those simulations progressively more expensive.

In summary, the direct simulations support the transient variance theory at several levels.  The matrix expression at finite population size reproduces how the simulated critical genome length changes with \(M\), \(\mu\), and \(q_{\min}\), and in Figures~\ref{fig:simulation-qmin-sweep}, \ref{fig:simulation-mu-sweep}, and \ref{fig:simulation-regime-II} its numerical values are also close to the simulation estimates.  Regime~I shows the predicted weak dependence on population size, although the tested populations are not large enough for the slope in logarithmic coordinates to reach zero or for the ratio to the asymptotic formula to reach one.  Regime~II gives the strongest agreement in scaling: the simulation slopes closely follow the slopes of \(L_c^{\mat}\) over the same parameter ranges and lie near the \(M^{3/2}\) prediction.  In Regime~V, the simulation slope is steeper than the slope from \(L_c^{\mat}\), but its confidence interval extends to values close to \(-1\) and remains far from \(-2\); the matrix estimates are also consistent with the simulation intervals at the smallest tested values of \(\varepsilon\).  The remaining differences show that the available simulations have not fully reached the asymptotic limits and that the practical definition of \(L_c^{\mathrm{sim}}\) introduces additional uncertainty.  The main conclusion is not that the closed form exactly reproduces every simulation result, but that it identifies the main mechanism that determines the critical genome length even in the more complicated restricted model.

\clearpage

\section{Discussion}
\label{sec:discussion}

This paper converts the transient variance criterion for speciation in the finite genome Derrida--Higgs model into an explicit prediction for the critical genome length.  Starting from corrected moment equations for the unrestricted homogeneous population model and an exact decomposition of the overlap variance, we derived a matrix formula for \(L_c\) that displays its dependence on population size \(M\), mutation rate \(\mu\), and mating threshold \(q_{\min}\).  The matrix expression retains the covariance between overlaps that share an individual and is the appropriate result for quantitative calculations at finite \(M\).  The scalar reduction omits this covariance feedback, reproduces the leading behavior for large \(M\), and permits a transparent asymptotic analysis.

The main conceptual result is that the critical genome length is set by two quantities evaluated at the deterministic time \(\tau\) when the unrestricted mean overlap reaches the threshold.  The first is the deterministic one generation decrease
\(\delta q=m_{\tau-1}-m_\tau\) through which the unrestricted mean overlap reaches \(q_{\min}\) from above. As shown in Eq.~\eqref{eq:delta-q-definition}, \(\delta q=(q_{\min}-q_0)/[q_0(M-1)]\), so its magnitude depends on the separation between the mating threshold and the unrestricted equilibrium overlap, together with \(q_0\) and the population size \(M\). The second quantity is the transient genealogical width \(\sqrt{v_\infty(\tau)}\) of the overlap distribution.  A condition based only on the mean, such as \(q_{\min}>q_0\), identifies when fragmentation is possible in the infinite genome limit.  At finite genome length, however, the population can remain connected because the overlap distribution is still broad.  The critical length is the genome size at which the deterministic crossing distance becomes large enough relative to this width for compatible paths between groups to disappear.

The asymptotic analysis shows how this balance changes across parameter ranges.  When the deterministic distance is dominant, \(L_c\) becomes essentially independent of population size and scales as \(\mu^{-2}\).  When the genealogical width is dominant, the critical length grows as \(M^{3/2}\) for \(M\mu=O(1)\) and as \(\sqrt{M}/\mu\) in the range \(M^{-1}\ll\mu\ll M^{-1/2}\).  Near the infinite genome boundary, the width changes the divergence from \((q_{\min}-q_0)^{-2}\), obtained when the width is omitted, to \((q_{\min}-q_0)^{-1}\).  These limits are different reductions of one formula and agree in the parameter regions where their assumptions overlap.

The result closes a gap in the Derrida--Higgs literature.  The original analyses established the infinite genome transition \citep{HiggsDerrida1991,HiggsDerrida1992}, and later simulations showed that finite genome length can prevent fragmentation even when the infinite genome condition is satisfied \citep{Aguiar2017}.  The present theory explains that barrier in terms of the transient distribution of overlaps and provides a formula for its location.  It also supplies a well mixed reference case for extensions in which gene flow is reduced by space, dispersal, limited mating pools, or migration \citep{AguiarEtAl2009,CostaEtAl2019,Nelson2024,CaetanoEtAl2020,ManzoPeliti1994,PrincepeEtAl2022,PrincepeEtAl2024}.  In those settings, the moment equations will change, but the same question remains: how far must the population move beyond the compatibility threshold relative to the width generated by genealogy and finite genomes?

The compatibility graph provides another useful interpretation.  Structural studies of bit string similarity networks identify conditions under which a giant connected component exists \citep{SchneiderZanette2025}.  In the present model the graph changes together with the population, and parent choice depends on its current edges.  The closed form does not replace that dynamic graph by a static random network.  Instead, it predicts when the evolving overlap distribution becomes narrow enough that the graph can separate.  This connection suggests that analytical results from random similarity networks may help describe the structure that appears after the onset predicted here.

The present work extends the heuristic theory of Marquioni and de Aguiar \citep{MarquioniAguiar2025} in three ways.  First, the prediction no longer requires iteration of a coupled system to the crossing generation; it is given directly by an explicit expression.  Second, the underlying moment equations have been rederived without the previous simplifying assumption, which changes the variance and covariance coefficients and permits an exact solution of the reduced matrix system.  Third, the prediction has been tested against stochastic simulations of the restricted mating model.  The matrix formula is almost indistinguishable from the full unrestricted moment hierarchy and follows \(L_c^{\mathrm{sim}}\) over broad ranges of \(q_{\min}\) and \(\mu\).  The simulations also show the predicted weak dependence on \(M\) at fixed \(\mu\), growth close to \(M^{3/2}\) when \(M\mu\) is fixed, and an approximately inverse dependence on \(q_{\min}-q_0\) near the boundary.  The finite simulation ranges do not fully reach every limiting value, especially for the fixed mutation and boundary tests, but they support the mechanisms identified by the analytical theory.

Several limitations should be kept in mind.  The transient variance criterion estimates the onset of fragmentation from the width at one characteristic time.  It does not calculate the full probability distribution of the time at which the compatibility graph separates.   More fundamentally, the resulting prediction of \(L_c\) in the restricted mating model remains heuristic because it is based on the transient variance criterion rather than derived from the full restricted dynamics. A complete analytical characterization of the
evolving empirical overlap distribution under restricted mating remains an important open problem. The simulation estimate \(L_c^{\mathrm{sim}}\) is also operational: it is based on the average number of connected components near the end of the simulation, normalized by a reference richness scale.  Its value can depend on simulation duration, the tested length grid, and the definition used to identify fragmentation.  The reduced analytical system neglects feedback from the covariance between disjoint pairs, which enters at order \(M^{-2}\), and the scalar formula also omits the covariance between pairs that share an individual.  The model itself is neutral, well mixed, haploid, and biallelic, with independent loci and no selection or geographic structure.  The conclusions therefore describe a baseline mechanism rather than a direct model of a particular organism.

Several extensions are natural.  Larger populations and smaller values of \(q_{\min}-q_0\) would test the limiting predictions more deeply.  Alternative simulation definitions could use the probability that the population has fragmented by a specified time or the genome length at which persistent reproductive groups first appear.  Selection, spatially restricted mating, realistic dispersal, diploidy, correlations between loci, limited mating pools, and migration between subpopulations could all be incorporated by rederiving the corresponding moment equations.  These directions connect directly to existing extensions of the Derrida--Higgs framework \citep{HiggsWoodcock1995,AguiarEtAl2009,CaetanoEtAl2020,Nelson2024,PrincepeEtAl2022,PrincepeEtAl2024}.  Finally, because the formula can be inverted, an observed genome length and similarity threshold can be used to identify ranges of \(M\) and \(\mu\) for which the population would lie near the predicted fragmentation boundary.  This inverse use may be especially valuable when the theory is adapted to models with biological calibration.

\bigskip

\noindent Acknowledgments: 

\noindent M.A.M.A. would like to thank FAPESP, grant 2021/14335-0, and CNPq, grant 303814/2023-3 for financial support.

\bigskip

\appendix

\section{Scope and notation}
\label{sec:scope}

In the paper we reduce the finite-genome speciation transition to the evaluation of the infinite-genome genealogical variance $v_\infty(\tau)$ of the unrestricted homogeneous-population model at the deterministic crossing time $\tau$. That evaluation rests on a coupled system of moment recursions. The purpose of this supplement is to derive those recursions in full, to make explicit the combinatorial origin of every coefficient, and to establish the structural point that the system requires three covariance classes rather than two.

\subsection{Genomes, overlaps, and mutation}

The population has fixed size $M$ and genome length $L$. Individual $\alpha$ at generation $t$ carries a haploid biallelic genome
\[
S_t^\alpha=\bigl(s_{t,1}^\alpha,\ldots,s_{t,L}^\alpha\bigr),
\qquad
s_{t,i}^\alpha\in\{-1,+1\}.
\]
The pairwise overlap and four-individual overlap are
\begin{equation}
	q_t^{\alpha\beta}=\frac1L\sum_{i=1}^L s_{t,i}^\alpha s_{t,i}^\beta,
	\qquad
	q_t^{\alpha\beta\gamma\delta}=\frac1L\sum_{i=1}^L
	s_{t,i}^\alpha s_{t,i}^\beta s_{t,i}^\gamma s_{t,i}^\delta.
	\label{eq:overlap-defs}
\end{equation}
Because $(s_{t,i}^\alpha)^2=1$ at every locus, repeated labels collapse:
\begin{equation}
	q_t^{\alpha\alpha}=1,
	\qquad
	q_t^{\alpha\alpha\gamma\delta}=q_t^{\gamma\delta},
	\qquad
	q_t^{\alpha\alpha\gamma\gamma}=1.
	\label{eq:collapse}
\end{equation}
Each locus mutates independently after inheritance. With $\mu$ the mutation rate, write
\begin{equation}
	u=\frac{1-\ee^{-2\mu}}{2},
	\qquad
	\eta=\ee^{-2\mu}=1-2u,
	\qquad
	a=\ee^{-4\mu}=\eta^2,
	\label{eq:mutation-constants}
\end{equation}
so that a copied allele equals its pre-mutation value $s$ with probability $1-u$ and $-s$ with probability $u$, and
\begin{equation}
	\E[s_{\mathrm{after}}\given s_{\mathrm{before}}]=\eta\,s_{\mathrm{before}}.
	\label{eq:single-mutation}
\end{equation}
A copied allele therefore retains expected correlation $\eta$ with its parental allele, and two independently inherited and mutated alleles carry a joint attenuation $a=\eta^2$ (see Section~2 of the main text).

\subsection{Reproduction rule and parent sampling}

Each offspring of generation $t+1$ chooses two distinct parents from generation $t$. For the combinatorial enumeration below, we treat a parent pair as an \emph{ordered} pair $(p_1,p_2)$ with $p_1\ne p_2$, drawn uniformly from the $M(M-1)$ ordered distinct pairs; the ordering is immaterial to any overlap and only simplifies the counting. Distinct offspring choose their parent pairs independently. Consequently, two parents within one offspring's pair are always distinct, but parents chosen by \emph{different} offspring may coincide. These finite-population shared-parent events among different offspring are precisely what generate the genealogical covariances below.

At a fixed locus, an offspring $\alpha$ with parents $(p_1,p_2)$ copies one parental allele with probability $1/2$ each and then mutates. Conditional on the parental alleles,
\begin{equation}
	\E[s_{t+1,i}^\alpha\given p_1,p_2]
	=\frac{\eta}{2}\bigl(s_{t,i}^{p_1}+s_{t,i}^{p_2}\bigr).
	\label{eq:conditional-spin-mean}
\end{equation}
Equation~\eqref{eq:conditional-spin-mean} is the single basic identity from which everything else follows: the factor $1/2$ is recombination, the factor $\eta$ is mutation.

\subsection{The moment quantities}

We work with five population moments. The mean overlap and the four-individual overlap are
\begin{equation}
	m_t=\E[q_t^{\alpha\beta}],
	\qquad
	h_t=\E[q_t^{\alpha\beta\gamma\delta}],
	\label{eq:m-h-def}
\end{equation}
for distinct labels. The three second-moment quantities are, again for distinct labels,
\begin{equation}
	\VL(t)=\Var\!\bigl(q_t^{\alpha\beta}\bigr),
	\quad
	\CL(t)=\Cov\!\bigl(q_t^{\alpha\beta},q_t^{\alpha\gamma}\bigr),
	\quad
	\DL(t)=\Cov\!\bigl(q_t^{\alpha\beta},q_t^{\gamma\delta}\bigr).
	\label{eq:V-C-D-def}
\end{equation}
The subscript $L$ records that these are finite-genome quantities; the corresponding infinite-genome limits are written $v_t,c_t,d_t$ in Section~\ref{sec:infinite-genome}. Equivalently,
\begin{equation}
	\E[(q_t^{\alpha\beta})^2]=m_t^2+\VL(t),
	\quad
	\E[q_t^{\alpha\beta}q_t^{\alpha\gamma}]=m_t^2+\CL(t),
	\quad
	\E[q_t^{\alpha\beta}q_t^{\gamma\delta}]=m_t^2+\DL(t).
	\label{eq:second-moments-decomposed}
\end{equation}
$\VL$ is the variance of a randomly sampled off-diagonal overlap; $\CL$ is the covariance of two overlaps that share exactly one individual; $\DL$ is the covariance of two overlaps that share no individual. A two-variable closure of the kind used previously effectively sets $\DL\equiv0$. Section~\ref{sec:why-d} shows why this is not exact at finite $M$, and the remainder of the supplement keeps $\DL$.

\section{The mean and four-individual recursions}
\label{sec:mean-h}

The variance forcing terms depend on $m_t$ and $h_t$, so we record their recursions first.

\subsection{Mean overlap}

Let $\alpha$ have parents $(A,B)$ and $\beta$ have parents $(C,D)$. Conditional on the four parental genomes and labels, the reproduction and mutation events that build $\alpha$ and $\beta$ are independent, so by~\eqref{eq:conditional-spin-mean},
\begin{align}
	\E[s_{t+1,i}^\alpha s_{t+1,i}^\beta\given A,B,C,D]
	&=\frac{\eta^2}{4}\bigl(s_{t,i}^A+s_{t,i}^B\bigr)\bigl(s_{t,i}^C+s_{t,i}^D\bigr)
	\notag\\
	&=\frac{a}{4}\bigl(s_{t,i}^A+s_{t,i}^B\bigr)\bigl(s_{t,i}^C+s_{t,i}^D\bigr).
\end{align}
Averaging over loci gives the conditional mean overlap
\begin{equation}
	\E[q_{t+1}^{\alpha\beta}\given A,B,C,D]=\frac{a}{4}\,R_{AB,CD},
	\qquad
	R_{AB,CD}=q_t^{AC}+q_t^{AD}+q_t^{BC}+q_t^{BD}.
	\label{eq:conditional-mean-R}
\end{equation}
The object $R_{AB,CD}$ is the sum of the four cross-overlaps between the two parent pairs; it is the basic genealogical variable for the whole derivation. A single cross-overlap satisfies
\begin{equation}
	\E[q_t^{AC}]=\frac1M\cdot1+\Bigl(1-\frac1M\Bigr)m_t,
	\label{eq:cross-overlap-mean}
\end{equation}
because $A=C$ with probability $1/M$ (giving $q_t^{AA}=1$) and otherwise the two are distinct with expected overlap $m_t$. Since $R_{AB,CD}$ contains four such terms,
\begin{equation}
	m_{t+1}=\frac{a}{4}\,\E[R_{AB,CD}]
	=a\left[\frac1M+\Bigl(1-\frac1M\Bigr)m_t\right].
	\label{eq:mean-recursion}
\end{equation}
Its fixed point $q_0$ and relaxation rate $r$ are
\begin{equation}
	q_0=\frac{1}{M\ee^{4\mu}-(M-1)},
	\qquad
	r=a\Bigl(1-\frac1M\Bigr),
	\qquad
	m_t=q_0+(1-q_0)r^t
	\label{eq:q0-r-m}
\end{equation}
for the clonal initial condition $m_0=1$. We use the abbreviation
\begin{equation}
	s_t\equiv\frac1M+\Bigl(1-\frac1M\Bigr)m_t,
	\qquad
	\E[R_{AB,CD}]=4s_t,
	\label{eq:s-def}
\end{equation}
for the expected single cross-overlap.

\subsection{Four-individual overlap}

Let four distinct offspring have parent pairs $(A_1,A_2),\ldots,(D_1,D_2)$. At a fixed locus, conditional independence of the four offspring and $\eta^4=a^2$ give
\begin{equation}
	\E\Bigl[\textstyle\prod_{X\in\{\alpha,\beta,\gamma,\delta\}}s_{t+1,i}^X\,\Big|\,\Pcal_4\Bigr]
	=\frac{a^2}{16}\prod_{X\in\{A,B,C,D\}}\bigl(s_{t,i}^{X_1}+s_{t,i}^{X_2}\bigr).
\end{equation}
To pass from this conditional expression to the unconditional recursion, we average over the random parent-pair choices. By the law of total expectation and linearity,
\[
\E\!\left[
\prod_{X\in\{\alpha,\beta,\gamma,\delta\}}s_{t+1,i}^X
\right]
=
a^2\,\E\!\left[
\frac{1}{16}
\sum_{r_A,r_B,r_C,r_D\in\{1,2\}}
s_{t,i}^{A_{r_A}}s_{t,i}^{B_{r_B}}s_{t,i}^{C_{r_C}}s_{t,i}^{D_{r_D}}
\right].
\]
For any fixed choice of $r_A,r_B,r_C,r_D$, the selected labels
$A_{r_A},B_{r_B},C_{r_C},D_{r_D}$ have the same joint distribution as four independent uniform labels in the generation-$t$ population. Indeed, each parent pair is sampled uniformly from ordered distinct pairs, so although the two parents within a pair are not independent, each coordinate is marginally uniform on $\{1,\ldots,M\}$; because the four offspring parent pairs are sampled independently, one selected coordinate from each pair gives independent uniform labels. Thus, if $U,V,W,Z$ denote independent uniform labels in $\{1,\ldots,M\}$, then averaging over loci gives
\begin{equation}
	h_{t+1}=a^2\,\E[q_t^{UVWZ}].
	\label{eq:h-source}
\end{equation}
The equality pattern of $(U,V,W,Z)$ determines the contribution through the collapse identities~\eqref{eq:collapse}:
\begin{center}
	\begin{tabular}{p{0.36\linewidth}p{0.28\linewidth}p{0.18\linewidth}}
		\toprule
		Pattern of $(U,V,W,Z)$ & Probability & Contribution \\
		\midrule
		all four equal & $1/M^3$ & $1$ \\
		two equal pairs & $3(M-1)/M^3$ & $1$ \\
		three equal, one different & $4(M-1)/M^3$ & $m_t$ \\
		one equal pair, two singletons & $6(M-1)(M-2)/M^3$ & $m_t$ \\
		all four distinct & $(M-1)(M-2)(M-3)/M^3$ & $h_t$ \\
		\bottomrule
	\end{tabular}
\end{center}
Summing and substituting into~\eqref{eq:h-source} gives the four-individual recursion
\begin{equation}
	h_{t+1}=\frac{a^2}{M^3}
	\Bigl[(3M-2)+(M-1)(6M-8)m_t+(M-1)(M-2)(M-3)h_t\Bigr],
	\label{eq:h-recursion}
\end{equation}
with $h_0=1$ for the clonal start. The quantity $h_t$ enters only the finite-locus forcing terms below; it drops out of the infinite-genome limit (see Section~\ref{sec:infinite-genome}).

\section{Why the disjoint-pair covariance $\DL(t)$ cannot be dropped}
\label{sec:why-d}

For random variables $X,Y$ and conditioning information $\mathcal I$, the law of total covariance reads
\begin{equation}
	\Cov(X,Y)=\E\bigl[\Cov(X,Y\given \mathcal I)\bigr]+\Cov\bigl(\E[X\given \mathcal I],\E[Y\given \mathcal I]\bigr).
	\label{eq:total-cov}
\end{equation}
Take $X=q_{t+1}^{\alpha\beta}$ and $Y=q_{t+1}^{\gamma\delta}$ with all four offspring distinct, and let $\mathcal I$ denote the generation-$t$ population together with all parent assignments. The two overlaps are built from disjoint sets of offspring, so conditional on $\mathcal I$ they are independent and the first term vanishes. The second term does not: both conditional means are $\tfrac a4 R$ for genealogical variables $R$ that are functions of the \emph{same} random generation-$t$ population, and they are correlated through finite-population parent-label collisions among the eight parents. Hence $\DL(t)$ is generated and is generically nonzero even when $\DL(0)=0$. Conditional independence is not unconditional independence; this single observation is why the hierarchy needs three covariance classes, as shown below.

The same mechanism also explains why the corrected coupled system below is not obtained by merely appending a $\DL$ equation to the original two-variable $(\VL,\CL)$ system. Once the disjoint-pair covariance $\DL(t)$ is retained, the finite-$M$ parent-collision cases must be treated consistently for all three second-moment quantities, $\VL(t)$, $\CL(t)$, and $\DL(t)$. Consequently, the $\VL$- and $\CL$-recursions themselves change; the $\DL$ equation cannot be added as a separate component while leaving the rest of the system unchanged. The detailed analysis is carried out in Sections~\ref{sec:cond-second-moment}--\ref{sec:disjoint}.

\section{Conditional second moment of one overlap}
\label{sec:cond-second-moment}

Fix offspring $\alpha$ with parents $(A,B)$ and $\beta$ with parents $(C,D)$, and write $X_i=s_{t+1,i}^\alpha s_{t+1,i}^\beta$, so $q_{t+1}^{\alpha\beta}=L^{-1}\sum_i X_i$. Conditioning on $(A,B,C,D)$ and the parental genomes,
\begin{equation}
	\E[(q_{t+1}^{\alpha\beta})^2\given A,B,C,D]
	=\frac1{L^2}\sum_{i}\E[X_i^2\given\cdot]
	+\frac1{L^2}\sum_{i\ne j}\E[X_iX_j\given\cdot].
	\label{eq:q2-split}
\end{equation}
Since $X_i=\pm1$, the diagonal sum equals $1/L$. Distinct loci are conditionally independent, so
\begin{equation}
	\frac1{L^2}\sum_{i\ne j}\E[X_iX_j\given\cdot]
	=\Bigl(\frac1L\sum_i\E[X_i\given\cdot]\Bigr)^2
	-\frac1{L^2}\sum_i\E[X_i\given\cdot]^2.
	\label{eq:offdiag}
\end{equation}
By~\eqref{eq:conditional-mean-R}, the first term is the square of the conditional mean, $(\tfrac a4 R_{AB,CD})^2$. For the correction term, we use $\E[X_i\given\cdot]=\tfrac a4(s_{t,i}^A+s_{t,i}^B)(s_{t,i}^C+s_{t,i}^D)$ together with the identity $(s^u+s^v)^2=2(1+s^us^v)$ for $s^u,s^v\in\{-1,+1\}$:
\begin{align}
	\E[X_i\given\cdot]^2
	&=\frac{a^2}{16}(s_{t,i}^A+s_{t,i}^B)^2(s_{t,i}^C+s_{t,i}^D)^2
	\notag\\
	&=\frac{a^2}{4}\bigl(1+s_{t,i}^As_{t,i}^B\bigr)\bigl(1+s_{t,i}^Cs_{t,i}^D\bigr).
\end{align}
Averaging over loci and using~\eqref{eq:overlap-defs},
\begin{equation}
	\frac1L\sum_i\E[X_i\given\cdot]^2
	=\frac{a^2}{4}\bigl(1+q_t^{AB}+q_t^{CD}+q_t^{ABCD}\bigr).
	\label{eq:diag-correction}
\end{equation}
Collecting~\eqref{eq:q2-split}--\eqref{eq:diag-correction},
\begin{equation}
	\E[(q_{t+1}^{\alpha\beta})^2\given A,B,C,D]
	=\frac1L+\frac{a^2}{16}R_{AB,CD}^2
	-\frac{a^2}{4L}\bigl(1+q_t^{AB}+q_t^{CD}+q_t^{ABCD}\bigr).
	\label{eq:conditional-second-moment}
\end{equation}
This is a key identity used for the variance recursion established in Section~\ref{sec:variance}. The $1/L$ pieces are the finite-locus terms; the $R^2$ piece is the genealogical term.

\section{The variance recursion}
\label{sec:variance}

By the law of total variance,
\begin{equation}
	\VL(t+1)
	=\underbrace{\E\bigl[\Var(q_{t+1}^{\alpha\beta}\given A,B,C,D)\bigr]}_{F_t^{(v)}\ \text{(finite-locus)}}
	+\underbrace{\Var\bigl(\E[q_{t+1}^{\alpha\beta}\given A,B,C,D]\bigr)}_{G_t^{(v)}\ \text{(genealogical)}}.
	\label{eq:total-variance}
\end{equation}

\subsection{Finite-locus forcing \texorpdfstring{$F_t^{(v)}$}{Fv}}

Subtracting $\E^2[q_{t+1}^{\alpha\beta}\given A,B,C,D]=(\tfrac a4R_{AB,CD})^2$ from~\eqref{eq:conditional-second-moment} leaves the conditional variance
\begin{equation}
	\Var(q_{t+1}^{\alpha\beta}\given A,B,C,D)
	=\frac1L-\frac{a^2}{4L}\bigl(1+q_t^{AB}+q_t^{CD}+q_t^{ABCD}\bigr).
	\label{eq:conditional-variance}
\end{equation}
Now we average over parent labels. Since $A\ne B$ and $C\ne D$, $\E[q_t^{AB}]=\E[q_t^{CD}]=m_t$. For the four-individual term, fix the ordered pair $(A,B)$ and draw $(C,D)$ uniformly over the $M(M-1)$ ordered distinct pairs:
\begin{center}
	\begin{tabular}{p{0.40\linewidth}p{0.28\linewidth}p{0.18\linewidth}}
		\toprule
		Relation of $(C,D)$ to $(A,B)$ & Probability & Contribution \\
		\midrule
		$(C,D)=(A,B)$ or $(B,A)$ & $2/[M(M-1)]$ & $1$ \\
		exactly one label shared & $4(M-2)/[M(M-1)]$ & $m_t$ \\
		no label shared & $(M-2)(M-3)/[M(M-1)]$ & $h_t$ \\
		\bottomrule
	\end{tabular}
\end{center}
so that
\begin{equation}
	\E[q_t^{ABCD}]
	=\frac{2}{M(M-1)}+\frac{4(M-2)}{M(M-1)}m_t+\frac{(M-2)(M-3)}{M(M-1)}h_t.
	\label{eq:ABCD-mean}
\end{equation}
Hence
\begin{equation}
	F_t^{(v)}=\frac1L-\frac{a^2}{4L}
	\left[\Bigl(1+\frac{2}{M(M-1)}\Bigr)
	+\Bigl(2+\frac{4(M-2)}{M(M-1)}\Bigr)m_t
	+\frac{(M-2)(M-3)}{M(M-1)}h_t\right].
	\label{eq:Fv}
\end{equation}

\subsection{Genealogical forcing \texorpdfstring{$G_t^{(v)}$}{Gv}: the sixteen edge products}
\label{subsec:Gv}

From~\eqref{eq:conditional-mean-R}, $G_t^{(v)}=\tfrac{a^2}{16}\Var(R_{AB,CD})$. We can view the four cross-overlaps in
\[
R_{AB,CD}=q^{AC}+q^{AD}+q^{BC}+q^{BD}
\]
as the four edges of the complete bipartite graph between $\{A,B\}$ and $\{C,D\}$ (for convenience, we drop the time subscript on overlaps inside the products). Then $R^2$ is a sum of sixteen ordered edge products, falling into three structural groups:
\begin{enumerate}[leftmargin=2em,itemsep=2pt]
	\item \textbf{same-edge} products $q^{AC}q^{AC}$ etc.: $4$ products;
	\item \textbf{adjacent-edge} products sharing one endpoint, e.g.\ $q^{AC}q^{AD}$ or $q^{AC}q^{BC}$: $8$ products;
	\item \textbf{opposite-edge} products sharing no endpoint, e.g.\ $q^{AC}q^{BD}$ or $q^{AD}q^{BC}$: $4$ products.
\end{enumerate}
We evaluate below one representative of each.

\paragraph{Same-edge.}
In $q^{AC}q^{AC}=(q^{AC})^2$ the labels $A$ and $C$ come from different offspring and are independent, so $A=C$ with probability $1/M$, contributing $(q^{AA})^2=1$, and otherwise $A,C$ are distinct with $\E[(q^{AC})^2]=m_t^2+\VL(t)$ by~\eqref{eq:second-moments-decomposed}. Hence
\begin{equation}
	E_{\rm same}=\E[(q^{AC})^2]
	=\frac1M+\frac{M-1}{M}\bigl(m_t^2+\VL(t)\bigr).
	\label{eq:E-same}
\end{equation}

\paragraph{Adjacent-edge.}
In $q^{AC}q^{AD}$ the labels $C\ne D$ (parents of one offspring), while $A$ is independent:
\begin{center}
	\begin{tabular}{p{0.30\linewidth}p{0.22\linewidth}p{0.40\linewidth}}
		\toprule
		Case & Probability & Contribution \\
		\midrule
		$A=C$ & $1/M$ & $q^{CD}$, mean $m_t$ \\
		$A=D$ & $1/M$ & $q^{CD}$, mean $m_t$ \\
		$A\ne C,D$ & $(M-2)/M$ & shared label, mean $m_t^2+\CL(t)$ \\
		\bottomrule
	\end{tabular}
\end{center}
Summing the three cases in the table gives
\begin{equation}
	E_{\rm adj}=\E[q^{AC}q^{AD}]
	=\frac2M m_t+\frac{M-2}{M}\bigl(m_t^2+\CL(t)\bigr).
	\label{eq:E-adj}
\end{equation}

\paragraph{Opposite-edge.}
In $q^{AC}q^{BD}$, we fix $(A,B)$ with $A\ne B$ and draw $(C,D)$ over the $M(M-1)$ ordered distinct pairs. The collision cases are:
\begin{center}
	\begin{tabular}{p{0.34\linewidth}p{0.24\linewidth}p{0.34\linewidth}}
		\toprule
		Case & Probability & Contribution \\
		\midrule
		$C=A,\,D=B$ & $\frac{1}{M(M-1)}$ & $q^{AA}q^{BB}=1$ \\
		$C=B,\,D=A$ & $\frac{1}{M(M-1)}$ & $(q^{AB})^2$, mean $m_t^2+\VL(t)$ \\
		$C=A,\,D\ne A,B$ & $\frac{M-2}{M(M-1)}$ & $q^{BD}$, mean $m_t$ \\
		$D=B,\,C\ne A,B$ & $\frac{M-2}{M(M-1)}$ & $q^{AC}$, mean $m_t$ \\
		$D=A,\,C\ne A,B$ & $\frac{M-2}{M(M-1)}$ & shared label, mean $m_t^2+\CL(t)$ \\
		$C=B,\,D\ne A,B$ & $\frac{M-2}{M(M-1)}$ & shared label, mean $m_t^2+\CL(t)$ \\
		all four distinct & $\frac{(M-2)(M-3)}{M(M-1)}$ & disjoint, mean $m_t^2+\DL(t)$ \\
		\bottomrule
	\end{tabular}
\end{center}
Thus, summing the contributions, we obtain
\begin{align}
	E_{\rm opp}=\E[q^{AC}q^{BD}]
	&=\frac{1+(m_t^2+\VL)+2(M-2)m_t+2(M-2)(m_t^2+\CL)}{M(M-1)}
	\notag\\
	&\quad+\frac{(M-2)(M-3)(m_t^2+\DL)}{M(M-1)}.
	\label{eq:E-opp}
\end{align}
This is where $\DL$ first enters: the ``all four distinct'' opposite-edge case is a disjoint-pair product.

\paragraph{Assembly.}
With $4$ same-edge, $8$ adjacent-edge, and $4$ opposite-edge products, and $\E[R_{AB,CD}]=4s_t$ by~\eqref{eq:s-def},
\begin{equation}
	\Var(R_{AB,CD})=4E_{\rm same}+8E_{\rm adj}+4E_{\rm opp}-16s_t^2.
	\label{eq:VarR-pre}
\end{equation}
Substituting~\eqref{eq:s-def},~\eqref{eq:E-same},~\eqref{eq:E-adj}, and~\eqref{eq:E-opp} and collecting terms, the constant, $m_t$, and $m_t^2$ contributions cancel against $16s_t^2$ apart from a residual proportional to $(1-m_t)^2$ produced by the self-overlap collisions described below, leaving
\begin{equation}
	\Var(R_{AB,CD})
	=\frac{4(M-2)^2}{M^2(M-1)}(1-m_t)^2
	+\frac{4(M^2-2M+2)}{M(M-1)}\VL
	+\frac{8(M-2)}{M-1}\CL
	+\frac{4(M-2)(M-3)}{M(M-1)}\DL.
	\label{eq:VarR-final}
\end{equation}
The inhomogeneous term $\propto(1-m_t)^2$ comes from the self-overlap collisions ($A=C$ in a same-edge product, $C=A,D=B$ in an opposite-edge product), where $q^{AA}=1$ replaces a random distinct-pair overlap. The coefficient of $\VL$ is the sum of two contributions, which is worth isolating because it is an illustration of how the three-class re-enumeration changes the variance coefficients relative to the two-class coefficients of~\citep{MarquioniAguiar2025}. The same-edge products give $4\cdot\tfrac{M-1}{M}$, which is the value a two-class accounting (with $\DL$ set to zero) assigns to the coefficient of $\VL$; the additional single opposite-edge swap $C=B,\,D=A$, which collapses to $(q^{AB})^2$, contributes $\tfrac{4}{M(M-1)}$, so that together
\begin{equation}
	4\,\frac{M-1}{M}+\frac{4}{M(M-1)}
	=\frac{4(M-1)^2+4}{M(M-1)}
	=\frac{4(M^2-2M+2)}{M(M-1)}.
	\label{eq:v-coeff-decomposition}
\end{equation}
The extra opposite-edge term is precisely what the three-class enumeration adds to the two-class value $4(M-1)/M$.

\subsection{The variance recursion}

Adding $F_t^{(v)}$ from~\eqref{eq:Fv} and $G_t^{(v)}=\tfrac{a^2}{16}\Var(R_{AB,CD})$,
\begin{align}
	\VL(t+1)
	&=\frac1L-\frac{a^2}{4L}
	\left[\Bigl(1+\frac{2}{M(M-1)}\Bigr)
	+\Bigl(2+\frac{4(M-2)}{M(M-1)}\Bigr)m_t
	+\frac{(M-2)(M-3)}{M(M-1)}h_t\right]
	\notag\\
	&\quad+\frac{a^2(M-2)^2}{4M^2(M-1)}(1-m_t)^2
	+\frac{a^2(M^2-2M+2)}{4M(M-1)}\VL(t)
	\notag\\
	&\quad+\frac{a^2(M-2)}{2(M-1)}\CL(t)
	+\frac{a^2(M-2)(M-3)}{4M(M-1)}\DL(t).
	\label{eq:V-recursion}
\end{align}

\section{The one-shared-individual covariance recursion}
\label{sec:covariance}

Consider two overlaps sharing offspring $\beta$,
\[
q_{t+1}^{\alpha\beta}\quad\text{and}\quad q_{t+1}^{\beta\gamma},
\qquad
\alpha:(A,B),\ \ \beta:(C,D),\ \ \gamma:(E,F),
\]
with the associated genealogical variables
\[
R_{AB,CD}=q^{AC}+q^{AD}+q^{BC}+q^{BD},
\qquad
R_{CD,EF}=q^{CE}+q^{CF}+q^{DE}+q^{DF}.
\]
By the law of total covariance,
\begin{equation}
	\CL(t+1)
	=\underbrace{\E\bigl[\Cov(q_{t+1}^{\alpha\beta},q_{t+1}^{\beta\gamma}\given\Pcal)\bigr]}_{F_t^{(c)}}
	+\underbrace{\Cov\bigl(\E[q_{t+1}^{\alpha\beta}\given\Pcal],\E[q_{t+1}^{\beta\gamma}\given\Pcal]\bigr)}_{G_t^{(c)}}.
	\label{eq:c-total-cov}
\end{equation}

\subsection{Finite-locus forcing \texorpdfstring{$F_t^{(c)}$}{Fc}}

Write $X_i=s_{t+1,i}^\alpha s_{t+1,i}^\beta$ and $Y_i=s_{t+1,i}^\beta s_{t+1,i}^\gamma$. Because the two overlaps share $\beta$, at a common locus $X_iY_i=s_{t+1,i}^\alpha s_{t+1,i}^\gamma$. Distinct loci are conditionally independent, so only same-locus terms survive in the conditional covariance:
\begin{equation}
	\Cov(q_{t+1}^{\alpha\beta},q_{t+1}^{\beta\gamma}\given\Pcal)
	=\frac1{L^2}\sum_i\Bigl(\E[X_iY_i\given\Pcal]-\E[X_i\given\Pcal]\E[Y_i\given\Pcal]\Bigr).
	\label{eq:c-cond-cov-raw}
\end{equation}
The cross term $\E[X_iY_i\given\Pcal]$ is, locus by locus, the conditional mean of the product $s_{t+1,i}^\alpha s_{t+1,i}^\gamma$ defining the overlap $q^{\alpha\gamma}$ between offspring with parents $(A,B)$ and $(E,F)$; averaged over loci it contributes $\tfrac a4 R_{AB,EF}$, where $R_{AB,EF}=q^{AE}+q^{AF}+q^{BE}+q^{BF}$. For the product of conditional means, using $(s^C+s^D)^2=2(1+s^Cs^D)$ with $s^C,s^D\in\{-1,+1\}$ and applying~\eqref{eq:conditional-spin-mean},
\begin{align}
	\E[X_i\given\Pcal]\E[Y_i\given\Pcal]
	&=\frac{a^2}{16}(s^A+s^B)(s^C+s^D)^2(s^E+s^F)
	\notag\\
	&=\frac{a^2}{8}(s^A+s^B)(1+s^Cs^D)(s^E+s^F),
\end{align}
whose locus average splits into $R_{AB,EF}$ and the four-spin remainder
\[
T_{AB,CD,EF}=q^{ACDE}+q^{ACDF}+q^{BCDE}+q^{BCDF}.
\]
Therefore
\begin{equation}
	\Cov(q_{t+1}^{\alpha\beta},q_{t+1}^{\beta\gamma}\given\Pcal)
	=\frac1L\left[\frac a4 R_{AB,EF}-\frac{a^2}{8}\bigl(R_{AB,EF}+T_{AB,CD,EF}\bigr)\right].
	\label{eq:c-cond-cov}
\end{equation}
Averaging over labels: $\E[R_{AB,EF}]=4s_t$. For a representative term of $T$, hold $C\ne D$ fixed and let $A,E$ be independent:
\begin{center}
	\begin{tabular}{p{0.42\linewidth}p{0.22\linewidth}p{0.18\linewidth}}
		\toprule
		Case for $(A,E)$ relative to $\{C,D\}$ & Probability & Contribution \\
		\midrule
		$(A,E)=(C,D)$ or $(D,C)$ & $2/M^2$ & $1$ \\
		$A,E,C,D$ all distinct & $(M-2)(M-3)/M^2$ & $h_t$ \\
		all other collisions & $(5M-8)/M^2$ & $m_t$ \\
		\bottomrule
	\end{tabular}
\end{center}
The last row collects every assignment of $(A,E)$ in which at least two of the four labels $A,C,D,E$ coincide, other than the two ordered choices $(A,E)=(C,D)$ and $(A,E)=(D,C)$ already counted in the first row; each such collision collapses the four-individual overlap $q^{ACDE}$ to an ordinary pairwise overlap of mean $m_t$, and there are $M^2-2-(M-2)(M-3)=5M-8$ of them. Hence
\begin{equation}
	\E[q^{ACDE}]=\frac{2+(5M-8)m_t+(M-2)(M-3)h_t}{M^2},
	\qquad
	\E[T]=4\,\E[q^{ACDE}].
	\label{eq:ACDE-mean}
\end{equation}
Substituting into~\eqref{eq:c-cond-cov} and simplifying,
\begin{equation}
	F_t^{(c)}=\frac{a}{L}\left[\frac1M+\Bigl(1-\frac1M\Bigr)m_t\right]
	-\frac{a^2}{2L}
	\left[\frac{2}{M^2}+\frac1M
	+\Bigl(1+\frac4M-\frac8{M^2}\Bigr)m_t
	+\Bigl(1-\frac2M\Bigr)\Bigl(1-\frac3M\Bigr)h_t\right].
	\label{eq:Fc}
\end{equation}

\subsection{Genealogical forcing \texorpdfstring{$G_t^{(c)}$}{Gc}}

From~\eqref{eq:conditional-mean-R} and~\eqref{eq:c-total-cov}, we conclude that $G_t^{(c)}=\tfrac{a^2}{16}\Cov(R_{AB,CD},R_{CD,EF})$. We first evaluate the product expectation $\E[R_{AB,CD}R_{CD,EF}]$ and then subtract $\E[R_{AB,CD}]\E[R_{CD,EF}]$. The two genealogical variables share the middle pair $\{C,D\}$. Thus, of the sixteen products between the four left edges and four right edges, $8$ use the same member of $\{C,D\}$ on both sides (\emph{same-middle}, e.g.\ $q^{AC}q^{CE}$) and $8$ use different members (\emph{opposite-middle}, e.g.\ $q^{AC}q^{DE}$).

\paragraph{Same-middle.}
In $q^{AC}q^{CE}$, fix $C$ and let $A,E$ be independent:
\begin{center}
	\begin{tabular}{p{0.34\linewidth}p{0.26\linewidth}p{0.30\linewidth}}
		\toprule
		Case & Probability & Contribution \\
		\midrule
		$A=C,\,E=C$ & $1/M^2$ & $1$ \\
		exactly one of $A,E$ equals $C$ & $2(M-1)/M^2$ & $m_t$ \\
		$A=E\ne C$ & $(M-1)/M^2$ & $m_t^2+\VL$ \\
		$A,E,C$ all distinct & $(M-1)(M-2)/M^2$ & $m_t^2+\CL$ \\
		\bottomrule
	\end{tabular}
\end{center}
Hence,
\begin{equation}
	E_{\rm sm}=\E[q^{AC}q^{CE}]=\frac{1+2(M-1)m_t+(M-1)(m_t^2+\VL)+(M-1)(M-2)(m_t^2+\CL)}{M^2}.
	\label{eq:E-sm}
\end{equation}

\paragraph{Opposite-middle.}
In $q^{AC}q^{DE}$, fix $C\ne D$ and let $A,E$ be independent:
\begin{center}
	\begin{tabular}{p{0.42\linewidth}p{0.22\linewidth}p{0.26\linewidth}}
		\toprule
		Case & Probability & Contribution \\
		\midrule
		$A=C,\,E=D$ & $1/M^2$ & $1$ \\
		$A=C,\,E\ne D$ or $E=D,\,A\ne C$ & $2(M-1)/M^2$ & $m_t$ \\
		$A=D,\,E=C$ & $1/M^2$ & $m_t^2+\VL$ \\
		$A=D$ or $E=C$ or $A=E$, else distinct & $3(M-2)/M^2$ & $m_t^2+\CL$ \\
		$A,C,D,E$ all distinct & $(M-2)(M-3)/M^2$ & $m_t^2+\DL$ \\
		\bottomrule
	\end{tabular}
\end{center}
Hence,
\begin{equation}
	\begin{aligned}
		E_{\rm om}=\E[q^{AC}q^{DE}]
		&=\frac{1}{M^2}\big[1+2(M-1)m_t+(m_t^2+\VL)\\
		&\qquad+3(M-2)(m_t^2+\CL)+(M-2)(M-3)(m_t^2+\DL)\big].
	\end{aligned}
	\label{eq:E-om}
\end{equation}
Again the disjoint class appears through the all-distinct opposite-middle case.

\paragraph{Assembly.}
With $8$ same-middle and $8$ opposite-middle products and $\E[R]=4s_t$ for each factor,
\begin{align}
	\Cov(R_{AB,CD},R_{CD,EF})&=8E_{\rm sm}+8E_{\rm om}-16s_t^2
	\notag\\
	&=\frac{8}{M}\VL+\frac{8(M-2)(M+2)}{M^2}\CL+\frac{8(M-2)(M-3)}{M^2}\DL.
	\label{eq:CovR-shared}
\end{align}

\subsection{The covariance recursion}

Adding $F_t^{(c)}$ and $G_t^{(c)}=\tfrac{a^2}{16}\Cov(R_{AB,CD},R_{CD,EF})$, by~\eqref{eq:c-total-cov} we obtain
\begin{align}
	\CL(t+1)
	&=\frac{a}{L}\left[\frac1M+\Bigl(1-\frac1M\Bigr)m_t\right]
	\notag\\
	&\quad-\frac{a^2}{2L}
	\left[\frac{2}{M^2}+\frac1M
	+\Bigl(1+\frac4M-\frac8{M^2}\Bigr)m_t
	+\Bigl(1-\frac2M\Bigr)\Bigl(1-\frac3M\Bigr)h_t\right]
	\notag\\
	&\quad+\frac{a^2}{2M}\VL(t)
	+\frac{a^2(M^2-4)}{2M^2}\CL(t)
	+\frac{a^2(M-2)(M-3)}{2M^2}\DL(t),
	\label{eq:C-recursion}
\end{align}
using $(M-2)(M+2)=M^2-4$.

\section{The disjoint-pair covariance recursion}
\label{sec:disjoint}

Finally take two overlaps with no offspring in common,
\[
q_{t+1}^{\alpha\beta}\quad\text{and}\quad q_{t+1}^{\gamma\delta},
\qquad
\alpha:(A,B),\ \beta:(C,D),\ \gamma:(E,F),\ \delta:(G,H),
\]
with conditional means $\tfrac a4 R_{AB,CD}$ and $\tfrac a4 R_{EF,GH}$. As established in Section~\ref{sec:why-d}, the offspring pairs are disjoint, so the conditional covariance $\Cov(q_{t+1}^{\alpha\beta},q_{t+1}^{\gamma\delta}\given\Pcal)=0$ and only the genealogical term remains:
\begin{equation}
	\DL(t+1)=\frac{a^2}{16}\Cov(R_{AB,CD},R_{EF,GH}).
	\label{eq:d-start}
\end{equation}
Take a representative product $q^{AC}q^{EG}$; the four labels $A,C,E,G$ are independent. Classifying by whether each pair is a self-pair ($A=C$ or $E=G$, contributing the diagonal value $1$) and, for nonself pairs, whether the two unordered pairs coincide, share one label, or are disjoint:
\begin{center}
	\begin{tabular}{p{0.44\linewidth}p{0.26\linewidth}p{0.16\linewidth}}
		\toprule
		Case & Probability & Contribution \\
		\midrule
		both pairs self-pairs & $1/M^2$ & $1$ \\
		exactly one pair a self-pair & $2(M-1)/M^2$ & $m_t$ \\
		both nonself, same unordered pair & $2(M-1)/M^3$ & $m_t^2+\VL$ \\
		both nonself, share one label & $4(M-1)(M-2)/M^3$ & $m_t^2+\CL$ \\
		both nonself, disjoint & $(M-1)(M-2)(M-3)/M^3$ & $m_t^2+\DL$ \\
		\bottomrule
	\end{tabular}
\end{center}
Subtracting $s_t^2=\E[q^{AC}]\,\E[q^{EG}]$ cancels the constant, $m_t$, and $m_t^2$ pieces (which together reconstitute $s_t^2=[1+(M-1)m_t]^2/M^2$) and leaves only the covariance-class contributions for the representative product $q^{AC}q^{EG}$,
\begin{equation}
	\Cov(q^{AC},q^{EG})
	=\frac{2(M-1)}{M^3}\VL+\frac{4(M-1)(M-2)}{M^3}\CL+\frac{(M-1)(M-2)(M-3)}{M^3}\DL.
	\label{eq:one-disjoint}
\end{equation}
All sixteen products of the two disjoint genealogical variables $R_{AB,CD}$ and $R_{EF,GH}$ have the same covariance, so multiplying~\eqref{eq:one-disjoint} by $16$ and then by $\tfrac{a^2}{16}$ gives the purely genealogical recursion
\begin{equation}
	\DL(t+1)
	=\frac{2a^2(M-1)}{M^3}\VL(t)
	+\frac{4a^2(M-1)(M-2)}{M^3}\CL(t)
	+\frac{a^2(M-1)(M-2)(M-3)}{M^3}\DL(t).
	\label{eq:D-recursion}
\end{equation}
There is no $1/L$ finite-locus term at this order, because the disjoint conditional covariance is exactly zero. Equation~\eqref{eq:D-recursion} makes explicit that $\DL$ is sourced by $\VL$ and $\CL$ and is therefore generated from a clonal start even though $\DL(0)=0$.

\section{The complete finite-genome system}
\label{sec:complete}

Collecting~\eqref{eq:V-recursion},~\eqref{eq:C-recursion}, and~\eqref{eq:D-recursion}, with the deterministic inputs~\eqref{eq:mean-recursion} and~\eqref{eq:h-recursion}, the full finite-genome system of coupled moment equations is
\begin{align}
	m_{t+1}&=a\left[\frac1M+\Bigl(1-\frac1M\Bigr)m_t\right],
	\label{eq:final-m}\\[2pt]
	h_{t+1}&=\frac{a^2}{M^3}\Bigl[(3M-2)+(M-1)(6M-8)m_t+(M-1)(M-2)(M-3)h_t\Bigr],
	\label{eq:final-h}\\[2pt]
	\VL(t+1)&=\frac1L-\frac{a^2}{4L}
	\left[\Bigl(1+\tfrac{2}{M(M-1)}\Bigr)
	+\Bigl(2+\tfrac{4(M-2)}{M(M-1)}\Bigr)m_t
	+\tfrac{(M-2)(M-3)}{M(M-1)}h_t\right]
	\notag\\
	&\quad+\frac{a^2(M-2)^2}{4M^2(M-1)}(1-m_t)^2
	+\frac{a^2(M^2-2M+2)}{4M(M-1)}\VL(t)
	\notag\\
	&\quad+\frac{a^2(M-2)}{2(M-1)}\CL(t)
	+\frac{a^2(M-2)(M-3)}{4M(M-1)}\DL(t),
	\label{eq:final-V}\\[2pt]
	\CL(t+1)&=\frac{a}{L}\left[\frac1M+\Bigl(1-\frac1M\Bigr)m_t\right]
	-\frac{a^2}{2L}
	\left[\tfrac{2}{M^2}+\tfrac1M
	+\Bigl(1+\tfrac4M-\tfrac8{M^2}\Bigr)m_t
	+\Bigl(1-\tfrac2M\Bigr)\Bigl(1-\tfrac3M\Bigr)h_t\right]
	\notag\\
	&\quad+\frac{a^2}{2M}\VL(t)
	+\frac{a^2(M^2-4)}{2M^2}\CL(t)
	+\frac{a^2(M-2)(M-3)}{2M^2}\DL(t),
	\label{eq:final-C}\\[2pt]
	\DL(t+1)&=\frac{2a^2(M-1)}{M^3}\VL(t)
	+\frac{4a^2(M-1)(M-2)}{M^3}\CL(t)
	+\frac{a^2(M-1)(M-2)(M-3)}{M^3}\DL(t).
	\label{eq:final-D}
\end{align}
For a clonal initial population,
\begin{equation}
	m_0=1,\qquad h_0=1,\qquad \VL(0)=\CL(0)=\DL(0)=0.
	\label{eq:final-initial}
\end{equation}
Equations~\eqref{eq:final-V}--\eqref{eq:final-D} are the recursions quoted in the main text; every genealogical coefficient is fixed by the combinatorial enumerations of Sections~\ref{sec:variance}--\ref{sec:disjoint}.

\paragraph{Finite-locus decomposition.}
The system~\eqref{eq:final-m}--\eqref{eq:final-D} is linear in $(\VL,\CL,\DL)$, and the forcing separates into an $L$-independent genealogical part and an explicit $1/L$ finite-locus part. As shown in Section~\ref{subsec:linear-in-L}, the following decompositions hold
\begin{equation}
	\VL(t)=v_\infty(t)+\frac{v_1(t)}{L},
	\qquad
	\CL(t)=c_\infty(t)+\frac{c_1(t)}{L},
	\qquad
	\DL(t)=d_\infty(t)+\frac{d_1(t)}{L},
	\label{eq:three-linear-intro}
\end{equation}
where the genealogical component $v_\infty(t)$ is exactly the infinite-genome limit of Section~\ref{sec:infinite-genome}, the one solved in closed form in the main text. Moreover, at every $t$ the exact variance decomposition~\eqref{eq:three-linear-intro} implies the consistency identity
\begin{equation}
	v_1(t)=1-m_t^2-v_\infty(t),
	\label{eq:v1-identity}
\end{equation}
which is a useful numerical check on any implementation of the full system of coupled moment equations~\eqref{eq:final-m}--\eqref{eq:final-initial}.

\section{Infinite-genome limit}
\label{sec:infinite-genome}

Here we show how to calculate the genealogical term $v_\infty(t)$ from the full system~\eqref{eq:final-m}--\eqref{eq:final-initial}. Define the genealogical limits
\begin{equation}
	v_t\equiv v_\infty(t)=\lim_{L\to\infty}\VL(t),
	\qquad
	c_t\equiv c_\infty(t)=\lim_{L\to\infty}\CL(t),
	\qquad
	d_t\equiv d_\infty(t)=\lim_{L\to\infty}\DL(t).
\end{equation}
Letting $L\to\infty$ in~\eqref{eq:final-V}--\eqref{eq:final-D} removes the finite-locus terms (and with them the dependence on $h_t$), leaving the three-variable genealogical system analyzed in Section~4.1 of the main text
\begin{align}
	v_{t+1}&=\frac{a^2(M-2)^2}{4M^2(M-1)}(1-m_t)^2
	+\frac{a^2(M^2-2M+2)}{4M(M-1)}v_t
	+\frac{a^2(M-2)}{2(M-1)}c_t
	+\frac{a^2(M-2)(M-3)}{4M(M-1)}d_t,
	\label{eq:inf-v}\\
	c_{t+1}&=\frac{a^2}{2M}v_t
	+\frac{a^2(M^2-4)}{2M^2}c_t
	+\frac{a^2(M-2)(M-3)}{2M^2}d_t,
	\label{eq:inf-c}\\
	d_{t+1}&=\frac{2a^2(M-1)}{M^3}v_t
	+\frac{4a^2(M-1)(M-2)}{M^3}c_t
	+\frac{a^2(M-1)(M-2)(M-3)}{M^3}d_t.
	\label{eq:inf-d}
\end{align}
At finite $M$ this triple is the exact infinite-genome genealogical second-moment system. In matrix form, with $\bm z_t=(v_t,c_t,d_t)^\top$,
\begin{equation}
	\bm z_{t+1}=K_M\,\bm z_t+\bm b_t,
	\qquad
	\bm b_t=\frac{a^2(M-2)^2}{4M^2(M-1)}(1-m_t)^2\,\bm e_1,
	\label{eq:matrix-form}
\end{equation}
where $\bm e_1=(1,0,0)^\top$ and
\begin{equation}
	K_M=a^2
	\begin{pmatrix}
		\dfrac{M^2-2M+2}{4M(M-1)} & \dfrac{M-2}{2(M-1)} & \dfrac{(M-2)(M-3)}{4M(M-1)}\\[1.0em]
		\dfrac{1}{2M} & \dfrac{M^2-4}{2M^2} & \dfrac{(M-2)(M-3)}{2M^2}\\[1.0em]
		\dfrac{2(M-1)}{M^3} & \dfrac{4(M-1)(M-2)}{M^3} & \dfrac{(M-1)(M-2)(M-3)}{M^3}
	\end{pmatrix}.
	\label{eq:KM}
\end{equation}
Starting from $\bm z_0=0$, the solution is the forced sum $\bm z_t=\sum_{s=0}^{t-1}K_M^{\,t-1-s}\bm b_s$. Because $1-m_s=(1-q_0)(1-r^s)$, the forcing vector $\bm b_s$ is a linear combination of $1,\,r^s,\,r^{2s}$, so $\bm z_t$ admits a closed spectral form in the eigenpairs of $K_M$. The main text carries out that evaluation; it reduces~\eqref{eq:matrix-form} to the two-variable subsystem in $(v_t,c_t)$ by dropping $d_t$, whose feedback into the variance is of order $M^{-2}$, and solves the resulting $2\times2$ system in closed form to obtain $v_\infty(\tau)$ and hence the critical genome length $L_c^{\mat}$ (Section~4 of main text).

The reduction is justified by power counting from a clonal start. The source in~\eqref{eq:inf-v} is $O(M^{-1})$, so $v_t=O(M^{-1})$. Equation~\eqref{eq:inf-c} then generates $c_t$ through $(a^2/2M)v_t=O(M^{-2})$ against a contraction $a^2(M^2-4)/(2M^2)\to a^2/2<1$, so $c_t=O(M^{-2})$. Equation~\eqref{eq:inf-d} sources $d_t$ from $v_t$ and $c_t$ through coefficients of order $M^{-2}$ and $M^{-1}$ against a self-coupling approaching $a^2<1$ with an $O(M^{-1})$ gap, so $d_t=O(M^{-2})$. The $d_t$ feedback into~\eqref{eq:inf-v} carries an $O(1)$ coefficient, hence enters $v_t$ at order $M^{-2}$, one order below the leading variance. Dropping $d_t$ therefore changes $v_t$, and so the critical length, only at order $M^{-2}$.

\section{The nature of the correction}
\label{sec:correction}

The moment equations for the unrestricted model were first written down by Marquioni and de Aguiar~\citep{MarquioniAguiar2025}. Their reported genealogical variance and covariance coefficients require revision, and the corrected hierarchy is the system~\eqref{eq:final-V}--\eqref{eq:final-D} derived above. The revision is structural rather than a change to one isolated constant; the following paragraphs identify which terms change and the combinatorial reason for each change.

\paragraph{The hierarchy does not close on two classes.}
The second-order overlap hierarchy does not close on the pair $(\VL,\CL)$. As shown in Section~\ref{sec:why-d}, the disjoint-pair covariance $\DL$ is produced by the law of total covariance~\eqref{eq:total-cov}: two overlaps built from disjoint sets of offspring are conditionally independent given the parental generation, so the conditional-covariance term in~\eqref{eq:total-cov} vanishes, yet their conditional means $\tfrac a4 R_{AB,CD}$ and $\tfrac a4 R_{EF,GH}$ are, by~\eqref{eq:conditional-mean-R}, functions of the same random generation-$t$ population and are correlated through finite-population parent-label collisions among the eight parents. The disjoint-pair recursion~\eqref{eq:final-D} makes this explicit: $\DL$ is sourced by $\VL$ and $\CL$, and is therefore nonzero for every $t\ge1$ even from a clonal start with $\DL(0)=0$. A two-variable formulation that sets $\DL\equiv0$ is thus not an exact finite-$M$ closure.

\paragraph{Retaining \texorpdfstring{$\DL$}{D} changes the other coefficients.}
Keeping $\DL$ is not the same as appending a $\DL$ column to an otherwise unchanged two-variable system. The genealogical part of each recursion is built from a second moment of the genealogical variable $R_{AB,CD}$ of~\eqref{eq:conditional-mean-R}: the variance forcing is $G_t^{(v)}=\tfrac{a^2}{16}\Var(R_{AB,CD})$ (Section~\ref{subsec:Gv}), and the shared-individual forcing is $G_t^{(c)}=\tfrac{a^2}{16}\Cov(R_{AB,CD},R_{CD,EF})$ (Section~\ref{sec:covariance}). Once $\DL$ is retained as a third variable, the sixteen edge products that make up $R^2$ in $\Var(R_{AB,CD})$ must be sorted into three structural classes---same-edge, adjacent-edge, and opposite-edge (Section~\ref{subsec:Gv})---rather than two. This re-sorting reassigns finite-$M$ parent-collision cases that a two-class accounting would have folded into the variance or the shared-individual covariance, moving them into the disjoint class and conversely. The coefficient of $\VL$ in~\eqref{eq:VarR-final} shows the mechanism concretely: by~\eqref{eq:v-coeff-decomposition} it is the sum of the four same-edge contributions, $4(M-1)/M$, and the single opposite-edge swap $C=B,\,D=A$ that collapses to $(q^{AB})^2$, contributing $4/[M(M-1)]$; together these give $4(M^2-2M+2)/[M(M-1)]$, not the value $4(M-1)/M$ obtained when the opposite-edge swap is omitted. The corrected coefficient is therefore a consequence of the complete three-class enumeration, not an isolated additive term.

\paragraph{Practical size of the correction.}
The practical consequence for the companion analysis is mild, but the conceptual one is not. The power counting of Section~\ref{sec:infinite-genome} gives $\DL=O(M^{-2})$, so the feedback of $\DL$ enters the variance one order below the leading term in~\eqref{eq:inf-v}--\eqref{eq:inf-d}. The two-variable reduction used in the main text to obtain the closed-form critical genome length is therefore accurate to order $M^{-2}$, and the numerical shift from dropping $\DL$ is below one percent for $M\gtrsim50$. The exact finite-$M$ object, however, is the three-variable genealogical system~\eqref{eq:inf-v}--\eqref{eq:inf-d}; the two-variable matrix method is best understood as its leading large-$M$ reduction, not as the exact finite-$M$ closure.

\section{The finite-locus coefficient \texorpdfstring{$v_1(t)$}{v1(t)} from the coupled system}
\label{sec:finite-locus}

Recall the finite-locus decomposition $\VL(t)=v_\infty(t)+v_1(t)/L$ introduced in~\eqref{eq:three-linear-intro}. Section~\ref{sec:infinite-genome} extracted the infinite-genome variance $v_\infty(t)$ directly from the coupled system~\eqref{eq:final-V}--\eqref{eq:final-D}, by letting $L\to\infty$ and solving the resulting matrix recursion~\eqref{eq:matrix-form}. The finite-locus coefficient $v_1(t)$ can be obtained the same way, from the coupled system itself, rather than through the consistency identity~\eqref{eq:v1-identity}. More specifically, the genealogical terms $v_\infty(t)$, $c_\infty(t)$, and $d_\infty(t)$ are determined by the system of coupled equations in Section~\ref{sec:infinite-genome}; the finite-locus terms $v_1(t)$, $c_1(t)$, and $d_1(t)$ can similarly be determined by a closed three-variable recursion, driven by the explicit finite-locus forcing and propagated by the \emph{same} genealogical matrix $K_M$. This section derives that recursion directly from the full coupled system. The identity~\eqref{eq:v1-identity} then re-emerges as an algebraic consequence, and so becomes a check on every coefficient of the full system of moment equations rather than an input.

Throughout, we write the infinite-genome limits of Section~\ref{sec:infinite-genome} with an $\infty$ subscript,
\begin{equation}
	\bm z_t=\bigl(v_\infty(t),c_\infty(t),d_\infty(t)\bigr)^\top,
	\label{eq:zinf-recall}
\end{equation}
which is the vector $(v_t,c_t,d_t)^\top$ of~\eqref{eq:matrix-form}. This notation distinguishes the limiting quantities from the finite-$L$ covariance vector $\bm Z_t=(\VL(t),\CL(t),\DL(t))^\top$ and from the finite-locus coefficients introduced below.

\subsection{The covariance classes are exactly linear in \texorpdfstring{$1/L$}{1/L}}
\label{subsec:linear-in-L}

Each genome-level overlap is an average over $L$ loci. With $X_i^{\alpha\beta}(t)=s_{t,i}^{\alpha}s_{t,i}^{\beta}\in\{-1,+1\}$,
\begin{equation}
	q_t^{\alpha\beta}=\frac1L\sum_{i=1}^{L}X_i^{\alpha\beta}(t),
	\qquad
	\E\bigl[X_i^{\alpha\beta}(t)\bigr]=m_t .
	\label{eq:locus-average}
\end{equation}
The mutation rule acts identically and independently at every locus, and the reproduction rule treats all loci alike, so the locus indicators are exchangeable. Consequently, in the double sum over locus pairs, the covariance has one value when the two factors use the same locus and another value when they use two different loci,
\begin{equation}
	\Cov\bigl(q_t^{\alpha\beta},q_t^{\gamma\delta}\bigr)
	=\frac1{L^2}\sum_{i=1}^{L}\sum_{j=1}^{L}\Cov\bigl(X_i^{\alpha\beta},X_j^{\gamma\delta}\bigr)
	=\kappa_{\mathrm{diff}}+\frac{\kappa_{\mathrm{same}}-\kappa_{\mathrm{diff}}}{L},
	\label{eq:cov-linear}
\end{equation}
where $\kappa_{\mathrm{same}}=\Cov(X_i^{\alpha\beta},X_i^{\gamma\delta})$ collects the $L$ same-locus terms and $\kappa_{\mathrm{diff}}=\Cov(X_i^{\alpha\beta},X_j^{\gamma\delta})$, $i\ne j$, collects the $L(L-1)$ different-locus terms. Both values are independent of $L$: same-locus covariances depend only on the single-locus probability distribution, and different-locus covariances are fixed by the shared genealogy, under which distinct loci are conditionally independent. The deterministic inputs $m_t$ and $h_t$ obey the $L$-independent recursions~\eqref{eq:final-m}--\eqref{eq:final-h}; hence the explicit finite-locus forcing terms also have only the displayed $1/L$ dependence.

Equation~\eqref{eq:cov-linear} is exact, not asymptotic: each covariance class has the form ``constant term plus coefficient divided by $L$,'' with both coefficients independent of $L$. Applied to the three classes,
\begin{equation}
	\VL(t)=v_\infty(t)+\frac{v_1(t)}{L},
	\qquad
	\CL(t)=c_\infty(t)+\frac{c_1(t)}{L},
	\qquad
	\DL(t)=d_\infty(t)+\frac{d_1(t)}{L},
	\label{eq:three-linear}
\end{equation}
with the limits $v_\infty(t),c_\infty(t),d_\infty(t)$ of~\eqref{eq:zinf-recall} as the constant terms. For the variance the same-locus value is explicit: since $X_i^{\alpha\beta}=\pm1$ with mean $m_t$, one has $\kappa_{\mathrm{same}}=\Var(X_i^{\alpha\beta})=1-m_t^2$, while $\kappa_{\mathrm{diff}}$ is by definition the infinite-genome variance $v_\infty(t)$. Reading off the $1/L$ coefficient in~\eqref{eq:cov-linear},
\begin{equation}
	v_1(t)=1-m_t^2-v_\infty(t),
	\label{eq:v1-identity-rederived}
\end{equation}
which recovers~\eqref{eq:v1-identity}. The remainder of this section obtains the same $v_1(t)$ independently, as the first component of the $1/L$ part of the coupled system.

\subsection{Separating the coupled system by powers of \texorpdfstring{$1/L$}{1/L}}
\label{subsec:separation}

The finite-locus forcings~\eqref{eq:Fv} and~\eqref{eq:Fc}, included in the variance and one-shared-individual covariance recursions~\eqref{eq:final-V}--\eqref{eq:final-C}, carry the genome length only through an overall factor $1/L$. We remove this factor by setting
\begin{equation}
	F_t^{(v)}=\frac{\phi_t^{(v)}}{L},
	\qquad
	F_t^{(c)}=\frac{\phi_t^{(c)}}{L},
	\label{eq:phi-def}
\end{equation}
so that, from~\eqref{eq:Fv} and~\eqref{eq:Fc},
\begin{align}
	\phi_t^{(v)}
	&=1-\frac{a^2}{4}
	\left[\Bigl(1+\tfrac{2}{M(M-1)}\Bigr)
	+\Bigl(2+\tfrac{4(M-2)}{M(M-1)}\Bigr)m_t
	+\tfrac{(M-2)(M-3)}{M(M-1)}h_t\right],
	\label{eq:phiv}\\[2pt]
	\phi_t^{(c)}
	&=a\left[\frac1M+\Bigl(1-\frac1M\Bigr)m_t\right]
	-\frac{a^2}{2}
	\left[\tfrac{2}{M^2}+\tfrac1M
	+\Bigl(1+\tfrac4M-\tfrac8{M^2}\Bigr)m_t
	+\Bigl(1-\tfrac2M\Bigr)\Bigl(1-\tfrac3M\Bigr)h_t\right].
	\label{eq:phic}
\end{align}
Collecting the variance, one-shared-individual, and disjoint-pair recursions~\eqref{eq:final-V}--\eqref{eq:final-D} into the vector $\bm Z_t=(\VL(t),\CL(t),\DL(t))^\top$, the corrected finite-genome covariance system is
\begin{equation}
	\bm Z_{t+1}=K_M\,\bm Z_t+\bm b_t+\frac1L\,\bm\phi_t,
	\qquad
	\bm\phi_t=\bigl(\phi_t^{(v)},\phi_t^{(c)},0\bigr)^\top,
	\label{eq:Z-vector-system}
\end{equation}
with exactly the propagation matrix $K_M$ of~\eqref{eq:KM} and the genealogical forcing $\bm b_t=\tfrac{a^2(M-2)^2}{4M^2(M-1)}(1-m_t)^2\,\bm e_1$ of~\eqref{eq:matrix-form}. The disjoint-pair equation for $\DL(t)$ has no explicit finite-locus forcing, so the third component of $\bm\phi_t$ is zero. The matrix $K_M$ and the vector $\bm b_t$ are independent of $L$; the genome length enters~\eqref{eq:Z-vector-system} only through the displayed $1/L$.

Insert the exact decomposition~\eqref{eq:three-linear}, written in vector form as
\begin{equation}
	\bm Z_t=\bm z_t+\frac1L\,\bm z_t^{(1)},
	\qquad
	\bm z_t=\bigl(v_\infty(t),c_\infty(t),d_\infty(t)\bigr)^\top,
	\qquad
	\bm z_t^{(1)}=\bigl(v_1(t),c_1(t),d_1(t)\bigr)^\top,
	\label{eq:Zsplit}
\end{equation}
into~\eqref{eq:Z-vector-system}. After this substitution, both sides of~\eqref{eq:Z-vector-system} have the form $A_t+B_t/L$ with $L$-independent coefficients, by Section~\ref{subsec:linear-in-L}. Therefore the $L^0$ and $L^{-1}$ parts must agree separately. This gives two recursions sharing the propagator $K_M$,
\begin{align}
	\bm z_{t+1}&=K_M\,\bm z_t+\bm b_t,
	\label{eq:z-order0}\\[2pt]
	\bm z_{t+1}^{(1)}&=K_M\,\bm z_t^{(1)}+\bm\phi_t .
	\label{eq:z1-order1}
\end{align}
The $L^0$ equation~\eqref{eq:z-order0} is precisely the infinite-genome system~\eqref{eq:matrix-form} of Section~\ref{sec:infinite-genome}. The $L^{-1}$ equation~\eqref{eq:z1-order1} is the new object: it is not a separate approximation but the coefficient of $1/L$ in the full coupled recursion. The clonal initial condition~\eqref{eq:final-initial} holds at every $L$, so
\begin{equation}
	\bm z_0=\bm 0,
	\qquad
	\bm z_0^{(1)}=\bm 0 .
	\label{eq:z1-initial}
\end{equation}

Writing out the three components of~\eqref{eq:z1-order1} gives the finite-locus coefficient recursion in full,
\begin{align}
	v_1(t+1)&=\phi_t^{(v)}
	+\frac{a^2(M^2-2M+2)}{4M(M-1)}v_1(t)
	+\frac{a^2(M-2)}{2(M-1)}c_1(t)
	+\frac{a^2(M-2)(M-3)}{4M(M-1)}d_1(t),
	\label{eq:v1-recursion}\\[2pt]
	c_1(t+1)&=\phi_t^{(c)}
	+\frac{a^2}{2M}v_1(t)
	+\frac{a^2(M^2-4)}{2M^2}c_1(t)
	+\frac{a^2(M-2)(M-3)}{2M^2}d_1(t),
	\label{eq:c1-recursion}\\[2pt]
	d_1(t+1)&=\frac{2a^2(M-1)}{M^3}v_1(t)
	+\frac{4a^2(M-1)(M-2)}{M^3}c_1(t)
	+\frac{a^2(M-1)(M-2)(M-3)}{M^3}d_1(t).
	\label{eq:d1-recursion}
\end{align}
The variance coefficient $v_1(t)$ in the variance decomposition~\eqref{eq:three-linear} is the first component,
\begin{equation}
	v_1(t)=\bm e_1^\top\bm z_t^{(1)} .
	\label{eq:v1-extract}
\end{equation}
The coupling to $c_1$ and $d_1$ is essential even though only $v_1$ is wanted: the off-diagonal entries of $K_M$ feed the disjoint-pair and one-shared-individual coefficients back into the variance coefficient, exactly as for the infinite-genome variance in Section~\ref{sec:infinite-genome}.

By evolving the recursions~\eqref{eq:v1-recursion}--\eqref{eq:d1-recursion}, together with the two deterministic recursions for $m_t$ and $h_t$ in~\eqref{eq:final-m}--\eqref{eq:final-h}, and then extracting the first component using~\eqref{eq:v1-extract}, one obtains $v_1(\tau)$ at the crossing time $\tau$. When $\tau$ is non-integer, this evaluation uses the same continuous spectral interpolation used in the main text. The value of $v_1(\tau)$, together with $v_\infty(\tau)$, supplies the finite-locus contribution needed to determine the critical genome length. The recursions~\eqref{eq:v1-recursion}--\eqref{eq:d1-recursion}, with the extraction~\eqref{eq:v1-extract}, can be solved in closed form by applying the spectral-decomposition method of Section~4.3 of the main text; for brevity, that calculation is not reproduced here.

\subsection{Consistency with the direct identity}
\label{subsec:consistency}

There are now two routes to calculate $v_1(t)$. The first approach uses the locus-average argument of Section~\ref{subsec:linear-in-L}, which gives $v_1(t)$ directly from~\eqref{eq:v1-identity-rederived} as $v_1(t)=1-m_t^2-v_\infty(t)$. The second method uses the coupled finite-locus recursions~\eqref{eq:v1-recursion}--\eqref{eq:d1-recursion}, together with the extraction~\eqref{eq:v1-extract}, which gives $v_1(t)=\bm e_1^\top\bm z_t^{(1)}$. The two agree identically. If $\bm z_t$ solves~\eqref{eq:z-order0} and $\bm z_t^{(1)}$ solves~\eqref{eq:z1-order1}, then $\bm z_t+\tfrac1L\bm z_t^{(1)}$ solves the full finite-genome recursion~\eqref{eq:Z-vector-system} with the clonal initial condition, and by uniqueness of the linear recursion this is the covariance vector $\bm Z_t$ obtained from~\eqref{eq:final-V}--\eqref{eq:final-D}. Both expressions for $v_1(t)$ are therefore the coefficient of $1/L$ in the same finite-genome variance $\VL(t)$, and must coincide.

At the deterministic crossing time $\tau$ of the main text, where the genealogical variance $v_\infty(\tau)$ and the critical genome length are evaluated, the identity reads
\begin{equation}
	v_1(\tau)=1-m_\tau^2-v_\infty(\tau),
	\qquad
	m_\tau=q_0+(1-q_0)r^\tau .
	\label{eq:v1-crossing}
\end{equation}
Its right-hand side uses only the scalar quantities $m_\tau$ and $v_\infty(\tau)$ already computed in Section~\ref{sec:infinite-genome}. Equating it with the recursive value~\eqref{eq:v1-extract} thus provides a stringent algebraic check on the full system of coupled moment equations.

\section{Summary}
\label{sec:summary}

The full moment hierarchy for the unrestricted homogeneous-population Derrida--Higgs model couples five quantities: the deterministic quantities $m_t$ and $h_t$, and the three finite-genome second-moment quantities $\VL(t)$, $\CL(t)$, and $\DL(t)$. The deterministic quantities obey~\eqref{eq:final-m} and~\eqref{eq:final-h}, while the finite-genome second-moment quantities obey~\eqref{eq:final-V}--\eqref{eq:final-D}. All coefficients in these coupled equations are obtained from the combinatorial analysis in Sections~\ref{sec:mean-h} and~\ref{sec:cond-second-moment}--\ref{sec:disjoint}.

In the infinite-genome limit, the hierarchy reduces to the three-variable genealogical recursion~\eqref{eq:matrix-form}. Solving this recursion gives the genealogical coefficient $v_\infty(\tau)$ at the deterministic crossing time $\tau$. We then developed two methods for determining the finite-locus coefficient $v_1(\tau)$, and showed that the two methods agree identically. Together, the coefficients $v_\infty(\tau)$ and $v_1(\tau)$ determine the critical genome length $L_c$.

In the main text, we develop a large-$M$ reduction of the three-variable genealogical recursion~\eqref{eq:matrix-form}. This reduction provides closed-form expressions for $v_\infty(\tau)$, $v_1(\tau)$, and the resulting critical genome length $L_c$.

\bibliographystyle{unsrtnat}
\bibliography{transient_variance_references_ver83}

\end{document}